\documentclass{acmtrans2f}
\usepackage{proof}

\firstfoot{ACM Transactions on Computational Logic,
  Vol.\ TBD, No.\ TBD, TBD TBD, Pages \pages.}

\runningfoot{ACM Transactions on Computational Logic,
  Vol.\ TBD, No.\ TBD, TBD TBD.}

\title{Reasoning with Higher-Order Abstract Syntax in a\\ Logical Framework}

\author{RAYMOND C. MCDOWELL\\ Kalamazoo College 
   \and DALE A. MILLER     \\ Pennsylvania State University}

\newtheorem{theorem}{Theorem}[section]
\newtheorem{corollary}[theorem]{Corollary}
\newtheorem{proposition}[theorem]{Proposition}

\newdef{definition}[theorem]{Definition}   
\newdef{example}[theorem]{Example}         

\newcommand{\N}{{\rm I} \! {\rm N}}
\def\FOLDN{FO\lambda^{\Delta\N}}
\newcommand{\oo}{\hbox{\sl o}}
\newcommand{\nt}{\hbox{\sl nt}}
\newcommand\nat[1]{\hbox{\sl nat} \; #1}
\newcommand\z{\hbox{\sl z}}
\newcommand\suc[1]{\hbox{\sl s} \; #1}
\newcommand\defeq{\stackrel{\scriptscriptstyle\triangle}{=}}
\newcommand\oimp{\supset}
\newcommand\boimp{\subset}
\newcommand\mc{\hbox{\sl cut}}
\newcommand\init{\hbox{\sl init}}
\newcommand\topR{\top{\cal R}}
\newcommand\botL{\bot{\cal L}}
\newcommand\landR{\land{\cal R}}
\newcommand\landL{\land{\cal L}}
\newcommand\lorR{\lor{\cal R}}
\newcommand\lorL{\lor{\cal L}}
\newcommand\oimpR{\oimp\!{\cal R}}
\newcommand\oimpL{\oimp\!{\cal L}}
\newcommand\forallR{\forall{\cal R}}
\newcommand\forallL{\forall{\cal L}}
\newcommand\existsR{\exists{\cal R}}
\newcommand\existsL{\exists{\cal L}}
\newcommand\defR{\hbox{\sl def}{\cal R}}
\newcommand\defL{\hbox{\sl def}{\cal L}}
\newcommand\natR{\hbox{\sl nat}{\cal R}}
\newcommand\natL{\hbox{\sl nat}{\cal L}}
\newcommand\cL{\hbox{\sl c}{\cal L}}

\newcommand\Seq[2]{#1\longrightarrow #2}

\newcommand{\level}[1]{{\rm lvl}(#1)}

\def\Dscr{{\cal D}}
\def\Pscr{{\cal P}}

\def\Dnat{\Dscr({\it nat})}
\newcommand\sump[3]{\hbox{\sl sum} \; #1 \; #2 \; #3}

\newcommand{\Dlist}[1]{\Dscr({\it list}(#1))}
\newcommand{\lst}{\hbox{\sl lst}}
\newcommand{\nil}{\hbox{\sl nil}}
\newcommand{\cons}[2]{#1 \! :: \! #2}
\newcommand{\length}[2]{\hbox{\sl length} \; #1 \; #2}
\newcommand{\tlist}[1]{\hbox{\sl list} \; #1}
\newcommand{\element}[2]{\hbox{\sl element} \; #1 \; #2}

\newcommand{\split}[3]{\hbox{\sl split} \; #1 \; #2 \; #3}
\newcommand{\permute}[3]{\hbox{\sl permute} \; #1 \; #2 \; #3}

\newcommand{\Dlistl}[1]{\Dscr({\it list}^*(#1))}
\newcommand{\Dlistll}[1]{\Dscr({\it list}^{**}(#1))}
\newcommand{\Dalllists}{\Dscr({\it lists})}
\newcommand{\lstl}{\lst^*}
\newcommand{\lstll}{\lst^{**}}
\newcommand{\nill}{\nil^*}
\newcommand{\consl}[2]{#1 \! ::^* \! #2}

\newcommand{\Deq}{\Dscr({\it eq})}
\newcommand{\cphi}[1]{\phi^{\circ}(#1)}

\newcommand{\same}[2]{#1 \equiv #2}
\newcommand{\Dintuit}{\Dscr({\it intuit})}
\newcommand{\Dlinear}{\Dscr({\it linear})}
\newcommand{\Dprog}{\Dscr({\it prog})}
\newcommand{\Devars}[1]{\Dscr({\it evars}(#1))}
\newcommand{\Dallevars}{\Dscr({\it evars})}
\newcommand{\itm}{i}
\newcommand{\itml}{\itm^*}
\newcommand{\itmll}{\itm^{**}}
\newcommand{\atm}{\hbox{\sl atm}}
\newcommand{\atml}{\atm^*}
\newcommand{\prp}{\hbox{\sl prp}}
\newcommand{\prpl}{\prp^*}
\newcommand{\atmlst}{\hbox{\sl atmlst}}
\newcommand{\atmlstl}{\atmlst^*}
\newcommand{\prplst}{\hbox{\sl prplst}}
\newcommand{\prplstl}{\prplst^*}
\newcommand{\atom}[1]{\langle #1\rangle}
\newcommand{\atoml}[1]{\atom{#1}^*}
\newcommand{\ttrue}{\hbox{\sl tt}}
\newcommand{\ttruel}{\ttrue^*}

\newcommand{\with}{\mathbin{\&}}
\newcommand{\withl}{\with^*}

\newcommand{\iimp}{\Rightarrow}
\newcommand{\iimpl}{\iimp^*}
\newcommand{\limp} {\mathbin{-\hspace{-0.70mm}\circ}}
\newcommand{\limpl} {\limp^*}
\newcommand{\bigwedgel} {\bigwedge^*}
\newcommand\cut{\hbox{\sl cut}}

\newcommand{\prove}[1]{\hbox{\sl prove} \; #1}
\newcommand{\hyp}[1]{\hbox{\sl hyp} \; #1}
\newcommand{\conc}[2]{\hbox{\sl conc}_{#1} \; #2}
\newcommand{\seq}[3]{\hbox{\sl seq}_{#1} \; #2 \; #3}
\newcommand{\lseq}[4]{\hbox{\sl seq}_{#1} \; #2 \; #3 \; #4}
\newcommand{\subst}[4]{\hbox{\sl subst} \; #1 \; #2 \; #3 \; #4}
\newcommand{\substz}[4]{\hbox{\sl subst}_0 \; #1 \; #2 \; #3 \; #4}
\newcommand{\extend}[4]{\hbox{\sl extend\_evars} \; #1 \; #2 \; #3 \; #4}
\newcommand{\extendz}[4]{\hbox{\sl extend\_evars}_0 \; #1 \; #2 \; #3 \; #4}
\newcommand{\splitseq}[4]{\hbox{\sl split\_seq}_{#2} \; #3 \; #4 \; #1}
\newcommand{\proj}[2]{\pi_{#1} \; #2}
\newcommand{\evars}{\hbox{\sl evs}}
\newcommand{\fst}[1]{\hbox{\sl fst} \; #1}
\newcommand{\tfst}[2]{\hbox{\sl fst}_{#1} \; #2}
\newcommand{\tfstl}[2]{\hbox{\sl fst}_{#1}^* \; #2}
\newcommand{\rst}[1]{\hbox{\sl rst} \; #1}
\newcommand{\prog}[2]{\hbox{\sl prog} \; #1 \; #2}
\newcommand{\lprog}[3]{\hbox{\sl prog} \; #1 \; #2 \; #3}
\newcommand{\prv}[1]{\rhd #1}
\newcommand{\prvv}[2]{#1 \rhd #2}
\newcommand{\lprvv}[3]{#1;#2 \rhd #3}

\newcommand\dcdmap[1]{\langle\![#1]\!\rangle}

\newcommand{\tm}{\hbox{\sl tm}}
\newcommand{\ty}{\hbox{\sl ty}}
\newcommand{\lc}{\hbox{\sl lc}}

\newcommand{\tmi}{\itm_{tm}}

\newcommand{\tyi}{\itm_{ty}}

\newcommand{\lci}{\itm_{lc}}

\newcommand{\sti}{\itm_{st}}

\newcommand{\ansi}{\itm_{ans}}

\newcommand{\instri}{\itm_{instr}}

\newcommand{\cntni}{\itm_{cntn}}

\newcommand{\pcfref}{{\rm PCF}_{:=}}
\newcommand{\Dlambda}{\Dscr({\it lambda})}
\newcommand{\Dpcf}{\Dscr({\rm PCF})}
\newcommand{\Dpcfref}{\Dscr(\pcfref)}
\newcommand{\Dstore}{\Dscr({\it store})}
\newcommand{\abs}[1]{\hbox{\sl abs} \; #1}
\newcommand{\tabs}[2]{\hbox{\sl abs} \; #1 \; #2}
\newcommand{\tabsl}[2]{\hbox{\sl abs}^* \; #1 \; #2}
\newcommand{\app}[2]{\hbox{\sl app} \; #1 \; #2}
\newcommand{\appl}[2]{\hbox{\sl app}^* \; #1 \; #2}
\newcommand{\zero}{\hbox{\sl zero}}

\newcommand{\true}{\hbox{\sl true}}

\newcommand{\false}{\hbox{\sl false}}

\newcommand{\succtm}[1]{\hbox{\sl succ} \; #1}

\newcommand{\pred}[1]{\hbox{\sl pred} \; #1}

\newcommand{\iszero}[1]{\hbox{\sl is\_zero} \; #1}

\newcommand{\iftm}[3]{\hbox{\sl if} \; #1 \; #2 \; #3}

\newcommand{\rec}[2]{\hbox{\sl rec} \; #1 \; #2}
\newcommand{\recl}[2]{\hbox{\sl rec}^* \; #1 \; #2}
\newcommand{\loc}[1]{\hbox{\sl cell} \; #1}
\newcommand{\locl}[1]{\hbox{\sl cell}^* \; #1}
\newcommand{\reftm}[1]{\hbox{\sl ref} \; #1}
\newcommand{\reftml}[1]{\hbox{\sl ref}^* \; #1}

\newcommand{\derefl}[1]{\hbox{\sl deref}^* \; #1}

\newcommand{\assignl}[2]{\hbox{\sl assign}^* \; #1 \; #2}

\newcommand{\sequencel}[2]{\hbox{\sl sequence}^* \; #1 \; #2}
\newcommand{\arr}[2]{\hbox{\sl arr} \; #1 \; #2}
\newcommand{\arrl}[2]{\hbox{\sl arr}^* \; #1 \; #2}
\newcommand{\gnd}{\hbox{\sl gnd}}
\newcommand{\num}{\hbox{\sl num}}

\newcommand{\bool}{\hbox{\sl bool}}

\newcommand{\refty}[1]{\hbox{\sl refty} \; #1}
\newcommand{\reftyl}[1]{\hbox{\sl refty}^* \; #1}
\newcommand{\nullst}{\hbox{\sl null\_st}}
\newcommand{\nullstl}{\hbox{\sl null\_st}^*}
\newcommand{\extendst}[3]{\hbox{\sl extend\_st} \; #1 \; #2 \; #3}
\newcommand{\extendstl}[3]{\hbox{\sl extend\_st}^* \; #1 \; #2 \; #3}

\newcommand{\answerl}[2]{\hbox{\sl answer}^* \; #1 \; #2}

\newcommand{\newl}[1]{\hbox{\sl new}^* \; #1}
\newcommand{\initk}{\hbox{\sl init}}
\newcommand{\initkl}{\hbox{\sl init}^*}

\newcommand{\extendkl}[2]{#2 \succ^* #1}
\newcommand{\eval}[1]{\hbox{\sl eval} \; #1}
\newcommand{\evall}[1]{\hbox{\sl eval}^* \; #1}
\newcommand{\return}[1]{\hbox{\sl return} \; #1}
\newcommand{\returnl}[1]{\hbox{\sl return}^* \; #1}

\newcommand{\evalarg}[2]{\hbox{\sl eval\_arg} \; #1 \; #2}
\newcommand{\evalargl}[2]{\hbox{\sl eval\_arg}^* \; #1 \; #2}
\newcommand{\apply}[2]{\hbox{\sl apply} \; #1 \; #2}
\newcommand{\applyl}[2]{\hbox{\sl apply}^* \; #1 \; #2}
\newcommand{\newref}[1]{\hbox{\sl new\_ref} \; #1}
\newcommand{\newrefl}[1]{\hbox{\sl new\_ref}^* \; #1}
\newcommand{\lookup}[1]{\hbox{\sl lookup} \; #1}
\newcommand{\lookupl}[1]{\hbox{\sl lookup}^* \; #1}
\newcommand{\evalrval}[2]{\hbox{\sl eval\_rvalue} \; #1 \; #2}
\newcommand{\evalrvall}[2]{\hbox{\sl eval\_rvalue}^* \; #1 \; #2}
\newcommand{\update}[2]{\hbox{\sl update} \; #1 \; #2}
\newcommand{\updatel}[2]{\hbox{\sl update}^* \; #1 \; #2}
\newcommand{\typeof}[2]{\hbox{\sl typeof} \; #1 \; #2}
\newcommand{\typeofl}[2]{\hbox{\sl typeof}^* \; #1 \; #2}

\newcommand{\typeofcntnl}[2]{\hbox{\sl typeof}_{cntn}^* \;\; #1 \;\; #2}
\newcommand{\typeofinstr}[2]{\hbox{\sl typeof}_{instr} \;\; #1 \;\; #2}
\newcommand{\typeofinstrl}[2]{\hbox{\sl typeof}_{instr}^* \;\; #1 \;\; #2}
\newcommand{\typeofans}[2]{\hbox{\sl typeof}_{ans} \;\; #1 \;\; #2}
\newcommand{\typeofansl}[2]{\hbox{\sl typeof}_{ans}^* \;\; #1 \;\; #2}

\newcommand{\welltypedl}[1]{\hbox{\sl well\_typed}^* \; #1}
\newcommand{\natsem}[2]{#1 \Downarrow #2}
\newcommand{\natsemst}[3]{(#1,#2) \Downarrow #3}
\newcommand{\natsemstl}[3]{(#1,#2) \Downarrow^* #3}
\newcommand{\nsmachone}[4]{\hbox{\sl ns\_mach\_1} \;\; #1 \;\; #2 \;\; #3 \;\; #4}
\newcommand{\nsmachonel}[4]{\hbox{\sl ns\_mach\_1}^* \;\; #1 \;\; #2 \;\; #3 \;\; #4}
\newcommand{\nsmachtwo}[3]{\hbox{\sl ns\_mach\_2} \;\; #1 \;\; #2 \;\; #3}
\newcommand{\nsmachtwol}[3]{\hbox{\sl ns\_mach\_2}^* \;\; #1 \;\; #2 \;\; #3}
\newcommand{\contains}[2]{\hbox{\sl contains} \; #1 \; #2}
\newcommand{\containsl}[2]{\hbox{\sl contains}^* \; #1 \; #2}

\newcommand{\collectstl}[1]{\hbox{\sl collect\_state}^* \; #1}
\newcommand{\valuep}[1]{\hbox{\sl value} \; #1}

\newcommand{\transone}[2]{#1 \leadsto #2}
\newcommand{\transsem}[2]{#1 \leadsto^* #2}
\newcommand{\cont}[3]{\hat{#2}.#3,#1}
\newcommand{\kfocus}[4]{#1 \vdash (#2,#3) \hookrightarrow #4}
\newcommand{\kreturn}[4]{#1 \vdash (#2,#3) \dot{\hookrightarrow} #4}
\newcommand{\store}[1]{\hbox{\sl store} \; #1}

\newcommand{\storetyping}[1]{\hbox{\sl store\_typing} \; #1}

\newcommand{\storetypeof}[2]{\hbox{\sl store\_typeof} \; #1 \; #2}

\newcommand{\sameitml}[2]{#1 \equiv_{\itml} #2}
\newcommand{\sameatml}[2]{#1 \equiv_{\atml} #2}

\def\foldnpart{Part I}
\def\stylespart{Part II}
\def\pcfpart{Part III}

\newlength{\infwidthi}
\newlength{\infwidthii}

\begin{abstract}
Logical frameworks based on intuitionistic or linear logics with
higher-type quantification have been successfully used to give
high-level, modular, and formal specifications of many important
judgments in the area of programming languages and inference systems.
Given such specifications, it is natural to consider proving
properties about the specified systems in the framework: for example,
given the specification of evaluation for a functional programming
language, prove that the language is deterministic or that evaluation
preserves types.
One challenge in developing a framework for such reasoning is that
{\em higher-order abstract syntax} (HOAS), an elegant and declarative
treatment of object-level abstraction and substitution,
is difficult to treat in proofs involving induction.
In this paper, we present a meta-logic that can be used to reason
about judgments coded using HOAS; this meta-logic is an extension of a
simple intuitionistic logic that admits higher-order quantification
over simply typed $\lambda$-terms (key ingredients for HOAS) as well as
induction and a notion of {\em definition}.
The latter concept of definition is a proof-theoretic device that
allows certain theories to be treated as ``closed'' or as defining
fixed points.
We explore the difficulties of formal meta-theoretic analysis of HOAS 
encodings by considering encodings of intuitionistic and linear logics,
and formally derive the admissibility of cut for important subsets of 
these logics.
We then propose an approach to avoid the apparent tradeoff between the
benefits of higher-order abstract syntax and the ability to analyze
the resulting encodings.
We illustrate this approach through examples involving the simple
functional and imperative programming languages PCF and $\pcfref$.
We formally derive such properties as unicity of typing, subject
reduction, determinacy of evaluation, and the equivalence of
transition semantics and natural semantics presentations of evaluation.
\end{abstract}


\begin{document}
\bibliographystyle{acmtrans}

\begin{bottomstuff}
Authors' addresses: Raymond C. McDowell, Department of Mathematics and
Computer Science, Kalamazoo College, 1200 Academy Street, Kalamazoo,
MI 49006-3295 USA.
Dale A. Miller, Department of Computer Science and Engineering, 220 Pond
Laboratory, The Pennsylvania State University, University Park, PA
16802-6106 USA. 
\newline
The authors have been funded in part by the grants ONR
N00014-93-1-1324, NSF CCR-92-09224, NSF CCR-94-00907, NSF
CCR-98-03971, and ARO DAAH04-95-1-0092.
\permission{TBD}{TBD}
\end{bottomstuff}
\maketitle

\section*{INTRODUCTION}

Meta-logics and type systems have been used to specify the semantics of
a wide range of logics and computation systems
\cite{avron92jar,chirimar95phd,felty93jar,pfenning92cade}.
This is done by making judgments, such as ``the term $M$ denotes a
program,'' ``the program $M$ evaluates to the value $V$'', and ``the
program $M$ has type $T$'', into predicates that can be proved or
types for which inhabitants (proofs) are needed.
Since these specification languages often contain quantification at
higher-order types and term structures involving $\lambda$-terms,
succinct and elegant specifications can be written using
{\em higher-order abstract syntax}, a high-level and declarative
treatment of object-level bound variables and object-level
substitution \cite{miller87slp,pfenning88pldi}.
In other approaches to syntactic representation where bound variables are
managed directly using either names or deBruijn-style numbering, these
details must be carefully addressed and dealt with at most levels of a
specification.

Recently, logical specification languages have been used to not only
describe how to {\em perform} computations but also describe {\em
properties about} the encoded computations
\cite{basin93le,magnusson94tpp,matthews93le,vaninwegen96phd}.
By proving these properties in a formal framework,
we can benefit from automated proof assistance and gain greater
confidence in our results.
However, this work has been done in languages that do not support
higher-order abstract syntax and so has not been able to benefit
from this representation technique.
As a result, theorems about substitution and bound variables can
dominate the task \cite{vaninwegen96phd}.
But meta-theoretic reasoning about systems represented in
higher-order abstract syntax has been difficult since the
languages and logics that support this notion of syntax do not provide
facilities for the fundamental operations of case analysis and
induction.
Moreover, higher-order abstract syntax leads to types and
recursive definitions that do not give rise to monotone inductive
operators, making inductive principles difficult to find.

These apparent difficulties can be overcome, and in this paper we
present a meta-logic in which we can naturally reason about
specifications in higher-order abstract syntax.
This meta-logic is a higher-order intuitionistic logic with
partial inductive definitions and natural number induction.
Induction on natural numbers allows us to derive other induction
principles via the construction of an appropriate measure.
A partial inductive definition \cite{hallnas91tcs} is a
proof-theoretic formalization that allows certain theories to be
treated as ``closed'' or as defining fixed points.
This allows us to perform case analyses on the defined judgments.
We use this definition mechanism to specify a small, object-level
logic which in turn is used to specify the computation systems under
consideration.
In this way, we can talk directly about the
structure of object-logic sequents and their provability.
This technique of representing a logic within a logic is not new (see,
for example, \citeN{felty88cade} and \citeN{paulson86jlp} for some early
references) and corresponds to the structure of common informal reasoning.

The first part of this paper (Sections~\ref{sec:foldn-description}
and \ref{sec:foldn-examples}) presents the meta-logic $\FOLDN$
(pronounced ``fold-n'').
To illustrate the use of $\FOLDN$, we derive several theorems
expressing properties of natural numbers and lists.
In \stylespart\ (Sections~\ref{sec:styles}, \ref{sec:logics}, and
\ref{sec:related1}) we consider encodings of intuitionistic and
linear logics in $\FOLDN$ to illustrate some difficulties with
reasoning in the specification logic about higher-order abstract
syntax and to also demonstrate some strategies to deal with these
difficulties.
Unfortunately these strategies involve sacrificing some
benefits of higher-order abstract syntax in order to gain the ability
to perform some meta-theoretic analyses.
We avoid this tradeoff in \pcfpart\ (Sections~\ref{sec:motivate}, 
\ref{sec:functional}, \ref{sec:imperative}, and \ref{sec:related2})
by taking a different approach to formal reasoning.
The key to this approach is to encode the object system in a
specification logic that is separate from the logic $\FOLDN$ in which
we perform the reasoning; this specification logic is itself specified
in $\FOLDN$.
This separation of the specification logic and the meta-logic allows
us to reason formally about specification logic sequents and their
derivability, and also reflects the structure of informal reasoning
about higher-order abstract syntax encodings.
We illustrate this approach by considering the static and dynamic
semantics of small functional and imperative programming languages; we
are able to derive in $\FOLDN$ such properties as the unicity of
typing, determinacy of semantics, and type preservation (subject
reduction).
We conclude in Section~\ref{sec:conclusion} with a brief discussion of our
accomplishments and possible extensions of this work.

\section*{\foldnpart: THE META-LOGIC $\FOLDN$}

In this part we introduce the logic which we call $\FOLDN$, an
acronym for ``first-order logic for $\lambda$ with definitions and
natural numbers.''
We present the logic in the first section, and then proceed in the
next with some sample definitions and propositions.
We conclude the part by briefly comparing the strength of $\FOLDN$ with 
that of other logical systems.

\section{A Description of the Logic}
\label{sec:foldn-description}

The basic logic is an intuitionistic version of a subset of Church's
Simple Theory of Types \cite{church40jsl} in which formulas have
the type $o$.
The logical connectives are $\bot$, $\top$, $\land$, $\lor$,
$\supset$, $\forall_{\tau}$, and $\exists_{\tau}$.
The quantification types $\tau$ (and thus the types of variables)
are restricted to not contain $o$.
Thus $\FOLDN$ supports quantification over higher-order (non-predicate)
types, a crucial feature for higher-order abstract syntax, 
but has a first-order proof theory, since there is no quantification
over predicate types.
We will use sequents of the form $\Seq{\Gamma}{B}$, where $\Gamma$ is
a finite multiset of formulas and $B$ is a single formula.
The basic inference rules for the logic are shown in
Table~\ref{tab:core-rules}.
In the $\forallR$ and $\existsL$ rules, $y$ is an eigenvariable that
is not free in the lower sequent of the rule.
\begin{table}[btp]
\caption{Inference rules for the core of $\FOLDN$}
\label{tab:core-rules}
\vspace{2pt}
\begin{center}
$\begin{array}{c@{\quad\quad}c@{\quad\quad}c}
\hline
\multicolumn{2}{c}
{\infer[\botL]{\Seq{\bot,\Gamma}{B}}{\rule{0pt}{6pt}}}
& \infer[\topR]{\Seq{\Gamma}{\top}}{}
\\ \\
\infer[\landL]{\Seq{B \land C,\Gamma}{D}}
	{\Seq{B,\Gamma}{D}}
& \infer[\landL]{\Seq{B \land C,\Gamma}{D}}
	{\Seq{C,\Gamma}{D}}
& \infer[\forallL]{\Seq{\forall x.B,\Gamma}{C}}
	{\Seq{B[t/x],\Gamma}{C}}
\\ \\
\multicolumn{2}{c}
{\infer[\landR]{\Seq{\Gamma}{B \land C}}
	{\Seq{\Gamma}{B}
	& \Seq{\Gamma}{C}}}
& \infer[\forallR]{\Seq{\Gamma}{\forall x.B}}
	{\Seq{\Gamma}{B[y/x]}}
\\ \\
\multicolumn{2}{c}
{\infer[\lorL]{\Seq{B \lor C,\Gamma}{D}}
	{\Seq{B,\Gamma}{D}
	& \Seq{C,\Gamma}{D}}}
& \infer[\existsL]{\Seq{\exists x.B,\Gamma}{C}}
	{\Seq{B[y/x],\Gamma}{C}}
\\ \\
\infer[\lorR]{\Seq{\Gamma}{B \lor C}}
	{\Seq{\Gamma}{B}}
& \infer[\lorR]{\Seq{\Gamma}{B \lor C}}
	{\Seq{\Gamma}{C}}
& \infer[\existsR]{\Seq{\Gamma}{\exists x.B}}
	{\Seq{\Gamma}{B[t/x]}}
\\ \\
\multicolumn{2}{c}
{\infer[\oimpL]{\Seq{B \oimp C,\Gamma}{D}}
	{\Seq{\Gamma}{B}
	& \Seq{C,\Gamma}{D}}}
& \infer[\oimpR]{\Seq{\Gamma}{B \oimp C}}
	{\Seq{B,\Gamma}{C}}
\\ \\
\multicolumn{2}{c}{\infer[init$, where $A$ is atomic $]{\Seq{A,\Gamma}{A}}{}}
& \infer[\cL]{\Seq{B,\Gamma}{C}}
	{\Seq{B,B,\Gamma}{C}}
\\ \\
\multicolumn{3}{c}
{\infer[\cut]
	{\Seq{\Delta,\Gamma}{C}}
	{\Seq{\Delta}{B}
	& \Seq{B,\Gamma}{C}}}
\\[2pt]
\hline
\end{array}$
\end{center}
\end{table}

We introduce the natural numbers via the constants $\z:\nt$ for zero
and $\hbox{\sl s}:\nt \rightarrow \nt$ for successor and the predicate
$\hbox{\sl nat}:\nt \rightarrow \oo$.
The right and left rules for this new predicate are
\begin{displaymath}
\infer[\natR]{\Seq{\Gamma}{\nat{\z}}}{}
\qquad\qquad
\infer[\natR]{\Seq{\Gamma}{\nat{(\suc{I})}}}{\Seq{\Gamma}{\nat{I}}}
\end{displaymath}
\begin{displaymath}
\infer[\natL\enspace .]{\Seq{\nat I,\Gamma}{C}}
	{\Seq{}{B \, \z}
	& \Seq{B \, j}{B \, (\suc{j})}
	& \Seq{B \, I,\Gamma}{C}}
\end{displaymath}
In the left rule, the predicate $B:\nt \rightarrow \oo$ represents the
property that is proved by induction, and $j$ is an eigenvariable that
is not free in $B$.
The third premise of that inference rule witnesses the fact that, in
general, $B$ will express a property stronger than
$(\bigwedge \Gamma) \oimp C$. 
Notice that the first two premises of the $\natL$ rule involve no 
assumptions other than the induction hypothesis (in the second premise).  
This is not a restriction on induction since one can choose to do 
induction on, say, $\lambda w. (\bigwedge \Gamma) \oimp B \, w$, which
would effectively provide the first two premises with the
assumptions from the multiset $\Gamma$.

A {\em definitional clause} is written 
$\forall\bar{x}[p \, \bar{t} \defeq B]$, where $p$ is a predicate
constant, every free variable of the formula $B$ is also free in at
least one term in the list $\bar{t}$ of terms, and all variables free
in $\bar{t}$ are contained in the list $\bar{x}$ of variables.
Since all free variables in $p \, \bar{t}$ and $B$ are universally
quantified, we often leave these quantifiers implicit when displaying
definitional clauses.
The atomic formula $p \, \bar{t}$ is called the {\em head} of the
clause, and the formula $B$ is called the {\em body}.
The symbol $\defeq$ is used simply to indicate a definitional clause:
it is not a logical connective.
A {\em definition} is a (perhaps infinite) set of definitional
clauses.
The same predicate may occur in the head of multiple clauses of a
definition:  it is best to think of a definition as a mutually
recursive definition of the predicates in the heads of the clauses.

We must also restrict the use of implication in the bodies of
definitional clauses; otherwise cut-elimination does not hold
\cite{schroeder-heister92nlip}.
Toward that end we assume that each predicate symbol $p$ in the
language has associated with it a natural number $\level{p}$, the
{\em level} of the predicate.
We then extend the notion of level to formulas and derivations.
Given a formula $B$, its {\em level} $\level{B}$ is defined as follows:
\begin{enumerate}
\item $\level{p \, \bar{t}} = \level{p}$
\item $\level{\bot} = \level{\top} = 0$
\item $\level{B \land C} = \level{B \lor C} = \max(\level{B},\level{C})$
\item $\level{B \oimp C} = \max(\level{B}+1,\level{C})$
\item $\level{\forall x.B} = \level{\exists x.B} = \level{B}$.
\end{enumerate}
Given a derivation $\Pi$ of $\Seq{\Gamma}{B}$,
$\level{\Pi} = \level{B}$.
We now require that for every definitional clause 
$\forall\bar{x}[p \, \bar{t} \defeq B]$,
$\level{B} \leq \level{p \, \bar{t}}$.

The inference rules for defined atoms are given relative to some 
fixed definition.
The right-introduction rule for defined atoms is
\begin{displaymath}
\infer[\defR, $
	where $p \, \bar{u} = (p \, \bar{t})\theta$ for
	some clause $\forall\bar{x}.{[p \, \bar{t} \defeq B]}\enspace,]
   {\Seq{\Gamma}{p \, \bar{u}}}
   {\Seq{\Gamma}{B\theta}}
\end{displaymath}
where $\theta$ is a substitution of terms for
variables.
The left rule for defined concepts uses
complete sets of unifiers (CSU):
\begin{displaymath}
\infer[\defL\enspace,]{\Seq{p \, \bar{u},\Gamma}{C}}
    {\left\{\mbox{$\Seq{B\theta,\Gamma\theta}{C\theta}\;\left\vert\,\right.
		\theta \in CSU(p \, \bar{u},p \, \bar{t}) $ for
		some clause $\forall\bar{x}.[p \, \bar{t} \defeq B]$}
	      \right\}}
\end{displaymath}
where $\theta$ is a substitution of terms for variables, and the
variables $\bar{x}$ are chosen to be distinct from the variables free
in the lower sequent of the rule.
(A set $S$ of unifiers of $t$ and $u$ is {\em complete} if for every unifier 
$\rho$ of $t$ and $u$ there is a unifier $\theta \in S$ such that 
$\rho$ is $\theta\circ\sigma$ for some substitution $\sigma$
\cite{huet75tcs}.)
Specifying a set of sequents as the premise should be understood
to mean that each sequent in the set is a premise of the rule.
The right rule corresponds to the logic programming notion of
{\em backchaining} if we think of $\defeq$ in definitional clauses as
reverse implication.
The left rule is similar to {\em definitional reflection}
\cite{schroeder-heister93lics} (not to be confused with another notion
of reflection often considered between a meta-logic and object-logic)
and to an inference rule used by Girard in his note on fixed points
\cite{girard92mail}.
This particular presentation of the rule is due to Eriksson
\cite{eriksson91elp}.
Notice that in the $\defL$ rule, the free variables of the conclusion
can be instantiated in the premises.

The number of premises of the $\defL$ rule may be either
infinite or finite (including zero).  If the formula $p \, \bar{u}$
does not unify with the head of any definitional clause, then the
number of premises will be zero.
In this case $p \, \bar{u}$ is an unprovable formula logically
equivalent to $\bot$, and $\defL$ corresponds to the $\botL$ rule.
If the formula $p \, \bar{u}$ does unify with the head of a
definitional clause, CSUs may be infinite, as is the case with
unifications involving simply typed $\lambda$-terms and variables of
functional type ({\frenchspacing a.k.a. higher-order unification}).
Clearly an inference rule with an infinite number of premises is
impossible to automate directly.
There are many important situations where CSUs are not only
finite but are also singleton (containing a most general
unifier) whenever terms are unifiable.
One such case is, of course, the first-order case.
Another case is when the application of functional variables are
restricted to distinct bound variables in the sense of
{\em higher-order pattern} unification \cite{miller91jlc}.
In this paper, all unification problems will fall into this latter
case and, hence, we can count on the definition left-introduction rule
to have a finite (and small) number of premises.

Assuming that a definition is given and fixed, we have the following results.

\begin{proposition}[Cut-Elimination for $\FOLDN$]
\label{prp:cut-elim}
If a sequent is derivable in $\FOLDN$, then it is derivable without
using the $\cut$ rule.
\end{proposition}
\begin{proof}
The proofs of \citeN{schroeder-heister93lics}
regarding cut-elimination for definitions do not appear to extend
to our setting where induction is included.  A complete proof of
this theorem appears in \citeN{mcdowell97phd} and \citeN{mcdowell00tcs}
and is modeled on proofs by Tait and Martin-L\"of that use the
technical notions of normalizability and reducibility.
\end{proof}

The following corollary is an immediate consequence of this
cut-elimination theorem.

\begin{corollary}[Consistency of $\FOLDN$]
There is no derivation in $\FOLDN$ of the sequent $\Seq{}{\bot}$.
\end{corollary}

Although cut-elimination holds for this logic, we do not have the
subformula property since the induction predicate $B$ used in the
$\natL$ rule is not necessarily a subformula of the conclusion of
that inference rule.
In fact, the following inference rule is derivable from the induction
rule:
\begin{displaymath}
\infer[\enspace .]{\Seq{\nat I,\Gamma}{C}}
	{\Seq{}{B}
	& \Seq{B,\Gamma}{C}}
\end{displaymath}
This inference rule resembles the cut rule except that it requires a
{\sl nat} assumption.
Although we fail to have the subformula property, the cut-elimination
theorem still provides a strong basis for reasoning about proofs in
$\FOLDN$.
Also this formulation of the induction principle is natural and close
to the one used in actual mathematical practice:  that is, invariants
must be, at times, clever inventions that are not simply
rearrangements of subformulas.
Any automation of $\FOLDN$ will almost certainly need to be
interactive, at least for retrieving instantiations for the induction
predicate $B$.

\section{Some Simple Definitions and Propositions}
\label{sec:foldn-examples}

In this section we illustrate the use of the logic $\FOLDN$ with some
examples.
We first define some predicates over the natural numbers and reason
about them.
Then we introduce a list type and consider predicates for it.
As we prove properties about these types and predicates, we will
interleave informal descriptions of the proofs with their realization
as derivations in $\FOLDN$.
The formal derivations are by nature detailed and low-level, breaking
down proof principles into small pieces.
As a result, what can seem obvious or be described informally in a
small number of words may take a number of steps to accomplish in the
formal derivation.
But it is exactly this nature that makes formal derivations amenable
to automation; tools such as proof editors and theorem provers can
make the construction of formal derivations more natural as well as 
more robust.

We will describe derivations in a ``bottom-up'' manner -- that is, we
will start with the sequent we wish to derive, apply a rule with that
sequent as the conclusion, and continue in this manner with the rule
premises.
Thus unproved premises represent statements of what remains to be
proved to establish the original sequent.
Since the formal ($\FOLDN$) derivation is presented in pieces, intermixed
with descriptive text, pieces that occur later in the text will
generally be (partial) derivations of unproved premises from earlier
pieces.

\subsection{Natural Numbers}

As described in Section~\ref{sec:foldn-description}, $\FOLDN$ includes a
type $\nt$ encoding natural numbers and a membership predicate
{\sl nat}.
We now introduce predicates representing equality, the less-than
relation, the less-than-or-equal-to relation, and the addition
function.
The types for these predicates are as follows:
\begin{displaymath}
\begin{array}[b]{rcl@{\quad\quad\quad\quad}rcl}
=	& \colon & \nt \rightarrow \nt \rightarrow \oo 
& \hbox{\sl sum} & \colon & \nt \rightarrow \nt \rightarrow \nt \rightarrow \oo \\
<	& \colon & \nt \rightarrow \nt \rightarrow \oo
& \leq	& \colon & \nt \rightarrow \nt \rightarrow \oo
\enspace .
\end{array}
\end{displaymath}
The definitional clauses for these predicates are shown in
Table~\ref{tab:def-nat}; we shall refer to this set of clauses as
$\Dnat$.
We define two numbers to be equal if they are unifiable.
The clauses for {\sl sum} indicate that the sum of zero and any other
number $J$ is $J$, and the sum of $(\suc{I})$ and $J$ is the successor
of the sum of $I$ and $J$.
Zero is less than the successor of any number, and $(\suc{I})$ is less
than $(\suc{J})$ whenever $I$ is less than $J$.
Finally, $I \leq J$ if $I$ is equal to $J$ or if $I$ is less than $J$.
\begin{table}[btp]
\caption{Definitional clauses for predicates over natural numbers}
\label{tab:def-nat}
\vspace{2pt}
\begin{center}
$\begin{array}{rcl@{\quad\quad\quad\quad}rcl}
\hline\rule{0pt}{14pt}
I = I
	& \defeq & \top

& \sump{\z}{J}{J}
	& \defeq & \nat{J} \\

	&	&

& \sump{(\suc{I})}{J}{(\suc{K})}
	& \defeq & \sump{I}{J}{K} \\
\\
\z < (\suc{J})
	& \defeq & \nat{J}

& I \leq I
	& \defeq & \top \\

(\suc{I}) < (\suc{J})
	& \defeq & I < J

& I \leq J
	& \defeq & I < J
\\[2pt]
\hline
\end{array}$
\end{center}
\end{table}

We now proceed to reason in the logic $\FOLDN$ about natural numbers
and these predicates over them.
As our first example, we derive a case analysis rule for natural
numbers.
In general the $\defL$ rule is used to formalize case analysis, but
the predicate {\sl nat} is not a defined predicate, and so the $\defL$
rule does not apply in the case of natural numbers.
However, a case analysis may be viewed as an induction in which we do
not use the induction hypothesis in the induction step.
Thus we can derive a case analysis rule for natural numbers from the
induction ($\natL$) rule.
\begin{proposition}
\label{prp:nat-case-analysis}
For any formula $C: \oo$, predicate $B: \nt \rightarrow \oo$, term
$I: \nt$, multiset $\Gamma$ of formulas, and eigenvariable $i: \nt$
such that $i$ is not free in $B$, the following rule is derivable in
$\FOLDN$:
\begin{displaymath}
\infer[\enspace .]{\Seq{\nat{I}, \Gamma}{C}}
      {\Seq{}{B \, \z}
      & \Seq{\nat{i}}{B \, (\suc{i})}
      & \Seq{B \, I, \Gamma}{C}}
\end{displaymath}
\end{proposition}
\begin{proof}
This rule expresses the following idea:  
we want to show that $C$
follows from $\Gamma$ and the fact that $I$ is a natural number.
Since $I$ is a natural number, it must be either zero or the successor
of another natural number.
Thus if we can show that $B$ holds for zero and for the successor of
any natural number (the first two premises), then we know that $B$
holds for $I$.
It then remains to show that $C$ follows from $B \, I$ and $\Gamma$
(the third premise).

To derive this rule, we assume that we have derivations of the premises and
proceed to prove the conclusion.
That is, we construct in $\FOLDN$ a partial derivation of the sequent
$\Seq{\nat{I}, \Gamma}{C}$, leaving unproved premises of the form
$\Seq{}{B \, \z}$, $\Seq{\nat{i}}{B \, (\suc{i})}$,
and $\Seq{B \, I, \Gamma}{C}$.
This corresponds to working under the assumption that $B$ holds both
for zero and for the successor of any number and that $B \, I$ and
$\Gamma$ imply $C$.
We proceed by induction on $I$, using
$(\lambda i.\nat{i} \land B \, i)$ as our induction predicate. 
As a result, we must establish three things:
\begin{enumerate}
\item the base case: zero is a natural number and $B$ holds for it;
\item the induction step: if $i$ is a natural number and $B$ holds for
it, then the same is true for $(\suc{i})$;
\item the relevance of the induction predicate: if $I$ is a natural
number and $B$ holds for it, then $\Gamma$ implies $C$.
\end{enumerate}
This staging of the problem is represented in $\FOLDN$ by applying
the $\natL$ rule:
\begin{displaymath}
\infer[\natL\enspace .]
      {\Seq{\nat{I},\Gamma}{C}}
      {\Seq{}{\nat{\z} \land B \, \z}
      & \Seq{\nat{i} \land B \, i}{\nat{(\suc{i})} \land B \, (\suc{i})}
      & \Seq{\nat{I} \land B \, I, \Gamma}{C}}
\end{displaymath}
The three premises to the $\natL$ rule correspond to the three proof
obligations enumerated above.

Let us first consider the relevance of the induction predicate.
This is clear, since we are working under the assumption that $C$
follows from $B \, I$ and $\Gamma$.
This is formally represented by the partial derivation
\begin{displaymath}
\infer[\landL\enspace .]
      {\Seq{\nat{I} \land B \, I, \Gamma}{C}}
      {\Seq{B \, I, \Gamma}{C}}
\end{displaymath}

The base case is also simple:  zero is obviously a natural number,
and we are working under the assumption that $B$ holds for zero.
This is expressed in $\FOLDN$ by the partial derivation
\begin{displaymath}
\infer[\landR\enspace .]
      {\Seq{}{\nat{\z} \land B \, \z}}
      {\infer[\natR]
	     {\Seq{}{\nat{\z}}}
	     {}
      & \Seq{}{B \, \z}}
\end{displaymath}

It remains to prove the induction step.
Since $i$ is a natural number, $(\suc{i})$ is as well.
In addition, $B$ holds for $(\suc{i})$ by our working assumption.
The formal representation of this reasoning is
\begin{displaymath}
\infer[\landL\enspace .]
      {\Seq{\nat{i} \land B \, i}
	   {\nat{(\suc{i})} \land B \, (\suc{i})}}
      {\infer[\landR]
	     {\Seq{\nat{i}}
		  {\nat{(\suc{i})} \land B \, (\suc{i})}}
	     {\infer[\natR]
		    {\Seq{\nat{i}}{\nat{(\suc{i})}}}
		    {\infer[\init]
			   {\Seq{\nat{i}}{\nat{i}}}
			   {}}
	     & \Seq{\nat{i}}{B \, (\suc{i})}}}
\vspace{-1.5\baselineskip}
\end{displaymath}
\nobreak\hfill\nobreak
\end{proof}

We now use this derived case analysis rule to prove that zero is the
smallest natural number.
\begin{proposition}
\label{prp:z-lesseq}
The formula $\forall i(\nat{i} \oimp \z\leq i)$ is derivable in
$\FOLDN$ using the definition $\Dnat$.
\end{proposition}
\begin{proof}
The proof is a simple case analysis on $i$.
To represent this in $\FOLDN$, we apply the $\forallR$ and $\oimpR$
rules to get
\begin{displaymath}
\Seq{\nat{i}}{\z\leq i}
\enspace ,
\end{displaymath}
and then use the derived rule of Proposition~\ref{prp:nat-case-analysis},
which yields the three sequents
\begin{displaymath}
\Seq{}{\z\leq \z}
\quad\quad \Seq{\nat{i'}}{\z\leq(\suc{i'})}
\quad\quad \Seq{\z\leq i}{\z\leq i}
\enspace .
\end{displaymath}

In this case, the third premise is immediate:
\begin{displaymath}
\infer[\init\enspace .]
      {\Seq{\z\leq i}{\z\leq i}}
      {}
\end{displaymath}

If $i$ is zero, then it is immediate that zero is equal to itself and
thus less than or equal to itself:
\begin{displaymath}
\infer[\defR\enspace .]
      {\Seq{}{\z\leq \z}}
      {\infer[\topR]
	     {\Seq{}{\top}}
	     {}}
\end{displaymath}

If $i$ is the successor of some number $i'$, then $\z < (\suc{i'})$ by
definition, and so $\z \leq (\suc{i'})$ also by definition.
This is represented formally by the derivation
\begin{displaymath}
\infer[\defR\enspace .]
      {\Seq{\nat{i'}}{\z\leq(\suc{i'})}}
      {\infer[\defR]
	     {\Seq{\nat{i'}}{\z < (\suc{i'})}}
	     {\infer[\init]
		    {\Seq{\nat{i'}}{\nat{i'}}}
		    {}}}
\vspace{-1.5\baselineskip}
\end{displaymath}
\nobreak\hfill\nobreak
\end{proof}

It is also possible to derive in $\FOLDN$ a rule for complete induction
over the natural numbers \cite{mcdowell97phd}.
\begin{proposition}[Complete Induction]
\label{prp:complete-ind}
For any formula $C: \oo$, predicate $B: \nt \rightarrow \oo$, term
$I: \nt$, multiset $\Gamma$ of formulas, and eigenvariable $j: \nt$
such that $j$ is not free in $B$, the following rule is derivable in
$\FOLDN$ using the definition $\Dnat$:
\begin{displaymath}
\infer[\enspace .]{\Seq{\nat{I}, \Gamma}{C}}
      {\Seq{\nat{j}, \forall k (\nat{k} \oimp k<j \oimp B \, k)}{B \, j}
      & \Seq{B \, I, \Gamma}{C}}
\end{displaymath}
\end{proposition}

The following proposition presents additional properties of natural
numbers that we have derived in $\FOLDN$, although we do not show the
derivations here.
\begin{proposition}
\label{prp:nat-other}
The following formulas are derivable in $\FOLDN$ using the definition
$\Dnat$:
\begin{displaymath}
\forall i
	(\nat{(\suc{i})} \oimp \nat{i})
\end{displaymath}
\begin{displaymath}
\forall i
    (\nat{i} \oimp \forall j (i<j \oimp \nat{j}))
\end{displaymath}
\begin{displaymath}
\forall i (\nat{i} \oimp i<(\suc{i}))
\end{displaymath}
\begin{displaymath}
\forall i (\nat{i} \oimp \forall j (i<(\suc{j}) \oimp i\leq j))
\end{displaymath}
\begin{displaymath}
\forall i
    (\nat{i}
     \oimp \forall j \forall k
		(i<j \oimp j<k \oimp i<k))
\end{displaymath}
\begin{displaymath}
\forall i
    (\nat{i}
     \oimp \forall j
		(\nat{j}
		 \oimp \exists k
			(\nat{k} \land i<k \land j<k)))
\end{displaymath}
\begin{displaymath}
\forall i
    (\nat{i}
     \oimp \forall j \forall k
		(\sump{i}{(\suc{j})}{k} \oimp \sump{(\suc{i})}{j}{k}))
\end{displaymath}
\begin{displaymath}
\forall i
    (\nat{i}
     \oimp \forall j
		(\nat{j}
		 \oimp \exists k
			(\nat{k} \land \sump{i}{j}{k})))
\end{displaymath}
\begin{displaymath}
\forall i
    (\nat{i}
     \oimp \forall j \forall k
		(\nat{j} \oimp \sump{i}{j}{k} \oimp i\leq k))
\end{displaymath}
\begin{displaymath}
\forall i
    (\nat{i}
     \oimp \forall j \forall k
		(\nat{j} \oimp \sump{(\suc{i})}{j}{k} \oimp j<k))
\enspace .
\end{displaymath}
\end{proposition}

\subsection{Lists}
\label{sec:foldn-lists}

In this section we introduce a type $\lst$ for lists over an arbitrary
but fixed type $\tau$.
The type has two constructors, $\nil: \lst$ representing the empty list
and the infix operator $::$ of type $\tau \rightarrow \lst \rightarrow
\lst$ that adds an element to the front of a list.
Consider the list predicates
\begin{displaymath}
\begin{array}[b]{rcl@{\quad\quad}rcl}
\hbox{\sl length}	& \colon & \lst \rightarrow \nt \rightarrow \oo
& \hbox{\sl split}	& \colon & \lst \rightarrow \lst \rightarrow \lst \rightarrow \oo \\
\hbox{\sl list}	& \colon & \lst \rightarrow \oo
& \hbox{\sl permute}	& \colon & \lst \rightarrow \lst \rightarrow \oo \\
\hbox{\sl element}	& \colon & \tau \rightarrow \lst \rightarrow \oo\enspace ,
\end{array}
\end{displaymath}
whose definitional clauses are shown in Table~\ref{tab:def-list}; we
shall refer to this set of clauses as $\Dlist{\tau}$.
The predicate {\sl length} represents the function that returns the
length of its list argument.
The length of the empty list is zero, and the length of
$(\cons{X}{L})$ is one more than the length of $L$.
The predicate {\sl list} indicates that its argument has a finite
(natural number) length.
We shall find this predicate useful for constructing induction
principles over lists.
The predicate {\sl element} indicates that its first argument is a
member of its second argument.
$X$ is an element of $(\cons{Y}{L})$ if $X$ and $Y$ are the same or if
$X$ is an element of $L$.
The predicate {\sl split} holds if its first argument represents a
merging of the second and third in which the order of elements in
second and third lists is preserved in the first.
The empty list can only be split into two empty lists.
To split $(\cons{X}{L})$, we split $L$ and add $X$ to the front of
either of the resulting lists.
The predicate {\sl permute} holds if its two arguments contain the
same elements (including repetitions), though not necessarily in the
same order.
The empty list only permutes to itself.
A list $(\cons{X}{L_1})$ permutes to $L_2$ if removing $X$ from $L_2$
yields a permutation of $L_1$.
\begin{table}[btp]
\caption{Definitional clauses for predicates over lists}
\label{tab:def-list}
\vspace{2pt}
\begin{center}
$\begin{array}{rcl}
\hline\rule{0pt}{14pt}
\length{\nil}{\z}
	& \defeq & \top \\
\length{(\cons{X}{L})}{(\suc{I})}
	& \defeq & \length{L}{I} \\
\\
\tlist{L}
	& \defeq & \exists i (\nat{i} \land \length{L}{i}) \\
\\
\element{X}{(\cons{X}{L})}
	& \defeq & \top \\
\element{X}{(\cons{Y}{L})}
	& \defeq & \element{X}{L} \\
\\
\split{\nil}{\nil}{\nil}
	& \defeq & \top \\
\split{(\cons{X}{L_1})}{(\cons{X}{L_2})}{L_3}
	& \defeq & \split{L_1}{L_2}{L_3} \\
\split{(\cons{X}{L_1})}{L_2}{(\cons{X}{L_3})}
	& \defeq & \split{L_1}{L_2}{L_3} \\
\\
\permute{\nil}{\nil}
	& \defeq & \top \\
\permute{(\cons{X}{L_1})}{L_2}
	& \defeq & \exists l_{22} (\split{L_2}{(\cons{X}{\nil})}{l_{22}} \land \permute{L_1}{l_{22}})
\\[2pt]
\hline
\end{array}$
\end{center}
\end{table}

We now derive an induction rule for lists from the induction rule for
natural numbers ($\natL$) using the length of a list as our measure.
\begin{proposition}
\label{prp:list-ind}
For any formula $C: \oo$, predicate $B: \lst \rightarrow \oo$, term
$L: \lst$, multiset $\Gamma$ of formulas, and eigenvariables $x: \tau$
and $l: \lst$ such that $x$ and $l$ are not free in $B$, the following
rule is derivable in $\FOLDN$ using the definition $\Dlist{\tau}$:
\begin{displaymath}
\infer[\enspace .]{\Seq{\tlist{L}, \Gamma}{C}}
      {\Seq{}{B \, \nil}
      & \Seq{B \, l}{B \, (\cons{x}{l})}
      & \Seq{B \, L, \Gamma}{C}}
\end{displaymath}
\end{proposition}
\begin{proof}
To derive this rule, we construct a partial derivation of the sequent
$\Seq{\tlist{L}, \Gamma}{C}$, leaving unproved premises of the form
$\Seq{}{B \, \nil}$, $\Seq{B \, l}{B \, (\cons{x}{l})}$,
and $\Seq{B \, L, \Gamma}{C}$.
This corresponds to proving that $C$ follows from $\Gamma$ and the
fact that $L$ is a list under the assumptions
\begin{itemize}
\item $B$ holds for $\nil$;
\item for any $x'$ and $l'$, if $B$ holds for $l'$, then it also holds
for $(\cons{x'}{l'})$;
\item $B \, L$ and $\Gamma$ imply $C$.
\end{itemize}

The proof is by induction on the length of the list $L$.
Since $\tlist{L}$ holds, by definition $L$ has a length which is a
natural number:
\begin{displaymath}
\infer[\defL\enspace .]
      {\Seq{\tlist{L}, \Gamma}{C}}
      {\infer[\existsL]
	     {\Seq{\exists i (\nat{i} \land \length{L}{i}), \Gamma}{C}}
	     {\infer[\cL]
		    {\Seq{\nat{i} \land \length{L}{i}, \Gamma}{C}}
		    {\infer[\landL]
			   {\Seq{\nat{i} \land \length{L}{i},
				 \nat{i} \land \length{L}{i}, \Gamma}
				{C}}
			   {\infer[\landL]
				  {\Seq{\nat{i}, \nat{i} \land \length{L}{i}, \Gamma}
				       {C}}
				  {\Seq{\nat{i}, \length{L}{i}, \Gamma}{C}}}}}}
\end{displaymath}

We now claim that $B$ holds for lists of any length, and wish to prove
this claim by induction on the length of the list.
Thus we must prove
\begin{enumerate}
\item the base case: $B$ holds for lists of length zero;
\item the induction step:  if $B$ holds for lists of length $i'$, it
holds for lists of length $(\suc{i'})$;
\item the relevance of the claim:  $C$ follows from $\Gamma$, the fact
that $L$ has length $i$, and the fact that $B$ holds for lists of
length $i$.
\end{enumerate}
This is represented in $\FOLDN$ by applying the $\natL$ rule with the
induction predicate $\lambda i.\forall l (\length{l}{i}$ $\oimp B \ l)$,
which yields the three sequents
\begin{displaymath}
\Seq{}{\forall l (\length{l}{\z} \oimp B \, l)}
\end{displaymath}
\begin{displaymath}
\Seq{\forall l (\length{l}{i'} \oimp B \, l)}
    {\forall l (\length{l}{(\suc{i'})} \oimp B \, l)}
\end{displaymath}
\begin{displaymath}
\Seq{\forall l (\length{l}{i} \oimp B \, l),
     \length{L}{i}, \Gamma}
    {C\enspace .}
\end{displaymath}

Once we have proved that $B$ holds for lists of length $i$, then we
know it holds for $L$.
Thus we know that $C$ follows from $\Gamma$, since our third working
assumption says that $C$ follows from $B \, L$ and $\Gamma$.
This is represented formally by the partial derivation of the third premise
of the $\natL$ rule:
\begin{displaymath}
\infer[\forallL\enspace .]
      {\Seq{\forall l (\length{l}{i} \oimp B \, l), \length{L}{i}, \Gamma}
	   {C}}
      {\infer[\oimpL]
	     {\Seq{\length{L}{i} \oimp B \, L, \length{L}{i}, \Gamma}
		  {C}}
	     {\infer[\init]
		    {\Seq{\length{L}{i}, \Gamma}{\length{L}{i}}}
		    {}
	     & \Seq{B \, L, \length{L}{i}, \Gamma}{C}}}
\end{displaymath}
The unproved premise of this partial derivation is actually a weakening of
the third premise of the induction rule we are deriving.
We do not have an explicit weakening rule in $\FOLDN$, but it suffices
here to use the cut rule:
\begin{displaymath}
\infer[\mc\enspace .]
      {\Seq{B \, L, \length{L}{i}, \Gamma}{C}}
      {\Seq{B \, L, \length{L}{i}}{B \, L}
      & \Seq{B \, L, \Gamma}{C}}
\end{displaymath}
The first premise of the cut rule is derivable for any $B$ and $L$,
since the consequent $B \, L$  also occurs as an antecedent.
The second premise is the desired premise of the rule we are deriving.

In the base case of the induction, we must show that $B$ holds for
lists of length zero.
Since the only list of length zero is $\nil$, this follows from the
first working assumption, which says that $B \, \nil$ holds.
This case is formalized in the following partial derivation of the first
premise of the $\natL$ rule:
\begin{displaymath}
\infer[\forallR\enspace .]
      {\Seq{}{\forall l (\length{l}{\z} \oimp B \, l)}}
      {\infer[\oimpR]
	     {\Seq{}{\length{l}{\z} \oimp B \, l}}
	     {\infer[\defL]
		    {\Seq{\length{l}{\z}}{B \, l}}
		    {\Seq{\top}{B \, \nil}}}}
\end{displaymath}

The induction step requires us to prove that $B$ holds for all lists
of length $(\suc{i'})$, given that it holds for all lists of length $i'$.
Since a list of length $(\suc{i'})$ is constructed by adding an element to
the front of a list of length $i'$, this step follows from the second
working assumption, which says that if $B$ holds for a list $l$, then
for any $x : \tau$, $B$ holds for $\cons{x}{l}$.
This reasoning is represented in the partial derivation of the second
premise of the $\natL$ rule:
\begin{displaymath}
\infer[\forallR\enspace .]
      {\Seq{\forall l (\length{l}{i'} \oimp B \, l)}
	   {\forall l (\length{l}{(\suc{i'})} \oimp B \, l)}}
      {\infer[\oimpR]
	     {\Seq{\forall l (\length{l}{i'} \oimp B \, l)}
		  {\length{l}{(\suc{i'})} \oimp B \, l}}
	     {\infer[\defL]
		    {\Seq{\forall l (\length{l}{i'} \oimp B \, l), \length{l}{(\suc{i'})}}
			 {B \, l}}
		    {\infer[\forallL]
			   {\Seq{\forall l (\length{l}{i'} \oimp B \, l), \length{l'}{i'}}
				{B \, (\cons{x'}{l'})}}
			   {\infer[\oimpL]
				  {\Seq{\length{l'}{i'} \oimp B \, l', \length{l'}{i'}}
				       {B \, (\cons{x'}{l'})}}
				  {\infer[\init]
					 {\Seq{\length{l'}{i'}}
					      {\length{l'}{i'}}}
					 {}
				  & \Seq{B \, l', \length{l'}{i'}}
					{B \, (\cons{x'}{l'})}}}}}}
\end{displaymath}
In this use of the $\defL$ rule, the complete set of unifiers for the 
atomic formula $\length{l}{(\suc{i'})}$ and the head of the clause
$\forall x', l', j[\length{(\cons{x'}{l'})}{(\suc{j})} \defeq \length{l'}{j}]$
is the singleton set $\{[\cons{x'}{l'}/l,i'/j]\}$.
The unproved premise of the partial derivation above is a weakening of the
second premise of the induction rule we are deriving.
We can achieve this weakening using the cut rule in the same manner as
we did for the third premise:
\begin{displaymath}
\infer[\mc\enspace .]
      {\Seq{B \, l', \length{l'}{i'}}{B \, (\cons{x'}{l'})}}
      {\Seq{B \, l', \length{l'}{i'}}{B \, l'}
      & \Seq{B \, l'}{B \, (\cons{x'}{l'})}}
\vspace{-1.5\baselineskip}
\end{displaymath}
\nobreak\hfill\nobreak
\end{proof}

We will now use this derived induction rule for lists to prove a very
simple property, namely that we can split any list $L$ into $\nil$ and $L$.
\begin{proposition}
\label{prp:split-nil-refl}
The formula $\forall l (\tlist{l} \oimp \split{l}{\nil}{l})$ is
derivable in $\FOLDN$ using the definition $\Dlist{\tau}$.
\end{proposition}
\begin{proof}
We prove this by induction on $l$; using the right rules for $\forall$
and $\oimp$ and the derived rule of Proposition~\ref{prp:list-ind} with
the induction predicate $(\lambda l.\split{l}{\nil}{l})$, we get the three
sequents
\begin{displaymath}
\Seq{}{\split{\nil}{\nil}{\nil}}
\end{displaymath}
\begin{displaymath}
\Seq{\split{l'}{\nil}{l'}}{\split{(\cons{x'}{l'})}{\nil}{(\cons{x'}{l'})}}
\end{displaymath}
\begin{displaymath}
\Seq{\split{l}{\nil}{l}}{\split{l}{\nil}{l}}
\enspace .
\end{displaymath}
Since the induction predicate applied to $l$ is the same as the
consequent, the relevance of the induction predicate is immediate.
Thus the third sequent follows from the $\init$ rule.

The base case follows immediately from the definition of {\sl split},
and so the first sequent is derivable using the $\defR$ and $\topR$
rules.

The induction step also follows easily from the definition of
{\sl split}:
\begin{displaymath}
\infer[\defR\enspace.]
      {\Seq{\split{l'}{\nil}{l'}}{\split{(\cons{x'}{l'})}{\nil}{(\cons{x'}{l'})}}}
      {\infer[\init]
	    {\Seq{\split{l'}{\nil}{l'}}{\split{l'}{\nil}{l'}}}
            {}}
\vspace{-1.5\baselineskip}
\end{displaymath}
\nobreak\hfill\nobreak
\end{proof}

We conclude this section with a proposition that presents additional
properties of lists that we have derived in $\FOLDN$, though we omit
the derivations here.
\begin{proposition}
\label{prp:list-other}
The following formulas are derivable in $\FOLDN$ using the definition
$\Dlist{\tau}$:
\begin{displaymath}
\forall l
    (\tlist{l}
     \oimp \forall l_1 \forall l_2
		(\split{l}{l_1}{l_2}
		 \oimp (\tlist{l_1} \land \tlist{l_2})))
\end{displaymath}
\begin{displaymath}
\forall l_1
    (\tlist{l_1}
     \oimp \forall l_2
		(\tlist{l_2}
		 \oimp \forall l
			(\split{l}{l_1}{l_2} \oimp \tlist{l})))
\end{displaymath}
\begin{displaymath}
\forall l
    (\tlist{l}
     \oimp \forall l_1 \forall l_2
		(\split{l}{l_1}{l_2}
		 \oimp (\begin{array}[t]{@{}l}
		 	\forall x
				(\element{x}{l_1} \oimp \element{x}{l})
			\land \\
			\forall x
				(\element{x}{l_2} \oimp \element{x}{l}))))
			\end{array}
\end{displaymath}
\begin{displaymath}
\forall l
    (\tlist{l}
     \oimp \forall l_1 \forall l_2
		(\split{l}{l_1}{l_2} \oimp \split{l}{l_2}{l_1}))
\end{displaymath}
\begin{displaymath}
\forall l
    (\tlist{l}
     \oimp \forall l_{23} \forall l_1 \forall l_2 \forall l_3
		(\split{l}{l_1}{l_{23}}
		 \oimp \split{l_{23}}{l_2}{l_3}
	         \oimp \exists l_{12}
			(\begin{array}[t]{@{}l}
                            \split{l}{l_{12}}{l_3} \land\\
                            \split{l_{12}}{l_1}{l_2})))
                         \end{array}
\end{displaymath}
\begin{displaymath}
\forall l
    (\tlist{l}
     \oimp \forall l_{12} \forall l_1 \forall l_2 \forall l_3
		(\split{l}{l_{12}}{l_3}
		 \oimp \split{l_{12}}{l_1}{l_2}
                 \oimp \exists l_{23}
			(\begin{array}[t]{@{}l}
                            \split{l}{l_1}{l_{23}} \land \\
                            \split{l_{23}}{l_2}{l_3})))
                         \end{array}
\end{displaymath}
\begin{displaymath}
\forall l (\tlist{l} \oimp \permute{l}{l})
\end{displaymath}
\begin{displaymath}
\forall l
    (\tlist{l}
     \oimp \forall l' (\permute{l}{l'} \oimp \tlist{l'}))
\end{displaymath}
\begin{displaymath}
\forall l
    (\tlist{l}
     \oimp \forall l' \forall l_1 \forall l_2
		(\begin{array}[t]{@{}l}
		 \tlist{l'}
		 \oimp \permute{l}{l'}
		 \oimp \split{l}{l_1}{l_2}
		 \oimp \\
		 \exists l_1' \exists l_2'
			(\permute{l_1}{l_1'} \land \permute{l_2}{l_2'}
			 \land \split{l'}{l_1'}{l_2'})))
		 \end{array}
\end{displaymath}
\begin{displaymath}
\forall l
    (\tlist{l}
     \oimp \forall l' \forall l_1 \forall l_1' \forall l_2 \forall l_2'
		(\begin{array}[t]{@{}l}
		 \tlist{l'}
		 \oimp \split{l}{l_1}{l_2}
		 \oimp \split{l'}{l_1'}{l_2'}
		 \oimp \\
		 \permute{l_1}{l_1'}
		 \oimp \permute{l_2}{l_2'}
		 \oimp \permute{l}{l'}))
\enspace .
		 \end{array}
\end{displaymath}
\end{proposition}

\section{The Strength of $\FOLDN$}
\label{sec:strength}

Before proceeding to consider $\FOLDN$ as a logic for meta-theoretic
analysis, we comment here on how to relate $\FOLDN$ to other logical systems.  

First, we show that $\FOLDN$ captures the theorems of an 
intuitionistic version of Peano's arithmetic (IPA) using a definition 
consisting of one clause for equality. 
The formulas of IPA are those of a first-order logic with equality using the 
same logical connectives as those in $\FOLDN$ and the same symbols 
$\z$ for zero and $\hbox{\sl s}$ for successor.  
The axiom schemes for IPA can be grouped into the following 
collections.
\begin{enumerate}
    \item  Axioms for first-order intuitionistic logic.

    \item  Axioms for equality: reflexivity, symmetry, transitivity, 
    and substitution.

    \item  The two formulas
    $$\forall x\forall y(\suc{x} = \suc{y} \oimp x = y) 
      \hbox{\quad and \quad}
      \forall x ( \z = \suc{x} \oimp \perp )\enspace.
    $$

    \item The axioms of induction:  all formulas of the form 
    $$\varphi(\z)\land 
           \forall j(\varphi(j)\oimp \varphi(\suc{j}))\oimp 
	   \forall x\varphi(x)\enspace,$$
    where $\varphi(x)$ ranges over formulas with at most the variable 
    $x$ free. 
\end{enumerate}
There are two inference rules for IPA: {\em Modus Ponens} allows the formula 
$B$ to be inferred from the formulas $A\oimp B$ and $A$, while 
{\em Universal Generalization} allows the formula $\forall x B$ to
be inferred from $B$.
A list of formulas $C_{1},\ldots,C_{n}$ ($n\ge1$) is an {\em 
IPA derivation} if for every $i\in\{1,\ldots,n\}$, $C_{i}$ is either an 
axiom or is the conclusion of modus ponens or universal 
generalization from formulas in the list $C_{1},\ldots,C_{i-1}$.  We 
write $\vdash_{ipa}C$ if $C$ is the last formula of an IPA derivation.

In order to map an IPA formula, say $B$, to a $\FOLDN$ formula, say 
$(B)^{\circ}$, we must adjust for typing.  
The single sort used in IPA formulas will be 
mapped to the type \nt, and all instances of quantifiers in IPA 
formulas must be qualified using the {\sl nat} predicate: that 
is, $(\forall x.B)^{\circ}=\forall x.\nat{x}\oimp (B)^{\circ}$ and
$(\exists x.B)^{\circ}=\exists x.\nat{x}\land (B)^{\circ}$.  
Predicates in IPA will be mapped to the corresponding  
predicates in $\FOLDN$ similarly adjusted for type.
Let $\Deq$ be the definition consisting of the one clause:
$$
  I = I \defeq  \top\enspace.
$$

We now sketch a proof that $\vdash_{ipa}C$ implies that 
$(C)^{\circ}$ has a $\FOLDN$ derivation using $\Deq$.  The proof is by 
induction on the length of IPA derivation.  
The axioms of intuitionistic logic are derivable in $\FOLDN$ since it 
is complete for intuitionistic logic (the rules for definition and
natural numbers are not needed).
The axioms for equality are derivable using the definition rules with $\Deq$ 
(as noted in \citeN{girard92mail} and \citeN{schroeder-heister93lics}). 
The two formulas concerning $\z$ and {\sl s} are also derivable using 
the definition rules.  The only remaining axiom that needs to be 
considered is that for induction in IPA.  Let $\phi(x)$ be a formula 
with at most $x$ free and let $\cphi x$ be the translation of that 
formula into $\FOLDN$.   We then need to prove that the sequent
$$\Seq{}{\cphi \z\land
         \forall j(\nat j\oimp\cphi j\oimp\cphi{\suc j})\oimp
         \forall x(\nat x\oimp \cphi x)}
$$
is derivable in $\FOLDN$.  Using 
the inference rules $\oimpR$, $\forallR$, and $\cL$, the 
derivability of this sequent can be reduced to the derivability of the 
sequent 
$$\Seq{\cphi \z\land\forall j(\nat j\oimp\cphi j\oimp\cphi{\suc j}),
       \nat I,\nat I}
      {\cphi I}\enspace,
$$
where $I$ is a new eigenvariable.  Consider now deriving this sequent 
with $\natL$, using the induction predicate
$$\lambda w.\ (\cphi \z\land
               \forall j(\nat j\oimp\cphi j\oimp\cphi{\suc j})\land
              \nat w)\supset\cphi w\enspace.$$
The three premises of this instance of $\natL$ are now easily derived.

Second, it may be possible to base the logic $\FOLDN$ on classical instead of
intuitionistic logic.  Since $\FOLDN$ is intended to formalize
informal mathematical reasoning about computation, such a choice might
well be interesting and useful, although none of the many
example applications we have explored require leaving 
intuitionistic logic.  We do not explore a classical version of
$\FOLDN$ here and simply point out that if the classical variant
satisfies a cut-elimination property, a proof of that fact does not
seem to be a straightforward generalization of the proof given in
\citeN{mcdowell97phd} and \citeN{mcdowell00tcs}. 

Finally, we add a word about how $\FOLDN$ can be used to reason about 
computation.  Subsets of intuitionistic logic, such as {\em herditary 
Harrop formulas} or {\em Horn clauses} can be used to specify 
computation using goal-directed derivation search \cite{miller91apal}.  The 
logic $\FOLDN$, which is much stronger than these subsets, can be used 
to reason about logic programs in the following fashion.  Let $\Pscr$ 
be, for example,  a Horn clause program and let $G$ be some goal formula 
(a formula 
composed of conjunctions, disjunctions, and existential quantifiers) 
such that there is goal-directed derivation of the sequent 
$\Seq{\Pscr}{G}$ in intuitionistic logic.  
That derivation is also a cut-free intuitionistic logic 
derivation \cite{miller91apal}.  Thus the sequent $\Seq{}{G}$ has a cut-free 
derivation in $\FOLDN$ using $\Pscr$ as a definition (given the restrictions 
on $G$ and 
$\Pscr$, there are no occurrences of the $\defL$ and $\natL$ inference 
rules in such a derivation).  Now assume that we 
have also a derivation in $\FOLDN$ using $\Pscr$ as a definition of the 
sequent $\Seq{G}{G'}$, for some goal formula $G'$.  Using the 
cut-elimination theorem for $\FOLDN$ (Proposition~\ref{prp:cut-elim}), we know 
that the sequent $\Seq{}{G'}$ has a cut-free derivation in $\FOLDN$ 
using $\Pscr$ as a definition.  Since induction is encoded as a 
left-introduction rule, it is easy to see that the resulting derivation 
does not contain occurrences of induction.  Similarly, there can be no 
occurrences of the $\defL$ rule.  Hence, we can 
conclude that $\Seq{\Pscr}{G'}$ will have an intuitionistic logic 
derivation as well as a goal-directed derivation.  Thus, informally, we can 
conclude that if $G\supset G'$ is derivable in $\FOLDN$ and there 
is a computation proving $G$, then there is computation proving $G'$.  
Hence, implications in the stronger logic can be used to show that the 
existence of certain computations can lead to the existence of other 
computations.  For example, as we have mentioned in 
Proposition~\ref{prp:list-other}, the formula
\begin{displaymath}
\forall l
    (\tlist{l}
     \oimp \forall l_1 \forall l_2
		(\split{l}{l_1}{l_2} \oimp \split{l}{l_2}{l_1}))
\end{displaymath}
can be derived in $\FOLDN$ using $\Dlist{\tau}$.  If we also assume 
that we are given three lists $L_{0},L_{1}, L_{2}$ such that  
$\tlist{L_{0}}$ and $\split{L_{0}}{L_1}{L_2}$ follow from 
$\Dlist{\tau}$ (considered as a Horn clause logic program), then the 
above argument  can be used to show that $\split{L_{0}}{L_2}{L_1}$ 
must also follow from that logic program.

\section*{\stylespart: LOGIC REPRESENTATIONS FOR META-THEORETIC ANALYSIS}

Since $\FOLDN$ contains quantification at higher-order types and 
term structures involving $\lambda$-terms, it easily supports
higher-order abstract syntax.
Eriksson \citeyear{eriksson93elp} demonstrated the use of his finitary
calculus of partial inductive definitions (which is similar to
$\FOLDN$) for the specification of various logics and type systems
using higher-order abstract syntax.
Our goal is to go a step beyond that and also reason within $\FOLDN$
about the object systems.
As we set about to do so, we encounter some difficulties in reasoning
about higher-order abstract syntax specifications within the
specification logic and develop strategies for surmounting those
difficulties.

We begin the first section of this part by presenting the usual
higher-order abstract syntax representation of intuitionistic logic
and illustrating the problems alluded to above.
We then proceed through several modifications of this encoding which
improve our ability to perform meta-theoretic analyses, although
at some loss of the benefits of higher-order abstract syntax.
In Section~\ref{sec:logics} we further illustrate these
encoding techniques through two examples involving fragments of
intuitionistic and linear logic.
The specifications of these two logics will also be used in
\pcfpart\ as part of an alternative strategy for formal
reasoning with higher-order abstract syntax that retains the full
benefits of this representation style.
We conclude the present part with a section discussing related work.

To keep our discussion succinct, we do not prove the adequacy of the 
encodings presented in Section~\ref{sec:styles}.
The skeptical reader is referred to the discussion of similar encodings 
in the literature: see Section~\ref{sec:related1} for references.
The two encodings of Section~\ref{sec:logics}, however, play a key role 
in our work, and so we do include adequacy theorems for these.

\section{A spectrum of encoding styles}
\label{sec:styles}

\subsection{Natural deduction-style encoding}
\label{sec:nat-ded}

In order to examine our ability to reason about higher-order abstract
syntax encodings in $\FOLDN$, we present a definition of first-order
intuitionistic logic.
For brevity we will restrict our discussion here to a fragment of the logic 
containing implication and quantification.
The full logic is considered in \citeN{mcdowell97phd}, though the remaining
connectives do not provide any additional insight.
We use the type \itm\ for terms of the object logic, the type \atm\ for
atoms (atomic propositions) and the type \prp\ for general
propositions; we also introduce the following constants:
\begin{displaymath}
\begin{array}[b]{rcl@{\quad\quad}rcl}
\atom{\;}      &\colon & \atm \rightarrow \prp
& \bigwedge_\itm &\colon & (\itm \rightarrow \prp)\rightarrow \prp \\
\iimp          &\colon & \prp \rightarrow \prp \rightarrow \prp
& \bigvee_\itm   &\colon & (\itm \rightarrow \prp)\rightarrow \prp
\enspace .
\end{array}
\end{displaymath}
The constant $\atom{\;}$ coerces atoms into propositions: object-level
predicates will be constants that build meta-level terms of type \atm.
The constant $\iimp$ represents the implication connective and
$\bigwedge_\itm$ and $\bigvee_\itm$ encode universal and existential
quantification at type $\itm$.
Notice that we are using the $\lambda$-abstraction of $\FOLDN$'s term
language to represent the variable binding of the two object logic
quantifiers.
As a result, $\alpha$-equivalence of quantified object
logic formulas follows from the $\alpha$-equivalence of
$\lambda$-bound terms in $\FOLDN$, and substitution for object logic
variables can be accomplished by $\beta$-reduction at the level of
$\FOLDN$ terms.

Derivability in the object logic is encoded via the predicate
{\sl prove} of type $\prp \rightarrow \oo$; the usual higher-order
abstract syntax encoding of this predicate is the theory shown in
Table~\ref{tab:intuit-nat-ded}.
Here we use $\boimp$ for reverse implication in the meta-logic; 
the first clause, for example, can be rewritten as
\begin{displaymath}
(\prove{B} \oimp \prove{C}) \oimp \prove{(B \iimp C)} \enspace .
\end{displaymath}
The first three clauses correspond to the introduction rules for
natural deduction; the remaining three correspond to the elimination
rules.
\begin{table}[btp]
\caption{Natural deduction encoding of intuitionistic logic}
\label{tab:intuit-nat-ded}
\vspace{2pt}
\begin{center}
$\begin{array}{rcl}
\hline\rule{0pt}{14pt}
\prove{(B \iimp C)}
	& \boimp & \prove{B} \oimp \prove{C} \\
\prove{\bigwedge_\itm B}
	& \boimp & \forall_\itm x\;\prove{(B \, x)} \\
\prove{\bigvee_\itm B}
	& \boimp & \exists_\itm x\;\prove{(B \, x)} \\
\\
\prove{C}
	& \boimp & \exists b(\prove{(b \iimp C)} \; \land \; \prove{b}) \\
\prove{(B \, X)}
 	& \boimp & \prove{\bigwedge_\itm B} \\
\prove{C}
 	& \boimp & \exists b(\prove{\bigvee_\itm b} \; \land \;
		 (\exists_\itm x\;\prove{(b \, x)} \oimp \prove{C}))
\\[2pt]
\hline
\end{array}$
\end{center}
\end{table}

Although this encoding mirrors the rules for natural deduction, we may
view it as an encoding of the sequent calculus, with the
derivability of the sequent $\Seq{B_1,\ldots,B_n}{C}$ represented by
the $\FOLDN$ formula
\begin{displaymath}
\prove{B_1} \oimp \cdots \oimp \prove{B_n} \oimp \prove{C}
\enspace .
\end{displaymath}
This is in keeping with the higher-order abstract syntax principle of
using specification logic hypotheses to represent contexts (in this
case, the left side of the sequent).
The structural rules (exchange, weakening, and contraction) follow
immediately from this representation; for example, the derivation for
weakening is
\settowidth{\infwidthi}
	{$\Seq{}{\forall b \forall c(\prove{c} \oimp (\prove{b} \oimp \prove{c}))}$}
\settowidth{\infwidthii}
	{$\Seq{}{\prove{c} \oimp (\prove{b} \oimp \prove{c})}$}
\begin{displaymath}
\infer{\Seq{}{\forall b \forall c(\prove{c} \oimp (\prove{b} \oimp \prove{c}))}}
      {\infer[\forallR\enspace .]{\makebox[\infwidthi]{}}
	     {\infer{\Seq{}{\prove{c} \oimp (\prove{b} \oimp \prove{c})}}
		    {\infer[\oimpR]{\makebox[\infwidthii]{}}
			   {\infer[\init]
				  {\Seq{\prove{c},\prove{b}}{\prove{c}}}
				  {}}}}}
\end{displaymath}
We use double horizontal lines to represent multiple applications
of an inference rule.
In this case, both the $\oimpR$ rule and the $\forallR$ rule are
applied twice.
The admissibility of the cut rule, encoded by the formula
\begin{displaymath}
\forall b \forall c(
	(\prove{b} \oimp \prove{c}) \oimp
	\prove{b} \oimp \prove{c})
\enspace ,
\end{displaymath}
also follows easily from the $\oimpL$ rule.
The right rules are the same as the corresponding introduction rules,
and the left rules are easily derived from the clauses for the
corresponding elimination rules.
The left rule for $\bigwedge_\itm$, for instance, is encoded by the
$\FOLDN$ formula
\begin{displaymath}
\begin{array}{c}
\forall b \forall c(
	\exists x(\prove{(b \, x)} \oimp \prove{c}) \oimp
	(\prove{\bigwedge_\itm b} \oimp \prove{c}))
\enspace ,
\end{array}
\end{displaymath}
whose derivation is evident from the clause for the elimination rule
for $\bigwedge_\itm$.

However, this encoding is not appropriate for meta-theoretic analysis
of object logic derivations.
To do such analysis in $\FOLDN$, we need to be able to perform
induction over the derivations.
Recall that in Section~\ref{sec:foldn-lists} we used the natural
number measure in the {\sl length} predicate to derive an induction
principle for lists.
But there is no apparent way to add a natural number induction measure
to the {\sl prove} predicate because of the clause for the $\iimp$
introduction rule.
This reflects the fact that this clause gives rise to a non-monotone
operator; this is generally true of the types and theories in 
higher-order abstract syntax encodings, and makes inductive 
principles difficult to find.
We would also like to change the specification into a definition so
that we can use the $\defL$ rule for the analysis of derivations.
Simply replacing the $\boimp$ in each clause by $\defeq$ is
problematic for two reasons.
First, the clause resulting from the introduction rule for $\iimp$
would not satisfy the level restriction for any level we might assign
to {\sl prove}.
Second, the clause resulting from the elimination rule for
$\bigwedge_\itm$ would have a problematic head.
There are too many ways that $(B\,X)$ can match and unify with other
terms; this makes the practical application of the $\defR$ and $\defL$
rules difficult and would result in many cases that are not
productive.

\subsection{Sequent calculus-style encoding}
\label{sec:impl-seq}

We can solve the problems with the encoding of the introduction rule
for $\iimp$ by introducing separate predicates
\begin{displaymath}
\begin{array}{rcl@{\quad\quad\quad}rcl}
\hbox{\sl hyp}      &\colon & \prp \rightarrow \oo
& \hbox{\sl conc}     &\colon & \nt \rightarrow \prp \rightarrow \oo
\end{array}
\end{displaymath}
for the left and right sides of the sequent, respectively.
The predicate {\sl hyp} will not be a defined predicate, and so can
have level zero.
The negative occurrence of {\sl prove} in the introduction clause for
$\iimp$ becomes an occurence of {\sl hyp}, so the predicate {\sl conc}
can then have level one.
This also makes possible the assignment of a measure to {\sl conc}, as
suggested by its type.
To emphasize that the first argument to {\sl conc} is a measure, we
will write it as a subscript.
The problem introduced by the elimination clause for $\bigwedge_\itm$
is avoided by patterning the encoding after the sequent calculus rules
rather than natural deduction rules.
The resulting definition is shown in Table~\ref{tab:intuit-impl-seq}.
The first clause encodes the initial axiom, the next three correspond
to the right introduction rules, and the remaining three correspond to
the left introduction rules.
\begin{table}[btp]
\caption{Sequent calculus encoding of intuitionistic logic}
\label{tab:intuit-impl-seq}
\vspace{2pt}
\begin{center}
$\begin{array}{rcl}
\hline\rule{0pt}{14pt}
\conc{I}{\atom{A}}
	& \defeq & \hyp{\atom{A}} \\
\\
\conc{(\suc{I})}{(B \iimp C)}
 	& \defeq & \hyp{B} \oimp \conc{I}{C} \\
\conc{(\suc{I})}{\bigwedge_\itm B}
	& \defeq & \forall_\itm x\;\conc{I}{(B \, x)} \\
\conc{(\suc{I})}{\bigvee_\itm B}
	& \defeq & \exists_\itm x\;\conc{I}{(B \, x)} \\
\\
\conc{(\suc{I})}{D}
	& \defeq & \exists b\exists c
			(\hyp{(b \iimp c)} \; \land \;
			 (\hyp{c} \oimp \conc{I}{D}) \; \land \;
			 \conc{I}{b}) \\
\conc{(\suc{I})}{C}
 	& \defeq & \exists b
			(\hyp{\bigwedge_\itm b} \; \land \;
			 (\forall_\itm x \; \hyp{(b \, x)} \oimp \conc{I}{C})) \\
\conc{(\suc{I})}{C}
 	& \defeq & \exists b
			(\hyp{\bigvee_\itm b} \; \land \;
			 (\exists_\itm x \; \hyp{(b \, x)} \oimp \conc{I}{C}))
\\[2pt]
\hline
\end{array}$
\end{center}
\end{table}

Since we have not changed the representation of quantification, we
get $\alpha$-equi\-val\-ence of quantified object logic formulas and
substitution for object logic variables from the relevant features of
$\FOLDN$ as before.
We are still using $\FOLDN$ hypotheses to represent contexts, so
the structural rules also follow as before.
However, the admissibility of the cut rule, now encoded as
\begin{displaymath}
\forall b \forall c(
	\exists i(\hyp{b} \oimp \conc{i}{c}) \oimp
	\exists i\;\conc{i}{b} \oimp \exists i\;\conc{i}{c})
\enspace ,
\end{displaymath}
is no longer immediate:  there is no simple proof of 
$\Seq{\exists i\;\conc{i}{b}}{\hyp{b}}$.
We expect, though, that the admissibility of cut is still derivable in
$\FOLDN$ following the method of \citeN{pfenning95lics}.

This encoding has another limitation; to see it, consider the
following example.
Suppose we know that the sequent $\Seq{b \iimp a}{a}$ is derivable in
intuitionistic logic for some atom $a$ and proposition $b$.
Since $a$ is atomic, the derivation must end with a left rule, and
since the only formula on the left is $b \iimp a$, it must be the left
implication rule.
Thus there are derivations of $\Seq{b \iimp a}{b}$ and
$\Seq{a,b \iimp a}{a}$.
This second sequent is not so interesting, since it is an initial
sequent.
So we have shown that if $\Seq{b \iimp a}{a}$ is derivable then
$\Seq{b \iimp a}{b}$ is as well.

Now let us try to capture this reasoning in $\FOLDN$ using our current
encoding of intuitionistic logic.
We want to derive the sequent
\begin{displaymath}
\Seq{}{\forall a \forall b(\exists i(\hyp{(b \iimp \atom{a})} \oimp \conc{i}{\atom{a}}) \oimp
	\exists j(\hyp{(b \iimp \atom{a})} \oimp \conc{j}{b}))}
\enspace .
\end{displaymath}
After the obvious uses of $\forallR$ and $\oimpR$, we get
\begin{displaymath}
\Seq{\exists i(\hyp{(b \iimp \atom{a})} \oimp \conc{i}{\atom{a}})}
    {\exists j(\hyp{(b \iimp \atom{a})} \oimp \conc{j}{b})}
\enspace .
\end{displaymath}
From our informal reasoning, we know that the derivation of $b$ will
have a smaller measure than the derivation of $a$; thus in applying the
$\existsL$ and $\existsR$ rules it is conservative to substitute $i$
for $j$:
\begin{displaymath}
\Seq{\hyp{(b \iimp \atom{a})} \oimp \conc{i}{\atom{a}}}
    {\hyp{(b \iimp \atom{a})} \oimp \conc{i}{b}}
\enspace .
\end{displaymath}
To follow the informal proof, we now want to indicate that
$\hyp{(b \iimp \atom{a})} \oimp \conc{i}{\atom{a}}$ must be true by the
definitional clause encoding the left $\iimp$ rule.
However, we cannot apply the $\defL$ rule to this formula, since it is
not an atom.
The closest thing to this that we can do is to eliminate the $\oimp$
and then apply $\defL$ to $\conc{i}{\atom{a}}$.
We can eliminate the $\oimp$ by using $\oimpR$ and then $\oimpL$,
yielding the two sequents
\begin{displaymath}
\Seq{\hyp{(b \iimp \atom{a})}}
    {\hyp{(b \iimp \atom{a})}}
\end{displaymath}
\begin{displaymath}
\Seq{\conc{i}{\atom{a}},\hyp{(b \iimp \atom{a})}}
    {\conc{i}{b}}
\enspace .
\end{displaymath}
The first is immediate by the $\init$ rule.
Applying the $\defL$ rule to $\conc{i}{\atom{a}}$ in the second
sequent yields four sequents corresponding to the cases where the
derivation of $a$ ends with the initial rule or any of the three left
rules:
\begin{displaymath}
\Seq{\hyp{\atom{a}},\hyp{(b \iimp \atom{a})}}
    {\conc{i}{b}}
\end{displaymath}
\begin{displaymath}
\begin{array}[t]{@{}l}
\exists b'\exists c'(\hyp{(b' \iimp c')} \; \land \;
		(\hyp{c'} \oimp \conc{i'}{\atom{a}}) \; \land \;
		\conc{i'}{b'}),\hyp{(b \iimp \atom{a})}
\longrightarrow\qquad\hfill\\
\hfill\conc{(\suc{i'})}{b}
\end{array}
\end{displaymath}
\begin{displaymath}
\begin{array}{c}
\Seq{\exists b'(\hyp{\bigwedge_\itm b'} \; \land \;
		(\forall_\itm x \; \hyp{(b' \, x)} \oimp \conc{i'}{\atom{a}})),
     \hyp{(b \iimp \atom{a})}}
    {\conc{(\suc{i'})}{b}}
\end{array}
\end{displaymath}
\begin{displaymath}
\begin{array}{c}
\Seq{\exists b'(\hyp{\bigvee_\itm b'} \; \land \;
		(\exists_\itm x \; \hyp{(b' \, x)} \oimp \conc{i'}{\atom{a}})),
     \hyp{(b \iimp \atom{a})}}
    {\conc{(\suc{i'})}{b}}
\enspace .
\end{array}
\end{displaymath}
This is clearly not what we want.
Even in the case corresponding to the left $\iimp$ rule we do not know
that the rule was applied to the implication $b \iimp \atom{a}$.
There are really two problems here.
The first is that $\hyp{(b \iimp \atom{a})} \oimp \conc{i}{\atom{a}}$
expresses the idea that $b \iimp \atom{a}$ is a hypothesis available
in the derivation of $\conc{i}{\atom{a}}$, but it does not capture the
idea that it is the only hypothesis available.
Thus the $\defL$ rule forces us to consider derivations ending with
the initial rule or any of the left rules, since the appropriate
formula may be available as a hypothesis.
The second problem is that we do not have any way to examine the
different ways of deriving something from a specific set of
hypotheses.
Although the formula $\hyp{(b \iimp \atom{a})} \oimp \conc{i}{\atom{a}}$
indicates that the atom $a$ is derivable from the hypothesis
$b \iimp \atom{a}$, we cannot examine how that derivation might take
place.
All we can do is use the $\oimpL$ rule, which says that we know that
the hypothesis $b \iimp \atom{a}$ is available and so can conclude
that $a$ holds.

\subsection{Explicit sequent encoding}
\label{sec:expl-seq}

To remedy this situation, we explicitly represent the entire sequent
in a single atomic judgement.
As a result, the relevant object logic hypotheses are known to be
exactly those listed in the judgement, and the $\defL$ rule can be
applied to the judgement to examine how the corresponding sequent
might be derived.
Thus derivability is encoded via the predicate
\begin{displaymath}
\begin{array}{rcl}
\hbox{\sl seq}     & \colon & \nt \rightarrow \prplst \rightarrow \prp \rightarrow \oo
\enspace .
\end{array}
\end{displaymath}
The first argument is an induction measure and will be displayed as a
subscript.
The second argument is a list of terms of type $\prp$ and represents
the left side of the sequent.
We will assume that $\prplst$ is the same as the type $\lst$
introduced in Section~\ref{sec:foldn-lists}, using $\prp$ for the type
of elements.
In particular we will assume that we have constructors $\nil$ and
$::$, and a predicate {\sl element} as defined in $\Dlist{\prp}$.
The third argument to {\sl seq} corresponds to the right side of
the sequent.
The definition for this predicate is shown in Table~\ref{tab:intuit-expl-seq}.
\begin{table}[btp]
\caption{Explicit sequent encoding of intuitionistic logic}
\label{tab:intuit-expl-seq}
\vspace{2pt}
\begin{center}
$\begin{array}{@{}rcl@{}}
\hline\rule{0pt}{14pt}
\seq{I}{L}{\atom{A}}
        & \defeq & \element{\atom{A}}{L} \\
\\
\seq{(\suc{I})}{L}{(B \iimp C)}
        & \defeq & \seq{I}{(\cons{B}{L})}{C} \\
\seq{(\suc{I})}{L}{(\bigwedge_\itm B)}
        & \defeq & \forall_\itm x\;\seq{I}{L}{(B \, x)} \\
\seq{(\suc{I})}{L}{(\bigvee_\itm B)}
        & \defeq & \exists_\itm x\;\seq{I}{L}{(B \, x)} \\
\\
\seq{(\suc{I})}{L}{D}
	& \defeq & \exists b\exists c
			(\element{(b \iimp c)}{L} \; \land \;
			 \seq{I}{(\cons{c}{L})}{D} \; \land \;
			 \seq{I}{L}{b}) \\
\seq{(\suc{I})}{L}{C}
 	& \defeq & \exists b
			(\element{\bigwedge_\itm b}{L} \; \land \;
			 \exists_\itm x \; \seq{I}{(\cons{(b \, x)}{L})}{C}) \\
\seq{(\suc{I})}{L}{C}
 	& \defeq & \exists b
			(\element{\bigvee_\itm b}{L} \; \land \;
			 \forall_\itm x \; \seq{I}{(\cons{(b \, x)}{L})}{C})
\\[2pt]
\hline
\end{array}$
\end{center}
\end{table}

Since we have not changed the representation of quantification, we
get $\alpha$-equi\-val\-ence of quantified object logic formulas and
substitution for object logic variables from the relevant features of
$\FOLDN$ as before.
We are no longer using $\FOLDN$ hypotheses to represent contexts,
however, so the structural rules must now be derived by induction.
The admissibility of the cut rule must also be derived by induction,
as was the case with the previous encoding.
With the atomic encoding of sequents, we now can analyze derivations
of propositions from hypotheses.
To see this, we revisit the example from above.
To formalize this example with the encoding of
Table~\ref{tab:intuit-expl-seq}, we derive the sequent
\begin{displaymath}
\Seq{}{\forall a \forall b(\exists i\;\seq{i}{(\cons{(b \iimp \atom{a})}{\nil})}{\atom{a}} \oimp
	\exists j\;\seq{j}{(\cons{(b \iimp \atom{a})}{\nil})}{b})\enspace.}
\end{displaymath}
Applying the $\forallR$, $\oimpR$, and $\existsL$ rules yields the sequent
\begin{displaymath}
\Seq{\seq{i}{(\cons{(b \iimp \atom{a})}{\nil})}{\atom{a}}}
    {\exists j\;\seq{j}{(\cons{(b \iimp \atom{a})}{\nil})}{b}}
\enspace .
\end{displaymath}
Now we apply the $\defL$ rule to the judgement on the left, which
yields four sequents, again corresponding to the cases where the
derivation of $a$ ends with the initial rule or any of the three left
rules:
\begin{displaymath}
\Seq{\element{\atom{a}}{(\cons{(b \iimp \atom{a})}{\nil})}}
    {\exists j\;\seq{j}{(\cons{(b \iimp \atom{a})}{\nil})}{b}}
\end{displaymath}
\begin{displaymath}
\Seq{\begin{array}[b]{r@{}}
	\exists b'\exists c'(
		\element{(b' \iimp c')}{(\cons{(b \iimp \atom{a})}{\nil})} \; \land \\
		\seq{i'}{(\cons{c'}{\cons{(b \iimp \atom{a})}{\nil}})}{\atom{a}} \; \land \\
		\seq{i'}{(\cons{(b \iimp \atom{a})}{\nil})}{b'})
     \end{array}}
    {\exists j\;\seq{j}{(\cons{(b \iimp \atom{a})}{\nil})}{b}}
\end{displaymath}
\begin{displaymath}
\exists b'
	(\begin{array}[t]{@{}l@{}}
	 \element{\bigwedge_\itm b'}{(\cons{(b \iimp \atom{a})}{\nil})} \; \land \\
\Seq{
	 \exists_\itm x\;\seq{i'}{(\cons{(b' \, x)}{\cons{(b \iimp \atom{a})}{\nil}})}
				 {\atom{a}})}
    {\exists j\;\seq{j}{(\cons{(b \iimp \atom{a})}{\nil})}{b}}
	 \end{array}
\end{displaymath}
\begin{displaymath}
\exists b'
	(\begin{array}[t]{@{}l@{}}
	 \element{\bigvee_\itm b'}{(\cons{(b \iimp \atom{a})}{\nil})} \; \land \\
\Seq{
	 \forall_\itm x\;\seq{i'}{(\cons{(b' \, x)}{\cons{(b \iimp \atom{a})}{\nil}})}
				 {\atom{a}})}
    {\exists j\;\seq{j}{(\cons{(b \iimp \atom{a})}{\nil})}{b\enspace .}}
     \end{array}
\end{displaymath}
But this time we can easily eliminate three of the four possibilities,
since the {\sl element} assumption is obviously false.
In the first sequent, for example, we have the assumption
$\element{\atom{a}}{(\cons{(b \iimp \atom{a})}{\nil})}$.
Since $\atom{a}$ cannot unify with $(b \iimp \atom{a})$, $\atom{a}$
cannot be the first element of the list; therefore it must be an
element of the remainder.
But the remainder is the empty list, so $\atom{a}$ cannot be an element
of it either.
This is accomplished formally by applying the $\defL$ rule twice:
\begin{displaymath}
\infer[\defL\enspace .]
      {\Seq{\element{\atom{a}}{(\cons{(b \iimp \atom{a})}{\nil})}}
	   {\exists j\;\seq{j}{(\cons{(b \iimp \atom{a})}{\nil})}{b}}}
      {\infer[\defL]
	     {\Seq{\element{\atom{a}}{\nil}}
		  {\exists j\;\seq{j}{(\cons{(b \iimp \atom{a})}{\nil})}{b}}}
	     {}}
\end{displaymath}
The remaining cases are done similarly, except for the one valid case,
which corresponds to a use of the left $\iimp$ rule:
\begin{displaymath}
\Seq{\begin{array}[b]{r@{}}
	\exists b'\exists c'(
		\element{(b' \iimp c')}{(\cons{(b \iimp \atom{a})}{\nil})} \; \land \; \\
		\seq{i'}{(\cons{c'}{\cons{(b \iimp \atom{a})}{\nil}})}{\atom{a}} \; \land \; \\
		\seq{i'}{(\cons{(b \iimp \atom{a})}{\nil})}{b'})
     \end{array}}
    {\exists j\;\seq{j}{(\cons{(b \iimp \atom{a})}{\nil})}{b}}
\enspace .
\end{displaymath}
In this case, $b' \iimp c'$ does match the first element of the list,
so we must consider the case where the left $\iimp$ rule was applied
to $(b \iimp \atom{a})$:
\settowidth{\infwidthi}
	{$\Seq{\element{(b' \iimp c')}
		       {(\cons{(b \iimp \atom{a})}{\nil})} \land \ldots}
	      {\exists j\;\seq{j}{(\cons{(b \iimp \atom{a})}{\nil})}{b}}$}
\settowidth{\infwidthii}
	{$\Seq{\element{(b' \iimp c')}{(\cons{(b \iimp \atom{a})}{\nil})},
	       \seq{i'}{(\cons{(b \iimp \atom{a})}{\nil})}{b'}}
	      {\exists j\ldots}$}
\begin{displaymath}
\infer{\Seq{\ldots}{\exists j\;\seq{j}{(\cons{(b \iimp \atom{a})}{\nil})}{b}}}
      {\infer[\existsL\enspace .]{\makebox[\infwidthi]{}}
	     {\infer[\cL]
		    {\Seq{\element{(b' \iimp c')}
				  {(\cons{(b \iimp \atom{a})}{\nil})} \land \ldots}
			 {\exists j\;\seq{j}{(\cons{(b \iimp \atom{a})}{\nil})}{b}}}
		    {\infer{\Seq{\element{(b' \iimp c')}{\ldots} \land \ldots,
				 \element{(b' \iimp c')}{\ldots} \land \ldots}
				{\exists j\ldots}}
			   {\infer[\landL]{\makebox[\infwidthii]{}}
		 		  {\infer[\defL]
					 {\Seq{\element{(b' \iimp c')}
						       {(\cons{(b \iimp \atom{a})}{\nil})},
					       \seq{i'}{(\cons{(b \iimp \atom{a})}{\nil})}{b'}}
					      {\exists j\ldots}}
					 {\Seq{\top,\seq{i'}{(\cons{(b \iimp \atom{a})}{\nil})}{b}}
					      {\exists j\ldots}
					 & \infer[\defL]
						 {\Seq{\element{(b' \iimp c')}{\nil}, \ldots}
						      {\exists j\ldots}}
						 {}}}}}}}
\end{displaymath}
But the unproved sequent is easily derived by choosing $j$ to be $i'$:
\begin{displaymath}
\infer[\existsR\enspace .]
      {\Seq{\top,\seq{i'}{(\cons{(b \iimp \atom{a})}{\nil})}{b}}
	   {\exists j\;\seq{j}{(\cons{(b \iimp \atom{a})}{\nil})}{b}}}
      {\infer[\init]
	     {\Seq{\top,\seq{i'}{(\cons{(b \iimp \atom{a})}{\nil})}{b}}
		  {\seq{i'}{(\cons{(b \iimp \atom{a})}{\nil})}{b}}}
	     {}}
\end{displaymath}

Now let us consider another example.
Suppose we know that the sequent
\begin{displaymath}
\Seq{}{\bigwedge y_1 \bigwedge y_2
	(p \, y_1 \, t_1 \iimp p \, y_2 \, t_2 \iimp p \, y_2 \, t_3)}
\end{displaymath}
is derivable in intuitionistic logic for some predicate constant $p$
and some terms $t_1$, $t_2$, and $t_3$.
The derivation must end with applications of the right rules for
$\bigwedge$ and $\iimp$, since these are the only rules that apply.
Thus we know that the sequent
$\Seq{p \, y_1 \, t_1, p \, y_2 \, t_2}{p \, y_2 \, t_3}$
is derivable.
Since $p$ is a predicate constant, these formulas are all atomic, so
the only rule that applies is the initial rule.
The eigenvariable condition for the application of the right rule for
$\bigwedge$ guarantees that $y_1$ and $y_2$ are distinct, so the
initial rule must apply to the second hypothesis.
Therefore, it must be the case that $t_2$ and $t_3$ are the same term.

Now let us try to capture this reasoning in $\FOLDN$ using our current
encoding of intuitionistic logic.
To do this, we will need some way to indicate term identity, and so we
introduce the predicate $\equiv$ of type
$\itm \rightarrow \itm \rightarrow \oo$ defined by the clause
$\same{X}{X} \defeq \top$. 
We then want to derive the sequent
\begin{displaymath}
\begin{array}{c}
\mathord{\longrightarrow}\forall p \forall t_1 \forall t_2 \forall t_3
	(\exists i \; \seq{i}{\nil}{\bigwedge_\itm y_1 \bigwedge_\itm y_2
		(\atom{p \, y_1 \, t_1}
		 \iimp \atom{p \, y_2 \, t_2}
		 \iimp \atom{p \, y_2 \, t_3})}
	 \oimp \same{t_2}{t_3}).
\end{array}
\end{displaymath}
The only way to proceed is by applying $\forallR$ and $\oimpR$, yielding
\begin{displaymath}
\begin{array}{c}
\Seq{\exists i \; \seq{i}{\nil}{\bigwedge_\itm y_1 \bigwedge_\itm y_2
	(\atom{p \, y_1 \, t_1}
	 \iimp \atom{p \, y_2 \, t_2}
	 \iimp \atom{p \, y_2 \, t_3})}}
    {\same{t_2}{t_3}}
\enspace .
\end{array}
\end{displaymath}
There is nothing more that we can do on the right, since the
definitional clause for $\equiv$ does not apply.
Applying $\existsL$ gives us the sequent
\begin{displaymath}
\begin{array}{c}
\Seq{\seq{i}{\nil}{\bigwedge_\itm y_1 \bigwedge_\itm y_2
	(\atom{p \, y_1 \, t_1}
	 \iimp \atom{p \, y_2 \, t_2}
	 \iimp \atom{p \, y_2 \, t_3})}}
    {\same{t_2}{t_3}}
\enspace .
\end{array}
\end{displaymath}
Now we want to reason about the derivation of
$\bigwedge_\itm y_1 \bigwedge_\itm y_2 \ldots$
to conclude that $\same{t_2}{t_3}$.
In the informal proof, we reasoned that this derivation must end with
the right rule for $\bigwedge$; we do the same thing here using
$\defL$, which yields the sequent
\begin{displaymath}
\begin{array}{c}
\Seq{\forall y_1\;\seq{i_1}{\nil}{\bigwedge_\itm y_2
	(\atom{p \, y_1 \, t_1}
	 \iimp \atom{p \, y_2 \, t_2}
	 \iimp \atom{p \, y_2 \, t_3})}}
    {\same{t_2}{t_3}}
\enspace ,
\end{array}
\end{displaymath}
as well as three other sequents corresponding to the cases where the
object logic derivation ends with the application of one of the left
rules.
Since these latter three sequents represent cases that are not
applicable, they are easily derivable as shown in the previous
example; we thus focus on the sequent shown above.
Before we can proceed to apply $\defL$ again for the second use of the
right rule for $\bigwedge$, we must first apply $\forallL$, which
requires supplying a substitution term for $y_1$.
For this proof, it doesn't matter what term we use for $y_1$, as long
as it is something that does not unify with the term we supply for
$y_2$.
So let $x_1$ and $x_2$ be two distinct, non-unifiable terms of type
$\itm$.
If we use $x_1$ for $y_1$, and then apply $\defL$ and $\forallL$ again
using $x_2$ for $y_2$, we get
\begin{displaymath}
\Seq{\seq{i_2}{\nil}{(\atom{p \, x_1 \, t_1}
     \iimp \atom{p \, x_2 \, t_2}
     \iimp \atom{p \, x_2 \, t_3})}}
    {\same{t_2}{t_3}}
\enspace .
\end{displaymath}
We now apply $\defL$ two more times, each of which corresponds to
reasoning that the object logic derivation must proceed with a use of
the right rule for $\iimp$.
This yields the sequent
\begin{displaymath}
\Seq{\seq{i_3}{(\cons{\atom{p \, x_2 \, t_2}}
		     {\cons{\atom{p \, x_1 \, t_1}}{\nil}})}
	    {\atom{p \, x_2 \, t_3}}}
    {\same{t_2}{t_3}}
\enspace .
\end{displaymath}
Another application of $\defL$ reflects the fact that in the
object logic derivation only the initial rule now applies:
\begin{displaymath}
\Seq{\element{\atom{p \, x_2 \, t_3}}
	     {(\cons{\atom{p \, x_2 \, t_2}}
		    {\cons{\atom{p \, x_1 \, t_1}}{\nil}})}}
    {\same{t_2}{t_3}}
\enspace .
\end{displaymath}
For $\atom{p \, x_2 \, t_3}$ to be the first element of the list,
$t_2$ and $t_3$ must be the same, and this is what we want to prove.
We have chosen $x_1$ and $x_2$ to be terms that do not unify, so
$\atom{p \, x_2 \, t_3}$ cannot be the other element of the list.
This reasoning is represented formally by the $\FOLDN$ derivation
\begin{displaymath}
\infer[\defL\enspace .]
      {\Seq{\element{\atom{p \, x_2 \, t_3}}
		    {(\cons{\atom{p \, x_2 \, t_2}}
			   {\cons{\atom{p \, x_1 \, t_1}}{\nil}})}}
	   {\same{t_2}{t_3}}}
      {\infer[\defR]
	     {\Seq{\top}{\same{t_2}{t_2}}}
	     {\infer[\topR]
		    {\Seq{\top}{\top}}
		    {}}
      & \infer[\defL]
	     {\Seq{\element{\atom{p \, x_2 \, t_3}}
			   {(\cons{\atom{p \, x_1 \, t_1}}{\nil})}}
		  {\same{t_2}{t_3}}}
	     {\infer[\defL]
		    {\Seq{\element{\atom{p \, x_2 \, t_3}}{\nil}}
			 {\same{t_2}{t_3}}}
		    {}}}
\end{displaymath}

If we are able to construct the two non-unifiable terms $x_1$ and
$x_2$, we are able to conduct this analysis in $\FOLDN$.
But the need for these two terms is rather disturbing.
The informal proof is independent of the type of $y_1$ and $y_2$ and
the term structure of this type.
In fact, the informal proof is valid even for a type that is
uninhabited; this is obviously not the case for our representation in
$\FOLDN$.
The problem is that our representation of object-level quantification
in terms of $\FOLDN$ quantification doesn't allow us to examine a
derivation that is generic over certain terms.
Although the formula $\forall y\;\seq{i}{L}{(B \, y)}$
indicates that the proposition $B \, y$ is derivable from the
hypotheses in $L$ for any $y$, it does not indicate that the
derivation is the same for all $y$, and we cannot examine that
derivation generically.
All we can do is use the $\forallL$ rule, which requires us to
substitute a specific term for $y$, and then examine the derivation
for that specific term.
This is analagous to the problem we encountered before related to the
encoding of object logic implication in terms of $\FOLDN$ implication.

\subsection{Explicit eigenvariable encoding}
\label{sec:expl-eigen}

To solve this problem we must explicitly keep track of the
eigenvariables introduced by the quantifier rules.
We do not wish to abandon, however, our higher-order abstract syntax
representation of quantification.
In the earlier encodings of this section, we encoded the rules for
object logic quantification using $\FOLDN$ quantification; the
key idea of our solution is to replace that use of $\FOLDN$
quantification with the use of $\FOLDN$ $\lambda$-abstraction.
If we follow this idea naively and simply replace the quantification
by $\lambda$-abstraction, we get the following encoding of the right
rule for $\bigwedge$:
\begin{displaymath}
\begin{array}{rcl}
\seq{(\suc{I})}{L}{(\bigwedge_\itm B)}
        & \defeq & \lambda x\;\seq{I}{L}{(B \, x)}
\enspace .
\end{array}
\end{displaymath}
This does not work, of course, since the body of this clause now has
type $\itm \rightarrow \oo$ instead of type $\oo$.
To address this problem, it is important to first realize that as more
eigenvariables are added and propositions are moved between the left
and right sides of the sequent, we must deal more generally with
``judgements'' of the form 
\begin{displaymath}
\lambda x_1 \ldots \lambda x_n\;
	\seq{I}{(L \, x_1 \ldots x_n)}{(B \, x_1 \ldots x_n)}
\end{displaymath}
for arbitrary $n \geq 0$.
First consider ``uncurrying'' this expression by replacing the
$\lambda$-abstractions over $x_1, \ldots, x_n$ by a single
$\lambda$-abstraction over the $n$-tuple $(x_1, \ldots, x_n)$:
\begin{displaymath}
\lambda x.
	\seq{I}{(L \, (\proj{1}{x}) \ldots (\proj{n}{x}))}
	       {(B \, (\proj{1}{x}) \ldots (\proj{n}{x}))}
\enspace .
\end{displaymath}
Now we can deal with the arbitrary $n$ by replacing the $n$-tuple with
a list, and using $\fst{x}$ in place of $\proj{1}{x}$,
$\fst{(\rst{x})}$ in place of $\proj{2}{x}$,
$\fst{(\rst{(\rst{x})})}$ in place of $\proj{3}{x}$, etc.
Finally, we push the $\lambda$-abstraction into the {\sl seq}
predicate by changing its type:
\begin{displaymath}
\begin{array}{rcl}
\hbox{\sl seq}     & \colon & \nt \rightarrow
			 (\evars \rightarrow \prplst) \rightarrow
			 (\evars \rightarrow \prp) \rightarrow
			 \oo
\enspace .
\end{array}
\end{displaymath}
Here $\evars$ is a new type representing a list of eigenvariables.
We have already seen the two operators on this type,
$\hbox{\sl fst} \colon \evars \rightarrow \itm$ and
$\hbox{\sl rst} \colon \evars \rightarrow \evars$;
$\fst{l}$ represents the first eigenvariable in the list $l$,
and $\rst{l}$ represents the remainder of the list.
The right rule for $\bigwedge$ is now encoded as follows:
\begin{displaymath}
\begin{array}{rcl}
\seq{(\suc{I})}{L}{(\lambda l \bigwedge_\itm x (B \, l \, x))}
        & \defeq & \seq{I}{(\lambda l'\;L (\rst{l'}))}
			  {(\lambda l'\;B \, (\rst{l'}) \, (\fst{l'}))}
\enspace .
\end{array}
\end{displaymath}
The bound variable $l'$ in the body of the clause should be thought of as a
list whose length is one longer than the length of the bound variable $l$
in the head of the clause; $\fst{l'}$ represents the new
eigenvariable, and $\rst{l'}$ represents the eigenvariables in $l$.
The left rule for $\bigvee_\itm$ is similarly modified:
\begin{displaymath}
\begin{array}[b]{rcl}
\seq{(\suc{I})}{L}{C}
 	& \defeq & \exists b
			(\begin{array}[t]{@{}l@{}}
			 \element{(\lambda l \bigvee_\itm x (b \, l \, x))}{L} \; \land \; \\
			 \seq{I}{(\lambda l'\;\cons{(b \, (\rst{l'}) \, (\fst{l'}))}
						   {(L \, (\rst{l'}))})}
				{(\lambda l'\;C \, (\rst{l'}))}
\enspace .
			 \end{array}
\end{array}
\end{displaymath}
The remainder of the clauses are only modified to reflect the change
in the type of {\sl seq}.
Note in particular that $\FOLDN$ quantification can still be used in
the encodings of the left rule for $\bigwedge$ and the right rule for
$\bigvee$;
since these rules do not introduce eigenvariables, this use of
$\FOLDN$ quantification is not problematic.
The type of the predicate {\sl element} also changes:
\begin{displaymath}
\begin{array}[b]{@{}rcl@{\quad\quad}rcl@{}}
\hbox{\sl element}	& \colon & (\evars \rightarrow \prp)
			 \rightarrow (\evars \rightarrow \prplst)
			 \rightarrow \oo
\enspace .
\end{array}
\end{displaymath}
Table~\ref{tab:intuit-expl-eigen} presents the definition for the
entire logic.
\begin{table}[btp]
\caption{Explicit eigenvariable encoding of intuitionistic logic}
\label{tab:intuit-expl-eigen}
\vspace{2pt}
\begin{center}
$\begin{array}{@{}rcl@{}}
\hline\rule{0pt}{14pt}
\seq{I}{L}{\lambda l\;\atom{(A \, l)}}
        & \defeq & \element{\lambda l\;\atom{(A \, l)}}{L} \\
\\
\seq{(\suc{I})}{L}{\lambda l\;((B \, l) \iimp (C \, l))}
        & \defeq & \seq{I}{\lambda l\;(\cons{(B \, l)}{(L \, l)})}{C} \\
\seq{(\suc{I})}{L}{(\lambda l \bigwedge_\itm x (B \, l \, x))}
        & \defeq & \seq{I}{(\lambda l'\;L \, (\rst{l'}))}
			  {(\lambda l'\; B \, (\rst{l'}) \, (\fst{l'}))} \\
\seq{(\suc{I})}{L}{(\lambda l \bigvee_\itm x (B \, l \, x))}
        & \defeq & \exists x\;
			\seq{I}{L}{(\lambda l\;B \, l \, (x \, l))} \\
\\
\seq{(\suc{I})}{L}{D}
	& \defeq & \exists b\exists c
			(\begin{array}[t]{@{}l@{}}
			 \element{\lambda l\;((b \, l) \iimp (c \, l))}{L} \; \land \; \\
			 \seq{I}{\lambda l\;(\cons{(c \, l)}{(L \, l)})}{D} \; \land \; \\
			 \seq{I}{L}{b})
			 \end{array} \\
\seq{(\suc{I})}{L}{C}
	& \defeq & \exists b
			(\begin{array}[t]{@{}l@{}}
			 \element{(\lambda l \bigwedge_\itm x (b \, l \, x))}{L} \; \land \; \\
			 \exists x\;
				\seq{I}{\lambda l\;(\cons{(b \, l \, (x \, l))}{(L \, l)})}{C})
			 \end{array} \\
\seq{(\suc{I})}{L}{C}
	& \defeq & \exists b
			(\begin{array}[t]{@{}l@{}}
			 \element{(\lambda l \bigvee_\itm x (b \, l \, x))}{L} \; \land \; \\
			 \seq{I}{\lambda l'\;(\cons{(b \, (\rst{l'}) \, (\fst{l'}))}
					     {(L \, (\rst{l'}))})}
				{(\lambda l'\;C \, (\rst{l'}))})
			 \end{array} \\
\\
\element{X}{\lambda l\;(\cons{(X \, l)}{(L \, l)})}
	& \defeq & \top \\
\element{X}{\lambda l\;(\cons{(Y \, l)}{(L \, l)})}
	& \defeq & \element{X}{L}
\\[2pt]
\hline
\end{array}$
\end{center}
\end{table}

Since we have not changed the representation of quantification, we
get $\alpha$-equi\-val\-ence of quantified object logic formulas and
substitution for object logic bound variables from the relevant
features of $\FOLDN$ as before.
Substitution for eigenvariables is a little more involved, as
shown by its encoding via the predicates
\begin{displaymath}
\begin{array}{rcl}
\hbox{\sl subst} & \colon & nt \rightarrow
			(\evars \rightarrow \itm) \rightarrow
			(\evars \rightarrow \itm) \rightarrow
			(\evars \rightarrow \itm) \rightarrow \oo \\
\hbox{\sl subst}_0 & \colon & nt \rightarrow
			(\evars \rightarrow \evars \rightarrow \itm) \rightarrow
			(\evars \rightarrow \evars \rightarrow \itm) \rightarrow
			(\evars \rightarrow \evars \rightarrow \itm) \rightarrow \oo
\enspace .
\end{array}
\end{displaymath}
The judgement
$\subst{i}{t_1}{t_2}{t_2'}$ indicates that $t_2'$ is the result of substituting $t_1$
in $t_2$ for the $(i+1)^{\rm th}$ eigenvariable.
We could just as easily use the actual encoding $(\fst{(\hbox{\sl rst}^i \, l)})$
of the $(i+1)^{\rm th}$ eigenvariable in place of its index, but we
find it more convenient to use the index so that we can perform
induction on it.
(Here we use $(\hbox{\sl rst}^i \, l)$ for $n$ applications of {\sl rst} to
$l$, {\it i.e.,} $(\hbox{\sl rst}^0 \, l)$ is $l$, $(\hbox{\sl rst}^1 \, l)$ is
$(\hbox{\sl rst} \, l)$, $(\hbox{\sl rst}^2 \, l)$ is
$(\hbox{\sl rst} \, (\hbox{\sl rst} \, l))$, etc.)
The $\hbox{\sl subst}_0$ predicate is used in the definition of
{\sl subst}; the extra $\evars$ argument is used to keep track of
eigenvariables at the beginning of the list as we search down the list
for the substitution variable.
The encoding of these predicates is shown in
Table~\ref{tab:expl-eigen-subst}.
\begin{table}[btp]
\caption{Encoding of substitution for eigenvariables}
\label{tab:expl-eigen-subst}
\vspace{2pt}
\begin{center}
$\begin{array}{rcl}
\hline\rule{0pt}{14pt}
\subst{I}{T_1}{T_2}{T_2'}
	& \defeq & \substz{I}{(\lambda l'\,T_1)}
			 {(\lambda l'\,T_2)}{(\lambda l'\,T_2')} \\
\\
\multicolumn{3}{l}{\hbox{\sl subst}_0 \;\; \z
	\;\; T_1
	\;\; (\lambda l'\lambda l\,T_2 \, l' \, (\fst{l}) \, (\rst{l}))
	\;\; (\lambda l'\lambda l\,T_2 \, l' \, (T_1 \, l' \, l) \, (\rst{l}))} \\
	& \defeq & \top \\
\multicolumn{3}{l}{\hbox{\sl subst}_0 \;\;
	\begin{array}[t]{@{}l}
	(\suc{I})
	  \;\; (\lambda l'\lambda l\,T_1 \, l' \, (\fst{l}) \, (\rst{l})) \\
	(\lambda l'\lambda l\,T_2 \, l' \, (\fst{l}) \, (\rst{l}))
	  \;\; (\lambda l'\lambda l\,T_2' \, l' \, (\fst{l}) \, (\rst{l}))
	\end{array}} \\
	& \defeq &
	\hbox{\sl subst}_0 \;\;
	\begin{array}[t]{@{}l}
		I
		  \;\; (\lambda l'\lambda l\,T_1 \, (\rst{l'}) \, (\fst{l'}) \, l) \\
		(\lambda l'\lambda l\,T_2 \, (\rst{l'}) \, (\fst{l'}) \, l)
		  \;\; (\lambda l'\lambda l\,T_2' \, (\rst{l'}) \, (\fst{l'}) \, l)
	\end{array}
\\[2pt]
\hline
\end{array}$
\end{center}
\end{table}
Substitution for the first eigenvariable can be done directly; to
substitute for the $(i+2)^{\rm th}$ eigenvariable we move the first
eigenvariable from the list $l$ to the list $l'$ and substitute for
the $(i+1)^{\rm th}$ eigenvariable of $l$.

As with the previous encoding of intuitionistic logic, we must derive
the admissibility of the structural rules and the cut rule by
induction.
We have retained the atomic encoding of sequents, so we can still
analyze derivations of propositions from hypotheses.
In addition, the explicit encoding of eigenvariables allows us to
better analyze derivations of generic propositions.
To see this, we revisit the example from before; the sequent we wish
to derive is
\begin{displaymath}
\begin{array}[t]{@{}l}
\longrightarrow\forall p \forall t_1 \forall t_2 \forall t_3
	(\exists i \; \seq{i}{\lambda l\;\nil}
		{(\lambda l \bigwedge_\itm y_1 \bigwedge_\itm y_2
			(\atom{p \, y_1 \, t_1}
			 \iimp \atom{p \, y_2 \, t_2}
			 \iimp \atom{p \, y_2 \, t_3}))}
	 \oimp \quad\hfill\\
\hfill\same{t_2}{t_3})\enspace .
\end{array}
\end{displaymath}
As before, we begin by applying the $\forallR$, $\oimpR$, and
$\existsL$ rules to obtain the sequent
\begin{displaymath}
\begin{array}{c}
\Seq{\seq{i}{\lambda l\;\nil}{(\lambda l \bigwedge_\itm y_1 \bigwedge_\itm y_2
		(\atom{p \, y_1 \, t_1}
		 \iimp \atom{p \, y_2 \, t_2}
		 \iimp \atom{p \, y_2 \, t_3}))}}
    {\same{t_2}{t_3}}
\enspace .
\end{array}
\end{displaymath}
The derivation of the object logic formula
$\bigwedge y_1 \bigwedge y_2 \ldots$ must end with two applications of
the right rule for $\bigwedge$; we formalize this by applying $\defL$
twice, which results in the sequent
\begin{displaymath}
\Seq{\seq{i_1}{\lambda l\;\nil}{\lambda l\;
		(\atom{p \, (\fst{(\rst{l})}) \, t_1}
		 \iimp \atom{p \, (\fst{l}) \, t_2}
		 \iimp \atom{p \, (\fst{l}) \, t_3})}}
    {\same{t_2}{t_3}}
\enspace .
\end{displaymath}
The object logic derivation must proceed with two applications of
the right rule for $\iimp$; we deduce this formally by two more
applications of the $\defL$ rule, yielding
\begin{displaymath}
\Seq{\seq{i_2}{\lambda l\;(\cons{\atom{p \, (\fst{l}) \, t_2}}
				{\cons{\atom{p \, (\fst{(\rst{l})}) \, t_1}}
				      {\nil}})}
	      {\lambda l\; \atom{p \, (\fst{l}) \, t_3}}}
    {\same{t_2}{t_3}}
\enspace .
\end{displaymath}
An additional use of the $\defL$ rule corresponds to the realization
that the initial rule must complete the object logic derivation,
giving us the sequent
\begin{displaymath}
\Seq{\element{\lambda l\; \atom{p \, (\fst{l}) \, t_3}}
	     {\lambda l\;(\cons{\atom{p \, (\fst{l}) \, t_2}}
			       {\cons{\atom{p \, (\fst{(\rst{l})}) \, t_1}}
				     {\nil}})}}
    {\same{t_2}{t_3}}
\enspace .
\end{displaymath}
If $\atom{p \, (\fst{l}) \, t_3}$ is the first element of the list, then
$t_2$ and $t_3$ are the same, which is the result we are trying to
establish.
The formula $\atom{p \, (\fst{l}) \, t_3}$ cannot be the other element of
the list, because the first argument to $p$ differs; thus we are done.
This is all formally encoded by the derivation
\begin{displaymath}
\infer
      {\Seq{\element{\lambda l\; \atom{p \, (\fst{l}) \, t_3}}
		    {\lambda l\;(\cons{\atom{p \, (\fst{l}) \, t_2}}
				      {\cons{\atom{p \, (\fst{(\rst{l})}) \, t_1}}
					    {\nil}})}}
	   {\same{t_2}{t_3}}\enspace ,}
      {\infer[\defR]
	     {\Seq{\top}{\same{t_2}{t_2}}}
	     {\infer[\topR]
		    {\Seq{\top}{\top}}
		    {}}
      & \infer
	      {\Seq{\element{\lambda l\; \atom{p \, (\fst{l}) \, t_3}}
			    {\lambda l\;(\cons{\atom{p \, (\fst{(\rst{l})}) \, t_1}}
					      {\nil})}}
		   {\same{t_2}{t_3}}}
	      {\infer
		     {\Seq{\element{\lambda l\; \atom{p \, (\fst{l}) \, t_3}}
				   {(\lambda l\;\nil)}}
			  {\same{t_2}{t_3}}}
		     {}}}
\end{displaymath}
where the three inference rules without labels are $\defL$.

\subsection{Discussion}

Before going on to formally derive theorems about encodings of logics,
let us reflect on the encoding styles we have discussed.
What we have is a spectrum of styles, all of which share the same
higher-order abstract syntax encoding of formulas, but which vary in
the degree to which they use the higher-order abstract syntax encoding
of inference rules.
The first encoding used the typical higher-order abstract syntax
techniques, which made a number of significant properties of the
object logic fall out easily from the properties of $\FOLDN$.
Unfortunately this encoding did not lend itself to formal analysis
within $\FOLDN$, since it could not be expressed as a definition nor
given an induction measure.
We then progressed through three other encodings, each of which
compromised the use of higher-order abstract syntax a bit more.
The cost of each compromise was a decrease in the elegance and an
increase in the complexity of the encoding, and a reduction in the
extent to which fundamental properties of the object logic followed
from corresponding properties of $\FOLDN$.
The benefit, of course, was a greater ability to perform formal
meta-theoretic analysis.

In \pcfpart\ we will discuss an approach which lets us use
the typical higher-order abstract syntax encodings and also perform
meta-theoretic analyses on these encodings.
The key to this approach is the use of a specification logic that is
separate from $\FOLDN$, and in fact is itself specified in $\FOLDN$.
In the next section we present two logics which will be used for this
purpose, and which also serve as examples of the last two encoding
techniques discussed in this section.

\section{Representation and analysis of logics}
\label{sec:logics}

In this section we illustrate the use of the some of the encoding
techniques just presented.
In Section~\ref{sec:intuit} we use the explicit sequent
technique of Section~\ref{sec:expl-seq} to encode a fragment of
intuitionistic logic;
Section~\ref{sec:linear} discusses a fragment of linear logic
encoded with the explicit eigenvariable technique of
Section~\ref{sec:expl-eigen}.
In each case we prove the adequacy of the encoding and also derive in
$\FOLDN$ some properties of the object logic.

\subsection{Intuitionistic logic}
\label{sec:intuit}

Consider the fragment of second-order intuitionistic logic given by
the grammar
\begin{displaymath}
\begin{array}{rcl}
D & ::= & A \;\; | \;\; G \iimp A \;\; | \;\; \bigwedge_\alpha x.D
	\;\; | \;\; \bigwedge_{\alpha \rightarrow \alpha} x.D \\
G & ::= & A \;\; | \;\; \ttrue \;\; | \;\; G \with G
	 \;\; | \;\; A \iimp G \;\; | \;\; \bigwedge_\alpha x.G
\enspace ,
\end{array}
\end{displaymath}
where $A$ ranges over atomic formulas and $\alpha$ ranges over ground
types.
$D$ and $G$ represent definite clauses and goal formulas,
respectively.
Although this seems like a rather simple fragment, higher-order
abstract syntax encodings generally fall within the set of definite
clauses given by this grammar.
Full intuitionistic logic could be used here instead, but its
encoding is larger and that increase does not contribute to the
set of examples that we wish to use here.
The set of goal formulas can be encoded using the following constants:
\begin{displaymath}
\begin{array}[b]{rcl@{\quad\quad}rcl@{\quad\quad}rcl}
\atom{\;}      &\colon & \atm \rightarrow \prp
& \with          &\colon & \prp \rightarrow \prp \rightarrow \prp
& \bigwedge_\itm &\colon & (\itm \rightarrow \prp)\rightarrow \prp \\
\ttrue           &\colon & \prp 
& \iimp          &\colon & \atm \rightarrow \prp \rightarrow \prp\enspace .
\end{array}
\end{displaymath}
Notice that the antecendent of the implication is restricted to be
atomic.  

If we take any sequent calculus inference rule and restrict the
conclusion to be a sequent whose antecedents are definite clauses
and whose consequent is a goal formula, then the premises will also be
sequents of this form.
In fact, any antecedent in the premises will either be an antecedent of
the conclusion or an atomic formula.
Thus in a derivation in this fragment of intuitionistic logic, all
non-atomic antecedents in any sequent of the derivation appear as
antecedents in the end-sequent.
So we can divide the antecedents into the original theory, which
remains constant throughout the derivation, and some atomic
antecedents, which vary throughout the derivation.
Leaving the fixed theory aside for the moment, we can restrict our
sequents to have only atomic antecedents:
\begin{displaymath}
\begin{array}{rcl}
\hbox{\sl seq}     & \colon & \nt \rightarrow \atmlst \rightarrow \prp \rightarrow \oo
\enspace ,
\end{array}
\end{displaymath}
where $\atmlst$ is the same as the type $\lst$ introduced in
Section~\ref{sec:foldn-lists}, using $\atm$ for the type of elements.
Since the antecedents are atomic, only the initial and right rules are
necessary:
\begin{displaymath}
\begin{array}{rcl}
\seq{I}{(\cons{A'}{L})}{\atom{A}}
        & \defeq & \element{A}{(\cons{A'}{L})} \\
\seq{I}{L}{\ttrue}
        & \defeq & \top \\
\seq{(\suc{I})}{L}{(B \with C)}
        & \defeq & \seq{I}{L}{B} \; \land \; \seq{I}{L}{C} \\
\seq{(\suc{I})}{L}{(A \iimp B)}
        & \defeq & \seq{I}{(\cons{A}{L})}{B} \\
\seq{(\suc{I})}{L}{(\bigwedge_\itm B)}
        & \defeq & \forall_\itm x\;\seq{I}{L}{(B \, x)}
\enspace .
\end{array}
\end{displaymath}

We now turn to consider the set of definite clauses that make up the
theory for the derivation.
Notice that the atomic formula $A$ is equivalent to the formula
$\ttrue \iimp A$, so every definite clause can be written in the form
$\bigwedge x_1 \cdots \bigwedge x_n (G \iimp A)$.
In addition, the logic under consideration is a subset of the logic of
hereditary Harrop formulas.
As a result, for any derivable sequent there is a uniform derivation
of that sequent \cite{miller90lcs,miller91apal}.
In our setting, a derivation is uniform if every subderivation
ending in a left rule is of the form
\settowidth{\infwidthi}
	{$\Seq{(G \iimp A)[t_1,\ldots,t_n/x_1,\ldots,x_n],\Gamma}{A'}$}
\begin{displaymath}
\infer{\Seq{\bigwedge x_1 \cdots \bigwedge x_n(G \iimp A),\Gamma}{A'}}
      {\infer[\bigwedge {\cal L}\enspace ,]{\makebox[\infwidthi]{}}
	     {\infer[\iimp {\cal L}]{\Seq{(G \iimp A)[t_1,\ldots,t_n/x_1,\ldots,x_n],
					  \Gamma} 
					 {A'}}
		    {\infer*{\Seq{\Gamma}{G[t_1,\ldots,t_n/x_1,\ldots,x_n]}}
			    {}
		    & \infer[\init]{\Seq{A',\Gamma}{A'}} 
			{}}}}
\end{displaymath}
where $A'$ and $A[t_1,\ldots,t_n/x_1,\ldots,x_n]$ are the same.
If we group these steps together, our aggregate left rule encoding
needs to say that $\seq{(\suc{I})}{L}{\atom{A'}}$ holds if and only if
there is a clause $\bigwedge x_1 \cdots \bigwedge x_n (G \iimp A)$ in
the theory such that $A$ can be instantiated to match $A'$, and
$\seq{I}{L}{G'}$ holds, where $G'$ is the corresponding
instantiation of $G$.
We use the predicate
\begin{displaymath}
\begin{array}{rcl}
\hbox{\sl prog}    & \colon & \atm \rightarrow \prp \rightarrow \oo
\end{array}
\end{displaymath}
to encode the theory.
The fact that the definite clause
$\bigwedge x_1 \cdots \bigwedge x_n (G \iimp A)$ is in the theory
is represented by the definitional clause $\prog{A}{G} \defeq \top$;
the quantification of the definite clause is encoded by the (elided)
quantification of the definitional clause.
The encoding for the aggregate left rule is
\begin{displaymath}
\begin{array}{rcl}
\seq{(\suc{I})}{L}{\atom{A}}
        & \defeq & \exists b(\prog{A}{b} \land \seq{I}{L}{b})
\enspace ;
\end{array}
\end{displaymath}
notice that the matching between A and the head of the definite clause
is accomplished by the definition rules.
Different object-level theories can be considered by varying the
definition of {\sl prog}, as illustrated in \pcfpart.
The object-level formulas encoded using {\sl prog} are treated by the
object logic as a theory and not as a definition: there is no rule
corresponding to $\FOLDN$'s $\defL$ in the object logic.

We will refer to the six clauses for {\sl seq} given in this section
as $\Dintuit$.
For convenience we will abbreviate the formula
$\exists i(\nat{i} \land \seq{i}{L}{B})$ as $\prvv{L}{B}$ (or as
$\prv{B}$ when $L$ is \nil).
We now state the following properties about this presentation of the
object logic.
If $B$ is a term of type $\prp$, then let $\dcdmap{B}$ be its (obvious)
translation into a formula of intuitionistic logic.  If $L$ is a
term of type $\atmlst$, let $\dcdmap{L}$ be its (obvious) translation to
a multiset of atomic formulas of intuitionistic logic.

\begin{theorem}[Adequacy of Encoding Intuitionistic Logic]
Let $\Dprog$ be the definition
$\{\forall \bar{x}_1[\prog{A_{1}}{G_{1}} \defeq \top],\ldots,
   \forall \bar{x}_n[\prog{A_{n}}{G_{n}} \defeq \top]\}$
($n\ge0$) which represents an object-level theory, and
let $\Pscr$ be the corresponding theory in intuitionistic
logic ({\it i.e.}, the set of formulas
$\bigwedge \bar{x}_i(\dcdmap{G_{i}} \iimp \dcdmap{A_{i}})$, for all
$i \in \{1,\ldots,n\}$).
Let $\Dscr$ be a definition that extends
$\Dnat\cup\Dlist{\atm}\cup\Dintuit\cup\Dprog$ with
clauses that do not define {\sl nat}, {\sl seq}, {\sl element}, or
{\sl prog}.
Then the sequent $\Seq{}{\prvv{L}{B}}$ is derivable in $\FOLDN$
with definition $\Dscr$ if and only if $\dcdmap{B}$ is an
intuitionistic consequence of $\dcdmap{L}\cup\Pscr$.
\end{theorem}
\begin{proof}
The reverse direction follows easily from the definition
$\Dintuit$. 
For the forward direction, the use of the $\defR$ rule with $\Dintuit$
will cause the structure of the $\FOLDN$ derivation to closely follow
that of the corresponding derivation in intuitionistic logic.
However, we need to be sure that the $\natL$ and $\defL$ rules don't
allow us to derive anything that we can't derive in intuitionistic
logic.
In fact, we can show that a cut-free derivation of
$\Seq{}{\prvv{L}{B}}$ will consist only of sequents with empty
antecedents \cite{mcdowell97phd}.
Thus the $\natL$ and $\defL$ rules are not used, since they both
require a formula in the antecedent.
\end{proof}

The following theorem states that we can derive in $\FOLDN$ that the
specialization rule, the cut rule and the usual structural rules
(exchange, weakening, and contraction) are admissible for our object logic.

\def\adjx{\kern -0.95pt}
\begin{theorem}[Admissibility \adjx of \adjx Rules \adjx for \adjx Intuitionistic \adjx Object \adjx Logic]
\label{thm:intuit-spec}
\label{thm:intuit-cut}
\label{thm:intuit-struct}
The following formulas are derivable in $\FOLDN$ using the definition
$$\Dnat\cup\Dlist{\atm}\cup\Dintuit:$$

{\rm Specialization Rule:}
\begin{displaymath}
\forall i \forall b \forall l(
	\nat{i} \; \oimp \;
        \seq{(\suc{i})}{l}{\bigwedge b} \; \oimp \;
        \forall x\,\seq{i}{l}{(b\,x)})
\end{displaymath}

{\rm Cut Rule:}
\begin{displaymath}
\forall a \forall b \forall l(
        \prvv{(\cons{a}{l})}{b}
        \; \oimp \; \prvv{l}{\atom{a}}
        \; \oimp \; \prvv{l}{b})
\end{displaymath}

{\rm Structural Rules:}
\begin{displaymath}
\forall i \forall b \forall l \forall l'(
        \nat{i}
        \; \oimp \; \forall a(\element{a}{l} \; \oimp \; \element{a}{l'})
        \; \oimp \; \seq{i}{l}{b}
        \; \oimp \; \seq{i}{l'}{b})
\end{displaymath}
\end{theorem}

\subsection{Linear logic}
\label{sec:linear}

Now consider the fragment of second-order linear logic given by
the grammar
\begin{displaymath}
\begin{array}{rcl}
D & ::= & A \;\; | \;\; G \limp A \;\; | \;\; G \iimp A
	\;\; | \;\; \bigwedge_\alpha x.D
	\;\; | \;\; \bigwedge_{\alpha \rightarrow \alpha} x.D \\
G & ::= & A \;\; | \;\; \ttrue \;\; | \;\; G \with G
	\;\; | \;\; A \limp G \;\; | \;\; A \iimp G
	\;\; | \;\; \bigwedge_\alpha x.G
\enspace ,
\end{array}
\end{displaymath}
where $A$ ranges over atomic formulas and $\alpha$ ranges over ground
types.
As in Section~\ref{sec:intuit}, $D$ and $G$ represent definite
clauses and goal formulas, respectively.
The constants encoding these connectives have the same types as the corresponding constants used in Section~\ref{sec:intuit}; the new constant $\limp$ has
type $ \atm \rightarrow \prp \rightarrow \prp$.

We again separate the antecedents of sequents in a derivation into a
theory, which remains constant throughout the derivation and is
encoded via a predicate {\sl prog}, and some atomic antecedents, which
vary from sequent to sequent in the derivation and are shown
explicitly in the sequent.
The atomic antecedents are further divided into linear and
intuitionistic antecedents:
\begin{displaymath}
\begin{array}{rcl}
\hbox{\sl seq}     & \colon & \nt \rightarrow
			 (\evars \rightarrow \atmlst) \rightarrow
			 (\evars \rightarrow \atmlst) \rightarrow
			 (\evars \rightarrow \prp) \rightarrow
			 \oo
\enspace .
\end{array}
\end{displaymath}
The second and third arguments to {\sl seq} represent multisets of
intuitionistic and linear antecedents, respectively.
Notice that we follow the explicit eigenvariable encoding style of
Section~\ref{sec:expl-eigen} by encoding the antecedents and
consequent as functions whose domain is a list of eigenvariables.
We could use the explicit sequent technique to encode linear 
logic and still prove the adequacy and admissibility theorems 
of this section.
However, in \pcfpart\ we will use the linear logic encoding of 
this section as a specification logic; 
the proof of the unicity of typing theorem in 
Section~\ref{sec:imperative} uses meta-theoretic analysis that 
is not possible if we use the explicit sequent technique here.
This also gives us the opportunity to provide a detailed 
illustration of the explicit eigenvariable encoding style.
In order to highlight both the similarities and differences between
our current encoding and the encoding of Section~\ref{sec:intuit},
we will use a number of abbreviations;
we introduce the first of these now. 
For any type $\tau$, we will use $\tau^*$ as an abbreviation for
$\evars \rightarrow \tau$.
Thus the type of {\sl seq} above can be expressed as
\begin{displaymath}
\begin{array}{rcl}
\hbox{\sl seq}     & \colon & \nt \rightarrow
			 \atmlstl \rightarrow
			 \atmlstl \rightarrow
			 \prpl \rightarrow
			 \oo
\enspace .
\end{array}
\end{displaymath}

We must modify the definition $\Dlist{\tau}$ from
Section~\ref{sec:foldn-lists} to work over the type $\lstl$.
The predicates will now have the following types:
\begin{displaymath}
\begin{array}[b]{rcl@{\quad\quad\quad}rcl}
\hbox{\sl length}	& \colon & \lstl \rightarrow \nt \rightarrow \oo
& \hbox{\sl split}	& \colon & \lstl \rightarrow \lstl \rightarrow \lstl \rightarrow \oo \\
\hbox{\sl list}	& \colon & \lstl \rightarrow \oo
& \hbox{\sl permute}	& \colon & \lstl \rightarrow \lstl \rightarrow \oo \\
\hbox{\sl element}	& \colon & \tau^* \rightarrow \lstl \rightarrow \oo\enspace .
\end{array}
\end{displaymath}
The new definition $\Dlistl{\tau}$ is shown in
Table~\ref{tab:def-listl};
we use $\nill$ and $\consl{A}{L}$ as abbreviations for $\lambda l\,\nil$ and
$\lambda l\,(\cons{(A \, l)}{(L \, l)})$.
\begin{table}[btp]
\caption{Explicit eigenvariable encoding of lists}
\label{tab:def-listl}
\vspace{2pt}
\begin{center}
$\begin{array}{rcl}
\hline\rule{0pt}{14pt}
\length{\nill}{\z}
	& \defeq & \top \\
\length{(\consl{A}{L})}{(\suc{I})}
	& \defeq & \length{L}{I} \\
\\
\tlist{L}
	& \defeq & \exists i (\nat{i} \land \length{L}{i}) \\
\\
\element{A}{(\consl{A}{L})}
	& \defeq & \top \\
\element{A}{(\consl{A'}{L})}
	& \defeq & \element{A}{L} \\
\\
\split{\nill}{\nill}{\nill}
	& \defeq & \top \\
\split{(\consl{A}{L_1})}{(\consl{A}{L_2})}{L_3}
	& \defeq & \split{L_1}{L_2}{L_3} \\
\split{(\consl{A}{L_1})}{L_2}{(\consl{A}{L_3})}
	& \defeq & \split{L_1}{L_2}{L_3} \\
\\
\permute{\nill}{\nill}
	& \defeq & \top \\
\permute{(\consl{A}{L_1})}{L_2}
	& \defeq & \exists l_{22} (\split{L_2}{(\consl{A}{\nill})}{l_{22}} \land \permute{L_1}{l_{22}})
\\[2pt]
\hline
\end{array}$
\end{center}
\end{table}

We similarly introduce abbreviations corresponding to constructors of
$\prpl$:
$\atoml{A}$ abbreviates $\lambda l\,\atom{A \, l}$,
$\ttruel$ abbreviates $\lambda l \,\ttrue$,
$B \withl C$ abbreviates $\lambda l \,((B \, l) \with (C \, l))$,
${A}\limpl{B}$ abbreviates $\lambda l \,({(A \, l)}\limp{(B \, l)})$,
${A}\iimpl{B}$ abbreviates $\lambda l \,({(A \, l)}\iimp{(B \, l)})$,
and $\bigwedgel{B}$ abbreviates $\lambda l \,(\bigwedge x (B \, l \, x))$.

Any definite clause in our fragment of linear logic is equivalent to a
formula of the form
\begin{displaymath}
\bigwedge x_1 \cdots \bigwedge x_k
	(B_1 \iimp \cdots B_m \iimp C_1 \limp \cdots C_n \limp A)
\enspace ,
\end{displaymath}
for some $k,m,n \geq 0$ and goal formulas
$B_1, \ldots, B_m, C_1, \ldots, C_n$.
Uniform derivations have also been shown to be complete for this logic
\cite{hodas94ic};
thus we use the predicate
\begin{displaymath}
\begin{array}{rcl}
\hbox{\sl prog}    & \colon & \atml \rightarrow \prplstl \rightarrow \prplstl \rightarrow \oo
\end{array}
\end{displaymath}
to encode the set of definite clauses that make up the theory.
The first argument represents the atomic head of the definite clause;
the second and third arguments represent the lists 
$C_1, \ldots, C_n$ of linear hypotheses and $B_1, \ldots, B_m$ of
intuitionistic hypotheses, respectively.
The quantification of the definite clause is again encoded by the
(elided) quantification of the corresponding definitional clause 
for {\sl prog}.
Notice that the quantified variables of the definitional clause should be
able to match terms containing object-level eigenvariables and so should
have type $\itml$ (for first-order variables) or
$(\itm \rightarrow \itm)^*$ (for second-order variables).
On the other hand, the definite clause itself should be closed, so the
constants $\hbox{\sl fst}_\tau$ and $\hbox{\sl rst}$ (used to encode eigenvariables) 
should not occur in the corresponding definitional clause.
The predicate
\begin{displaymath}
\begin{array}{rcl}
\hbox{\sl split\_seq}     & \colon & \nt \rightarrow
			 \atmlstl \rightarrow
			 \atmlstl \rightarrow
			 \prplstl \rightarrow
			 \oo
\end{array}
\end{displaymath}
will be used to express the idea that the propositions in the last
argument are derivable from the intuitionistic and linear
antecedents in the second and third arguments.
Each linear antecedent must be used exactly once in the derivation of
all propositions in the last list.

The inference rules for this logic are encoded in the definition
$\Dlinear$ of Table~\ref{tab:def-linear}, which defines the
predicates {\sl seq} and {\sl split\_seq}.
\begin{table}[btp]
\caption{Explicit eigenvariable encoding of linear logic}
\label{tab:def-linear}
\vspace{2pt}
\begin{center}
$\begin{array}{rcl}
\hline\rule{0pt}{14pt}
\lseq{I}{{\it IL}}{(\consl{A}{\nill})}{\atoml{A}}
        & \defeq & \top \\
\lseq{I}{(\consl{A'}{{\it IL}})}{\nill}{\atoml{A}}
        & \defeq & \element{\atoml{A}}{(\consl{A'}{{\it IL}})} \\
\lseq{(\suc{I})}{{\it IL}}{{\it LL}}{\atoml{A}}
        & \defeq & \exists ll \exists il\;
		(\begin{array}[t]{@{}l}
		 \tlist{ll} \; \land \; \tlist{il} \; \land \;
		 \lprog{A}{ll}{il} \; \land \\
		 \splitseq{ll}{I}{{\it IL}}{{\it LL}} \; \land \;
		 \splitseq{il}{I}{{\it IL}}{\nill})
		 \end{array} \\
\lseq{I}{{\it IL}}{{\it LL}}{\ttruel}
        & \defeq & \top \\
\lseq{(\suc{I})}{{\it IL}}{{\it LL}}{(B \withl C)}
        & \defeq & \lseq{I}{{\it IL}}{{\it LL}}{B} \; \land \; \lseq{I}{{\it IL}}{{\it LL}}{C} \\
\lseq{(\suc{I})}{{\it IL}}{{\it LL}}{(A \limpl B)}
        & \defeq & \lseq{I}{{\it IL}}{(\consl{A}{{\it LL}})}{B} \\
\lseq{(\suc{I})}{{\it IL}}{{\it LL}}{(A \iimpl B)}
        & \defeq & \lseq{I}{(\consl{A}{{\it IL}})}{{\it LL}}{B} \\
\lseq{(\suc{I})}{{\it IL}}{{\it LL}}{(\bigwedgel_\itm B)}
        & \defeq & \lseq{I}{(\lambda l \; {\it IL} \, (\rst{l}))}
			  {(\lambda l \; {\it LL} \, (\rst{l}))}
			  {(\lambda l \; B \, (\rst{l}) \, (\tfst{\itm}{l}))} \\
\\
\splitseq{\nill}{I}{{\it IL}}{\nill}
	& \defeq & \top \\
\splitseq{(\consl{B}{L})}{I}{{\it IL}}{{\it LL}}
	& \defeq & \exists ll_1 \exists ll_2 \; 
		(\begin{array}[t]{@{}l}
		 \split{{\it LL}}{ll_1}{ll_2} \; \land \; \\
		 \seq{I}{{\it IL}}{ll_1}{B} \; \land \;
		   \splitseq{L}{I}{{\it IL}}{ll_2})
		 \end{array}
\\[2pt]
\hline
\end{array}$
\end{center}
\end{table}
The third clause in the definition says that an atomic formula $A$ is 
derivable from intuitionistic antecedents $IL$ and linear antecedents $LL$
if there is a definite clause in the object-level theory whose head is $A$,
whose linear hypotheses are derivable from the antecedents $IL$ and $LL$,
and whose intuitionistic hypotheses are derivable from the antecendents $IL$.
The other definitional clauses in Table~\ref{tab:def-linear} are similar
to those in the explicit eigenvariable encoding of intuitionistic logic
given in Section~\ref{sec:expl-eigen}, but modified to reflect the linearity
constraints.
In the clause for $\bigwedge_\itm$ we subscript the constant {\sl fst} with
the type \itm\ because we also need a constant
$\hbox{\sl fst}_{\itm \rightarrow \itm}: \evars \rightarrow \itm \rightarrow \itm$
for the representation of second-order eigenvariables in definite clauses.
As in the previous section, different object-level theories can be
considered by varying the definition of {\sl prog};
an example theory will be given in \pcfpart.
For convenience we will abbreviate the formula $\exists i(\nat{i}
\land \lseq{i}{{\it IL}}{{\it LL}}{B})$ as $\lprvv{{\it IL}}{{\it LL}}{B}$ (or as $\prv{B}$ when
${\it IL}$ and ${\it LL}$ are $\nill$).
If $B$ is a term of type $\prp$ and $L$ is a term of type $\atmlst$,
then let $\dcdmap{B}$ and $\dcdmap{L}$ be their
translations into a formula of linear logic and
a multiset of atomic formulas of linear logic, respectively.

\begin{theorem}[Adequacy of Encoding Linear Logic]
Fix a $\FOLDN$ signature whose only constants with
types involving $\evars$ are $\hbox{\sl fst}_\itm$,
$\hbox{\sl fst}_{\itm\rightarrow\itm}$, and $\hbox{\sl rst}$.
Let $\Dprog$ be the definition
\begin{displaymath}
\{\forall \bar{y}_1[\lprog{A_1}{{{\it LL}}_1}{{{\it IL}}_1} \defeq \top], \ldots,
   \forall \bar{y}_n[\lprog{A_n}{{{\it LL}}_n}{{{\it IL}}_n} \defeq \top]\}
\end{displaymath}
($n\ge0$), where the quantified variables in the list $\bar{y}_i$
each have type $\itml$ or
$(\itm \rightarrow \itm)^*$, and the constants
$\hbox{\sl fst}_\tau$ and {\sl rst} do not occur in $A_i$, ${\it LL}_i$, or
${\it IL}_i$, for all $i \in \{1,\ldots,n\}$.
Let $\Pscr$ be the theory in linear logic that corresponds to
$\Dprog$, and let $\Dscr$ be a definition that extends
$\Dnat\cup\Dlistl{\atm}\cup\Dlistl{\prp}\cup\Dlinear\cup\Dprog$ with
clauses that do not define {\sl nat}, {\sl length}, {\sl list},
{\sl element}, {\sl split}, {\sl split\_seq}, {\sl prog}, or {\sl seq}.
Finally, let ${\it IL}\colon\atmlstl$, ${\it LL}\colon\atmlstl$, and
$B\colon\prpl$ be terms that do not contain occurrences of the
constant $\hbox{\sl fst}_{\itm\rightarrow\itm}$.
Then the sequent $\Seq{}{\lprvv{{\it IL}}{{\it LL}}{B}}$ is derivable
in $\FOLDN$ with definition $\Dscr$ if and only if the sequent
$\Seq{\Pscr,\dcdmap{{\it IL}};\dcdmap{{\it LL}}}{\dcdmap{B}}$
is derivable in linear logic.
\end{theorem}
\begin{proof}
We can restrict our attention to uniform derivations in linear logic,
since they are complete for this fragment of linear logic
\cite{hodas94ic}.
As before a cut-free derivation of
$\Seq{}{\lprvv{{\it IL}}{{\it LL}}{B}}$ will consist only of sequents
with empty antecedents.
Thus the definition of {\sl seq} will ensure that the structure of the
$\FOLDN$ derivation will closely follow that of the corresponding
derivation in linear logic.
The proof of the forward direction goes by induction on the structure
of the $\FOLDN$ derivation, and the reverse direction by induction on
the structure of the linear logic derivation.
In general each case follows easily from the induction hypothesis.
A more detailed proof of this theorem, including a definition of the
$\dcdmap{\,}$ translations, can be found in \citeN{mcdowell97phd}.
\end{proof}

We now present the theorems that we have derived in $\FOLDN$ about our
object logic.
In order to express and prove these theorems, we need additional
predicates for operations related to the $\evars$ parameter.
The predicates
\begin{displaymath}
\begin{array}{rcl}
\hbox{\sl subst} & \colon & nt \rightarrow
			\itml \rightarrow
			\tau^* \rightarrow
			\tau^* \rightarrow \oo \\
\hbox{\sl subst}_0 & \colon & nt \rightarrow
			\itmll \rightarrow
			\tau^{**} \rightarrow
			\tau^{**} \rightarrow \oo
\end{array}
\end{displaymath}
will be used to represent substitution for eigenvariables; this is a
simple generalization of the predicate of
Section~\ref{sec:expl-eigen} to allow substitution in
expressions of an arbitrary type $\tau$.
The type $\tau^{**}$ should be understood to mean $(\tau^*)^*$, {\it i.e.},
an abbrevation for 
$(\evars \rightarrow \evars \rightarrow \tau)$.
We will also use the predicates
\begin{displaymath}
\begin{array}{rcl}
\hbox{\sl extend\_evars} & \colon & nt \rightarrow
			\tau^* \rightarrow
			\tau^* \rightarrow \oo \\
\hbox{\sl extend\_evars}_0 & \colon & nt \rightarrow
			\tau^{**} \rightarrow
			\tau^{**} \rightarrow \oo
\end{array}
\end{displaymath}
to add a new eigenvariable to the list at an offset.
Thus $\extend{i}{x}{x'}$ indicates that $x'$ is the result of adding a
new eigenvariable in $x$ at the $(i+1)^{\rm th}$ position in the list;
the eigenvariables that previously occupied positions $(i+1)$ or
greater are shifted to one position later in the list.
These predicates are defined in the definition $\Devars{\tau}$ of
Table~\ref{tab:def-evars}.
\begin{table}[btp]
\caption{Encoding of eigenvariable operations}
\label{tab:def-evars}
\vspace{2pt}
\begin{center}
$\begin{array}{rcl}
\hline\rule{0pt}{14pt}
\subst{I}{T}{X}{X'}
	& \defeq & \substz{I}{(\lambda l'\,T)}
			 {(\lambda l'\,X)}{(\lambda l'\,X')} \\
\\
\multicolumn{3}{l}{\hbox{\sl subst}_0 \;\; \z
	\;\; T
	\;\; (\lambda l'\lambda l\,X \, l' \, (\fst{l}) \, (\rst{l}))
	\;\; (\lambda l'\lambda l\,X \, l' \, (T \, l' \, l) \, (\rst{l}))} \\
	& \defeq & \top \\
\multicolumn{3}{l}{\hbox{\sl subst}_0 \;\;
	\begin{array}[t]{@{}l}
	(\suc{I})
	  \;\; (\lambda l'\lambda l\,T \, l' \, (\fst{l}) \, (\rst{l})) \\
	(\lambda l'\lambda l\,X \, l' \, (\fst{l}) \, (\rst{l}))
	  \;\; (\lambda l'\lambda l\,X' \, l' \, (\fst{l}) \, (\rst{l}))
	\end{array}} \\
	& \defeq &
	\hbox{\sl subst}_0 \;\;
	\begin{array}[t]{@{}l}
		I
		  \;\; (\lambda l'\lambda l\,T \, (\rst{l'}) \, (\fst{l'}) \, l) \\
		(\lambda l'\lambda l\,X \, (\rst{l'}) \, (\fst{l'}) \, l)
		  \;\; (\lambda l'\lambda l\,X' \, (\rst{l'}) \, (\fst{l'}) \, l)
	\end{array} \\
\\
\extend{I}{X}{X'}
	& \defeq & \extendz{I}
			 {(\lambda l'\,X)}{(\lambda l'\,X')} \\
\\
\multicolumn{3}{l}{\hbox{\sl extend\_evars}_0 \;\; \z
	\;\; (\lambda l'\lambda l\,X \, l' \, l)
	\;\; (\lambda l'\lambda l\,X \, l' \, (\rst{l}))} \\
	& \defeq & \top \\
\multicolumn{3}{l}{\hbox{\sl extend\_evars}_0 \;\; (\suc{I})
	  \;\; (\lambda l'\lambda l\,X \, l' \, (\fst{l}) \, (\rst{l}))
	  \;\; (\lambda l'\lambda l\,X' \, l' \, (\fst{l}) \, (\rst{l}))} \\
	& \defeq &
	\hbox{\sl extend\_evars}_0 \;\;
	\begin{array}[t]{@{}l}
		I
		\;\; (\lambda l'\lambda l\,X \, (\rst{l'}) \, (\fst{l'}) \, l) \\
		(\lambda l'\lambda l\,X' \, (\rst{l'}) \, (\fst{l'}) \, l)
	\end{array}
\\[2pt]
\hline
\end{array}$
\end{center}
\end{table}
We will also need an version of $\Dlist{\tau}$ to work over the type
$\lstll$; it is similar to $\Dlistl{\tau}$ and we will refer it
as $\Dlistll{\tau}$.

Since we want our theorems about the object logic to be independent of
any particular object logic theory, we need to include some
assumptions about the predicate {\sl prog}.
Specifically, we will need to know that if an atom matches the head of
a clause in the theory, then if we substitute for an eigenvariable in
the atom or extend the list of eigenvariables, then the resulting atom
will still match the head of the clause.
We encode these assumptions as the following two formulas:
\begin{displaymath}
\forall i\forall t\forall a\forall a'\forall ll\forall il
	(\begin{array}[t]{@{}l@{}}
	 \nat{i} \oimp \lprog{a}{ll}{il} \oimp \subst{i}{t}{a}{a'} \oimp \\
	 \exists ll'\exists il'
		(\lprog{a'}{ll'}{il'} \land \subst{i}{t}{ll}{ll'} \land \subst{i}{t}{il}{il'}))
\enspace ,
	 \end{array}
\end{displaymath}
which we will refer to as $\hbox{\sl P}_{subst}$, and
\begin{displaymath}
\forall i\forall a\forall a'\forall ll\forall il
	(\begin{array}[t]{@{}l@{}}
	 \nat{i} \oimp \lprog{a}{ll}{il} \oimp \extend{i}{a}{a'} \oimp \\
	 \exists ll'\exists il'
		(\lprog{a'}{ll'}{il'} \land \extend{i}{ll}{ll'} \land \extend{i}{il}{il'}))
\enspace ,
	 \end{array}
\end{displaymath}
which we will refer to as $\hbox{\sl P}_{extend}$.
The theory should not contain occurrences of eigenvariables, so
the definition of {\sl prog} should not contain occurrences of {\sl fst}
or {\sl rst}.
If this is the case, then $\hbox{\sl P}_{subst}$ and $\hbox{\sl P}_{extend}$
will be derivable in $\FOLDN$.

The following theorem states that we can derive in $\FOLDN$ that
the specialization rule, the cut rule, and the usual linear logic
structural rules are admissible for our object logic.
We refer to the definition
\begin{displaymath}
\Dlistl{\atm} \cup \Dlistl{\prp}
 \cup \Dlistll{\atm} \cup \Dlistll{\prp}
\end{displaymath}
as $\Dalllists$ and the definition
\begin{displaymath}
\Devars{\atm} \cup \Devars{\prp} \cup
 \Devars{\atmlst} \cup \Devars{\prplst}
\end{displaymath}
as $\Dallevars$.

\begin{theorem}[Rule Admissibility for Linear Logic]
\label{thm:linear-spec}
\label{thm:linear-cut}
\label{thm:linear-struct}
The formulas below are derivable in $\FOLDN$ using the definition
$\Dnat \cup \Dalllists \cup \Dallevars \cup \Dlinear$:

{\rm Specialization Rule:}
\begin{displaymath}
\begin{array}{l}
\hbox{\sl P}_{subst} \; \oimp \\
\forall i \forall b \forall il \forall ll(
	\begin{array}[t]{@{}l}
	\nat{i} \; \oimp \;
	\tlist{il} \; \oimp \;
	\tlist{ll} \; \oimp \; \\
        \lseq{(\suc{i})}{il}{ll}{\bigwedgel b} \; \oimp \;
        \forall x\,\lseq{i}{il}{ll}{(b\,x)})
	\end{array}
\end{array}
\end{displaymath}

{\rm Cut Rule:}
\begin{displaymath}
\begin{array}{l}
\hbox{\sl P}_{extend} \; \oimp \\
\forall a \forall b \forall il \forall ll(
	\begin{array}[t]{@{}l}
	\tlist{il} \; \oimp \;
	\tlist{ll} \; \oimp \; \\
        \lprvv{(\consl{a}{il})}{ll}{b} \; \oimp \;
	\lprvv{il}{\nill}{\atoml{a}} \; \oimp \;
        \lprvv{il}{ll}{b})
	\end{array}
\end{array}
\end{displaymath}

{\rm Structural Rules:}
\begin{displaymath}
\begin{array}{l}
\hbox{\sl P}_{extend} \; \oimp \\
\forall a \forall b \forall il \forall ll \forall ll_1 \forall ll_2(
	\begin{array}[t]{@{}l}
	\tlist{il} \; \oimp \;
	\tlist{ll} \; \oimp \;
	\split{ll}{ll_1}{ll_2} \; \oimp \; \\
        \lprvv{il}{(\consl{a}{ll_1})}{b} \; \oimp \;
	\lprvv{il}{ll_2}{\atoml{a}} \; \oimp \;
        \lprvv{il}{ll}{b})
	\end{array}
\end{array}
\end{displaymath}
\begin{displaymath}
\forall i \forall b \forall il \forall il' \forall ll \forall ll'(
	\begin{array}[t]{@{}l}
        \nat{i} \; \oimp \;
        \tlist{il} \; \oimp \;
        \tlist{il'} \; \oimp \;
        \tlist{ll} \; \oimp \; \\
        \forall a(\element{a}{il} \; \oimp \; \element{a}{il'}) \; \oimp \;
        \permute{ll}{ll'} \; \oimp \; \\
        \lseq{i}{il}{ll}{b} \; \oimp \;
        \lseq{i}{il'}{ll'}{b})
	\end{array}
\end{displaymath}
\end{theorem}

\section{Related work}
\label{sec:related1}

In this part of the paper we have presented several different encodings of
logics;
for each we discussed the extent to which reasoning about the encoded
logic can take place within the meta-logic $\FOLDN$.
None of the encoding techniques is completely original, but their
ability to support formal meta-theoretic analysis is a relatively new
concern.

The natural deduction-style encoding of
Section~\ref{sec:nat-ded} is the prototypical representation
style of higher-order abstract syntax.
For example, the seminal paper on the Edinburgh Logical Framework (LF)
\cite{harper93jacm} encodes first-order and higher-order logic in this
manner and proves the adequacy of these encodings.
The issue of meta-theoretic analysis of the encodings within the
meta-logic is not addressed there.

The use of separate predicates for formulas on the left and right
sides of the sequent, as was done in Section~\ref{sec:impl-seq}, is
also common.
Pfenning \citeyear{pfenning95lics}, for example, uses this representation
style to encode structural cut-elimination proofs for intuitionistic,
classical, and linear logics.
The induction cases of these proofs are represented in Elf,
so some amount of reasoning about the encoded logics is done in the
meta-logic.
However Elf does not itself contain any support for induction, so the
completeness of the cases must be checked outside of the formal
framework using techniques such as schema checking
\cite{pfenning92cade,rohwedder96esop}.
Miller \citeyear{miller96tcs} uses both this sequent style of encoding and the
natural deduction style.
The two encodings are used to show that natural deduction and sequent
calculus presentations of minimal logic have the same theorems.
The proof of this result combines informal reasoning with formal
reasoning in a linear logic meta-logic.

Section~\ref{sec:expl-seq} presented an encoding of logic which
encoded the derivability of a sequent in a single predicate.
This style of encoding was used in an early paper on the use of
higher-order abstract syntax \cite{miller87slp}.
That paper focuses on an operational interpretation of such a
specification, however, and does not discuss the potential for
reasoning about the encoded logic in the meta-logic.

The idea of representing free variables as a list, discussed in
Section~\ref{sec:expl-eigen}, was first used in the context of
higher-order abstract syntax by Despeyroux and Hirschowitz
\citeyear{despeyroux94lpar}.
Their intent was to develop a way to use higher-order abstract syntax
within the setting of the inductive definition facility of Coq.
A key difference between their technique and ours is that they use
both constructor and deconstructor operators for lists in the context
of an equality theory.
The encoding of the right rule for universal quantification in that
setting might look like the following:
\begin{displaymath}
\begin{array}{rcl}
\seq{(\suc{I})}{L}{(\lambda l \bigwedge_\itm x (B \, (\hbox{\sl cons} \, x \, l)))}
        & \defeq & \seq{I}{(\lambda l'\;L (\rst{l'}))}{B}
\enspace .
\end{array}
\end{displaymath}
Within terms, bound and free variables are accessed by selecting the
appropriate element from the list.
In our simpler setting (without an equality theory) we use unification
to get by with only deconstructors for variable lists.
The paper \citeN{despeyroux94lpar} was the first attempt to fully
support formal reasoning about higher-order abstract syntax encodings
within a meta-logic.
Their examples involved encodings of simply-typed $\lambda$-terms, so
we will discuss their work further at the end of \pcfpart.

\section*{\pcfpart: OBJECT LOGICS AS SPECIFICATION LOGICS}

In this part we consider reasoning about higher-order abstract
syntax encodings of programming languages.
We could choose one of the representation strategies used for logics
in the previous part;
instead we adopt a different strategy that allows us to use the
traditional higher-order abstract syntax representation to its full
advantage and still reason formally about the encoded system.
The key to accomplishing this is to not specify the programming
language directly in $\FOLDN$, but in a small object logic that is
itself specified in $\FOLDN$.
In this way we can reason in $\FOLDN$ about the structure of object
logic sequents and their derivability.

The use of object-level sequents may seem at first a rather drastic 
step to take to embed the kind of hypothetical judgements common with 
higher-order abstract syntax into a meta-logic.  Such a representation 
is, however, used in various areas of programming language semantics.  
For example, Mitchell, in his textbook \citeyear{mitchell96book}, uses 
typing judgements of the form $\Gamma \rhd M : \sigma$ and performs 
induction over their (sequent-style) derivation.
This separation of the (object) specification logic from the
meta-logic ($\FOLDN$) in which reasoning is performed also reflects
the usual structure of informal reasoning about higher-order abstract
syntax encodings.

In the next section we motivate this approach through an informal
proof of subject reduction for the untyped $\lambda$-calculus.
We proceed in Section~\ref{sec:functional} to formalize this proof by
encoding the static and dynamic semantics for untyped
$\lambda$-terms in the intuitionistic object logic of
Section~\ref{sec:intuit}.
We also list a variety of other theorems about the language that we
have derived in $\FOLDN$.
The remainder of the section extends the encoding to the Programming
language of Computable Functions (PCF) \cite{scott69}.
In Section~\ref{sec:imperative} we consider an encoding of PCF
with references ($\pcfref$) \cite{gunter92book} in the linear object
logic of Section~\ref{sec:linear}.
Finally, Section~\ref{sec:related2} compares the framework of this
part with other research in formal reasoning about higher-order
abstract syntax encodings.

\section{Motivation from informal reasoning}
\label{sec:motivate}

In order to motivate our framework for reasoning about higher-order
abstract syntax encodings, we consider a specification in
intuitionistic logic of call-by-name evaluation and simple typing for
the untyped $\lambda$-calculus.
We introduce two types, $\tm$ and $\ty$, to denote object-level terms
and types.
To represent the untyped $\lambda$-terms we introduce the two
constants {\sl abs} of type $(\tm \rightarrow \tm) \rightarrow \tm$
and {\sl app} of type $\tm \rightarrow \tm \rightarrow \tm$ to
denote object-level abstraction and application, respectively.
Object-level types will be built up from a single primitive type using
the arrow type constructor; these are denoted in the specification
logic by the constants $\gnd$ of type $\ty$ and {\sl arr} of type
$\ty \rightarrow \ty \rightarrow \ty$.

To specify call-by-name evaluation, we use an infix predicate
$\Downarrow$ of type $\tm \rightarrow \tm \rightarrow o$ and the two
formulas
\begin{displaymath}
\begin{array}{@{}c@{}}
\bigwedge r ((\abs r) \Downarrow (\abs r)) \\
\bigwedge m\bigwedge n\bigwedge v\bigwedge r
    ((m \Downarrow (\abs r) \; \with \; (r\; n) \Downarrow v) \; \iimp \;
     (\app m n) \Downarrow v)
\enspace .
\end{array}
\end{displaymath}

To specify simple typing at the object-level, we use the binary predicate
{\sl typeof} of type $\tm \rightarrow \ty \rightarrow o$
and the two formulas
\begin{displaymath}
\begin{array}{@{}c@{}}
\bigwedge m\bigwedge n\bigwedge t\bigwedge u
  ((\typeof m (\arr u t) \; \with \; \typeof n u) \; \iimp \;
   \typeof{(\app m n)} t) \\
\bigwedge r\bigwedge t\bigwedge u
  (\bigwedge x (\typeof x t \; \iimp \; \typeof{(r \, x)}{u}) \; \iimp \;
		\typeof{(\abs r)}{(\arr t u)})
\enspace .
\end{array}
\end{displaymath}
Proofs that these two predicates correctly capture the notions of
call-by-name evaluation and of simple typing can be found in various
places in the literature: see, for example, \citeN{avron92jar} and
\citeN{hannan90phd}.

Now consider the following subject reduction theorem and its proof.
We use $\vdash$ here to represent derivability in intuitionistic logic
from the above formulas encoding evaluation and typing;
we omit displaying these formulas on the left of the turnstile to
simplify the presentation.
\begin{proposition}
\label{prp:subj-red}
If $\vdash P \Downarrow V$ and $\vdash \typeof{P}{T}$,
then $\vdash \typeof{V}{T}$. 
\end{proposition}
\begin{proof}
We prove this theorem by induction on the height of the derivation of
$P\Downarrow V$.
Since $P\Downarrow V$ is atomic, its derivation must end with the use of
one of the formulas encoding evaluation.
If the $\Downarrow$ formula for {\sl abs} is used, then $P$ and $V$ are
both equal to $\abs R$, for some $R$, and the consequent is immediate.
If $P \Downarrow V$ was derived using the $\Downarrow$ formula
for {\sl app}, then $P$ is of the form $(\app M N)$, and for some
$R$ there are shorter derivations of $M \Downarrow (\abs R)$ and
$(R \, N) \Downarrow V$. 
Since $P$ is $(\app M N)$, $\typeof{P}{T}$ must have been derived
using the formula encoding the typing rule for {\sl app}. 
Hence, there is a $U$ such that $\vdash\typeof{M}{(\arr U T)}$ and
$\vdash\typeof{N}{U}$.
Applying the inductive hypothesis to the evaluation and typing judgements
for $M$, we have $\vdash\typeof{(\abs R)}{(\arr U T)}$.
This atomic formula must have been derived using	the {\sl typeof}
formula for {\sl abs}, and, hence, 
$\vdash\bigwedge x(\typeof x U \iimp \typeof{(R \, x)}{T})$.
Since our specification logic is intuitionistic logic, we can
instantiate this quantifier with $N$ and use cut and
cut-elimination to conclude that $\vdash \typeof{(R \, N)}{T}$.
Applying the inductive hypothesis to the judgements for $(R \, N)$ yields
$\vdash	\typeof{V}{T}$.
\end{proof}

This proof is clear and natural, and we would like to be able to
formally capture proofs quite similar to this in structure.
This suggests that the following features would be valuable in our
framework:

\begin{longenum}
\item  {\em Two distinct logics.}  One of the logics would
correspond to the one written with logical syntax above and 
would capture judgements, {\frenchspacing {\it e.g.}, about} typability and
evaluation.
The second logic would represent a formalization of the English text
in the proof above. 
Atomic formulas of this second (meta-) logic would encode judgements
in the first (object) logic. 

\item  {\em Induction} over at least natural numbers.

\item\label{var instantiation}
{\em Instantiation of meta-level eigenvariables.}
In the proof above, for example, the meta-level variable $P$ was
instantiated in one part of the proof to $(\abs R)$ and in another part
of the proof to $(\app M N)$.
Notice that this instantiation of eigenvariables within a proof does
not happen in a strictly intuitionistic sequent calculus.

\item  {\em Analysis of the derivation of an assumed judgement.}
In the proof above this was done a few times, leading, for example,
from the assumption
\begin{displaymath}
\vdash\typeof{(\abs R)}{(\arr U T)}
\end{displaymath}
to the assumption
\begin{displaymath}
\vdash\bigwedge x(\typeof x U \iimp \typeof{(R \, x)}{T})
\enspace .
\end{displaymath}
The specification of {\sl typeof} allows the implication to go in the
other direction, but given the structure of the specification of {\sl
typeof}, this direction can also be justified at the meta-level.
\end{longenum}
In our framework, we accommodate the first feature by specifying
an object logic within the meta-logic $\FOLDN$, as illustrated in
\stylespart.
The $\natL$ rule of $\FOLDN$ provides natural number induction.
The last two features are accommodated by the definition facilities of
$\FOLDN$, in particular the $\defL$ rule.
We demonstrate our approach in the remaining sections of the paper, 
beginning with a formalization of the example from this section.

\section{Representation and analysis of a functional programming\\ language}
\label{sec:functional}

\subsection{The language of untyped $\lambda$-terms}
\label{sec:lambda}

We first demonstrate our approach to formal reasoning about
higher-order abstract syntax encodings using the example of untyped
$\lambda$-terms.
This encoding will be similar to the one used to motivate the
framework in the preceding section.
The object logic used will be the fragment of second-order
intuitionistic logic encoded by the definition $\Dintuit$ of
Section~\ref{sec:intuit}.

The required constants to represent $\lambda$-terms are
$\hbox{\sl abs}: (\tmi \rightarrow \tmi) \rightarrow \tmi$ and
$\hbox{\sl app}: \tmi \rightarrow \tmi \rightarrow \tmi$;
for simple types (over one primitive type) we need
$\hbox{\sl gnd} \colon \tyi $ and
$\hbox{\sl arr} \colon \tyi \rightarrow
	\tyi \rightarrow \tyi $.
Since both types and terms in the language are represented by the
object logic type $\itm$, we have added subscripts \tm\ and \ty.
These subscripts should not be considered part of the encoding, but
are added to improve the readability of these declarations.

Our object logic predicate representing typability
is denoted by the $\FOLDN$ constant {\sl typeof} of type
$\tmi \rightarrow \tyi \rightarrow \atm$. 
The predicates for natural semantics and transition semantics are
denoted by the constants $\Downarrow$, $\leadsto$, and $\leadsto^*$,
all of type $\tmi \rightarrow \tmi \rightarrow \atm$.
The object logic specifications for these are the usual ones, written
in the $L_{\lambda}$ subset of higher-order logic \cite{miller91jlc}
and are those common to specifications written in, say,
$\lambda$Prolog \cite{hannan92mscs} and Elf \cite{pfenning89lics}.
This object-level specification is represented in $\FOLDN$ as the
definition $\Dlambda$ shown in Table~\ref{tab:lambda}.
\begin{table}[btp]
\caption{Object logic encoding of typing and evaluation of untyped $\lambda$-terms} 
\label{tab:lambda}
\begin{center}
$\begin{array}{l}
\hline\rule{0pt}{14pt}
\prog{\;\;(\typeof{(\abs{R})}{(\arr{T}{U})})}
	 {\qquad\bigwedge n((\typeof{n}{T}) \iimp \atom{\typeof{(R \, n)}{U}})} \\
\prog{\;\;(\typeof{(\app{M}{N})}{T})}
	 {\qquad\atom{\typeof{M}{(\arr{U}{T})}}
		\with \atom{\typeof{N}{U}}} \\
\\
\prog{\;\;(\natsem{(\abs{R})}{(\abs{R})})}
	{\qquad\ttrue} \\
\prog{\;\;(\natsem{(\app{M}{N})}{V})}
	 {\qquad\atom{\natsem{M}{(\abs{R})}}
                        \with \atom{\natsem{(R \, N)}{V}}} \\
\\
\prog{\;\;(\transone{(\app{(\abs{R})}{M})}{(R \, M)})}
	{\qquad\ttrue} \\
\prog{\;\;(\transone{(\app{M}{N})}{(\app{M'}{N})})}
	{\qquad\atom{\transone{M}{M'}}} \\
\\
\prog{\;\;(\transsem{M}{M})}{\qquad\ttrue} \\
\prog{\;\;(\transsem{M}{N})}
	{\qquad\atom{\transone{M}{M'}}
                        \with \atom{\transsem{M'}{N}}}
\\[2pt]
\hline
\end{array}$
\end{center}
\end{table}
(We have dropped the $\relax\defeq \top$ body of these clauses.)
This definition can be interpreted in a logic programming fashion to
compute object-level simple type checking and call-by-name evaluation
in both structural operational semantic and natural semantic styles.
Call-by-value is just as easily represented and used.

The following theorem lists the properties of the untyped
$\lambda$-calculus that we have derived in $\FOLDN$:  determinacy of
semantics, equivalence of semantics, and subject reduction.
The $\FOLDN$ derivations closely follow the informal proofs of these
properties.
\begin{theorem}
\label{thm:lambda-det-sem}
\label{thm:lambda-equiv-sem}
\label{thm:lambda-subj-red}
The following formulas are derivable in $\FOLDN$ from the definition
that accumulates $\Dnat$, $\Dlist{\atm}$, $\Dintuit$, $\Dlambda$ and
the clause $\same{X}{X} \defeq \top$ defining the predicate
$\equiv: \itm \rightarrow \itm \rightarrow \oo$.

{\rm Determinacy of semantics:}
\begin{displaymath}
\begin{array}{c}
\forall m\forall m_1\forall m_2
        (\prv{\atom{\natsem{m}{m_1}}}
         \oimp \prv{\atom{\natsem{m}{m_2}}}
         \oimp \same{m_1}{m_2}) \\
\forall m\forall m_1\forall m_2
        (\prv{\atom{\transone{m}{m_1}}}
         \oimp \prv{\atom{\transone{m}{m_2}}}
         \oimp \same{m_1}{m_2}) \\
\forall m\forall r_1\forall r_2
        (\prv{\atom{\transsem{m}{(\abs{r_1})}}}
         \oimp \prv{\atom{\transsem{m}{(\abs{r_2})}}}
         \oimp \same{(\abs{r_1})}{(\abs{r_2})})
\end{array}
\end{displaymath}

{\rm Equivalence of semantics:}
\begin{displaymath}
\begin{array}{c}
\forall m\forall r
        (\prv{\atom{\natsem{m}{(\abs{r})}}}
         \oimp \prv{\atom{\transsem{m}{(\abs{r})}}}) \\
\forall m\forall r
        (\prv{\atom{\transsem{m}{(\abs{r})}}}
         \oimp \prv{\atom{\natsem{m}{(\abs{r})}}})
\end{array}
\end{displaymath}

{\rm Subject reduction:}
\begin{displaymath}
\begin{array}{c}
\forall m\forall n
        (\prv{\atom{\natsem{m}{n}}}
         \oimp \forall t(\prv{\atom{\typeof{m}{t}}}
                        \oimp \prv{\atom{\typeof{n}{t}}})) \\
\forall m\forall n
        (\prv{\atom{\transone{m}{n}}}
         \oimp  \forall t(\prv{\atom{\typeof{m}{t}}}
                         \oimp \prv{\atom{\typeof{n}{t}}})) \\
\forall m\forall n
        (\prv{\atom{\transsem{m}{n}}}
         \oimp  \forall t(\prv{\atom{\typeof{m}{t}}}
                         \oimp \prv{\atom{\typeof{n}{t}}}))
\end{array}
\end{displaymath}
\end{theorem}
\begin{proof}
We show the derivation of the first subject reduction property,
which is a formalization of Proposition~\ref{prp:subj-red}.

We wish to show that evaluation preserves types:
\begin{displaymath}
\Seq{}{\forall p\forall v
	(\prv{\atom{\natsem{p}{v}}}
	 \oimp \forall t(\prv{\atom{\typeof{p}{t}}} \oimp
			  \prv{\atom{\typeof{v}{t}}}))}
\enspace .
\end{displaymath}
(We have changed the names of the quantified variables to agree with
those in the informal proof.)
Applying the $\forallR$, $\oimpR$, $\existsL$, $\cL$, and $\landL$
rules to the above sequent yields
\begin{displaymath}
\Seq{\nat{i},\seq{i}{\nil}{\atom{\natsem{p}{v}}},
     \prv{\atom{\typeof{p}{t}}}}
    {\prv{\atom{\typeof{v}{t}}}}
\enspace .
\end{displaymath}
(Recall that $\prv{\atom{\natsem{p}{v}}}$ is an abbreviation for
$\exists i(\nat{i} \land \seq{i}{\nil}{\atom{\natsem{p}{v}}})$.)

As in the informal proof, we proceed with an induction on the height
of the derivation of $\natsem{p}{v}$, which is represented here by $i$.
We will use the derived rule for complete induction
(Proposition~\ref{prp:complete-ind}) and our induction predicate will
be
\begin{displaymath}
\lambda i\forall p\forall v\forall t
    (\seq{i}{\nil}{\atom{\natsem{p}{v}}} \oimp
     \prv{\atom{\typeof{p}{t}}} \oimp
     \prv{\atom{\typeof{v}{t}}})
\enspace ,
\end{displaymath}
which we will denote by {\sl IP}.
The derivation of the conclusion from the induction predicate applied to
$i$ is trivial, so it only remains to derive the induction step
\begin{displaymath}
\Seq{\nat{j},\forall k(\nat{k} \oimp k < j \oimp (\hbox{\sl IP} \, k))}
    {(\hbox{\sl IP} \, j)}
\enspace .
\end{displaymath}
We use the $\forallR$ and $\oimpR$ rules to obtain
\begin{displaymath}
\Seq{\nat{j},\forall k\ldots,
	\seq{j}{\nil}{\atom{\natsem{p}{v}}},
	\prv{\atom{\typeof{p}{t}}}}
    {\prv{\atom{\typeof{v}{t}}}}
\enspace .
\end{displaymath}

In the informal proof we use the fact that the derivation of the
atomic formula $\natsem{p}{v}$ must end with the use of a clause from
the specification of evaluation.
We deduce this formally by applying the $\defL$ rule to
$\seq{j}{\nil}{\atom{\natsem{p}{v}}}$, which yields
\begin{displaymath}
\Seq{\nat{(\suc{j_0})},\forall k\ldots,
        \exists b(\prog{(\natsem{p}{v})}{b}
		  \land \seq{j_0}{\nil}{b}),
	\prv{\atom{\typeof{p}{t}}}}
   {\prv{\atom{\typeof{v}{t}}}}
\enspace .
\end{displaymath}
We next apply the $\existsL$, $\cL$, and $\landL$ rules, and then
apply the $\defL$ rule to $\prog{(\natsem{p}{v})}{b}$ which yields the
two sequents
\begin{displaymath}
\Seq{\nat{(\suc{j_0})},\forall k\ldots,
     \seq{j_0}{\nil}{\ttrue},
     \prv{\atom{\typeof{(\abs{r})}{t}}}}
    {\prv{\atom{\typeof{(\abs{r})}{t}}}}
\end{displaymath}
\begin{displaymath}
\begin{array}{@{}r@{}l@{}}
\nat{(\suc{j_0})},\forall k\ldots,
\seq{j_0}{\nil}
	{\atom{\natsem{m}{(\abs{r})}}
		\with \atom{\natsem{(r \, n)}{v}}}, \\
\prv{\atom{\typeof{(\app{m}{n})}{t}}}
 & \longrightarrow {\prv{\atom{\typeof{v}{t}}}}
\enspace .
\end{array}
\end{displaymath}
This use of the $\defL$ rule corresponds to the case analysis of the
formula used to derive $\natsem{p}{v}$.
As in the informal case, the {\sl abs} case (represented here by the
first sequent) is immediate.
The derivation of the second sequent, representing the {\sl app} case,
begins with the use of the $\defL$, $\cL$, and $\landL$, bringing
us to the sequent
\begin{displaymath}
\begin{array}{@{}r@{}l@{}}
\nat{(\hbox{\sl s}^2 \, j_1)},\forall k\ldots,
\seq{j_1}{\nil}{\atom{\natsem{m}{(\abs{r})}}},
\seq{j_1}{\nil}{\atom{\natsem{(r \, n)}{v}}}, \\
\prv{\atom{\typeof{(\app{m}{n})}{t}}}
 & \longrightarrow {\prv{\atom{\typeof{v}{t}}}}
\enspace .
\end{array}
\end{displaymath}
(We use the term $\hbox{\sl s}^2 \, j_1$ as an abbreviation for
$\suc{(\suc{j_1})}$.)

The informal proof continues with an analysis of the derivation of
$$\typeof{(\app{m}{n})}{t}\enspace.$$
Again we accomplish this through two uses of the $\defL$ rule, the first to
indicate that the derivation must end with the use of a specification
clause, and the second to determine the applicable clauses.
In this case there is only one applicable clause, so we are left to
derive the sequent
\begin{displaymath}
\Seq{\ldots,\nat{(\suc{j_0'})},
     \seq{j_0'}{\nil}
	{\atom{\typeof{m}{(\arr{u}{t})}}
			\with \atom{\typeof{n}{u}}}}
    {\prv{\atom{\typeof{v}{t}}}}
\enspace .
\end{displaymath}
Additional uses of the $\defL$, $\cL$ and $\landL$ rules bring
us to the sequent
\begin{displaymath}
\begin{array}{@{}r@{}l@{}}
\ldots,\nat{(\hbox{\sl s}^2 \, j_1')},
\seq{j_1'}{\nil}{\atom{\typeof{m}{(\arr{u}{t})}}}, \\
\seq{j_1'}{\nil}{\atom{\typeof{n}{u}}}
 & \longrightarrow {\prv{\atom{\typeof{v}{t}}}}
\enspace .
\end{array}
\end{displaymath}

In the informal proof we now apply the induction hypothesis to
the evaluation and typing judgments for $m$.
We accomplish this here by applying the appropriate left rules to the
elided induction hypothesis $\forall k\ldots$.
This requires the derivation of the five sequents
\begin{displaymath}
\begin{array}{c@{\qquad\qquad\qquad}c}
\Seq{\nat{(\hbox{\sl s}^2 \, j_1)},\ldots}
    {\nat{j_1}}
& \Seq{\nat{(\hbox{\sl s}^2 \, j_1)}, \ldots}
    {j_1 < (\hbox{\sl s}^2 \, j_1)}
\end{array}
\end{displaymath}
\begin{displaymath}
\Seq{\ldots,\seq{j_1}{\nil}{\atom{\natsem{m}{(\abs{r})}}},\ldots}
    {\seq{j_1}{\nil}{\atom{\natsem{m}{(\abs{r})}}}}
\end{displaymath}
\begin{displaymath}
\Seq{\ldots,\nat{(\hbox{\sl s}^2 \, j_1')},
     \seq{j_1'}{\nil}{\atom{\typeof{m}{(\arr{u}{t})}}}, \ldots}
    {\prv{\atom{\typeof{m}{(\arr{u}{t})}}}}
\end{displaymath}
\begin{displaymath}
\begin{array}{@{}r@{}l@{}}
\nat{(\hbox{\sl s}^2 \, j_1)},\forall k\ldots,
\seq{j_1}{\nil}{\atom{\natsem{(r \, n)}{v}}}, \\
\prv{\atom{\typeof{(\abs{r})}{(\arr{u}{t})}}}, \\
\nat{(\hbox{\sl s}^2 \, j_1')},
\seq{j_1'}{\nil}{\atom{\typeof{n}{u}}}
 & \longrightarrow {\prv{\atom{\typeof{v}{t}}}}
\enspace .
\end{array}
\end{displaymath}
The first two of these represent the fact that the measure of the
evaluation derivation for $m$ is a natural number that is smaller than the
measure of the original evaluation derivation for $p$.
By Proposition~\ref{prp:nat-other} these are derivable in $\FOLDN$
from $\Dnat$.
The third sequent is immediate, and the fourth also follows easily from
Proposition~\ref{prp:nat-other}.

The derivation of the fifth sequent proceeds with another two
applications of the $\defL$ rule, corresponding to the analysis of the
proof of $\typeof{(\abs{r})}{(\arr{u}{t})}$ in the informal proof.
This yields the sequent
\begin{displaymath}
\begin{array}{@{}r@{}l@{}}
\ldots,\nat{(\suc{j_0''})},
\seq{j_0''}{\nil}
	{\bigwedge x ((\typeof{x}{u}) \iimp \atom{\typeof{(r \, x)}{t}})}, \\
\ldots
 & \longrightarrow {\prv{\atom{\typeof{v}{t}}}}
\enspace .
\end{array}
\end{displaymath}
This is followed by applications of the $\defL$ and $\forallL$ rules
to give us
\begin{displaymath}
\Seq{\ldots,\nat{(\hbox{\sl s}^3 \, j_1'')},
     \seq{j_1''}{(\cons{(\typeof{n}{u})}{\nil})}
	 {\atom{\typeof{(r \, n)}{t}}}, \ldots}
    {\prv{\atom{\typeof{v}{t}}}}
\enspace .
\end{displaymath}

The informal proof proceeds with a use of the cut rule, and here we
use the derived object-level cut rule (Theorem~\ref{thm:intuit-cut})
with the elided assumption $\seq{j_1'}{\nil}{\atom{\typeof{n}{u}}}$
to obtain
\begin{displaymath}
\begin{array}{@{}r@{}l@{}}
\ldots,\nat{(\hbox{\sl s}^3 \, j_1'')},
\seq{j_1''}{(\cons{(\typeof{n}{u})}{\nil})}
    {\atom{\typeof{(r \, n)}{t}}}, \\
\ldots \longrightarrow
    {\prvv{(\cons{(\typeof{n}{u})}{\nil})}
	 {\atom{\typeof{(r \, n)}{t}}}}
\end{array}
\end{displaymath}
\begin{displaymath}
\Seq{\ldots,\nat{(\hbox{\sl s}^2 \, j_1')},
     \seq{j_1'}{\nil}{\atom{\typeof{n}{u}}}}
    {\prv{\atom{\typeof{n}{u}}}}
\end{displaymath}
\begin{displaymath}
\Seq{\ldots, \prv{\atom{\typeof{(r \, n)}{t}}}}
    {\prv{\atom{\typeof{v}{t}}}}
\enspace .
\end{displaymath}
The first two of these follow easily from Proposition~\ref{prp:nat-other}.

The informal proof concludes by applying the induction hypothesis to
the evaluation and typing judgments for $(r \, n)$.
Again we accomplish this by applying the appropriate left rules to the
induction hypothesis $\forall k\ldots$, which requires the derivation
of the five sequents
\begin{displaymath}
\begin{array}{c@{\qquad\qquad\qquad}c}
\Seq{\nat{(\hbox{\sl s}^2 \, j_1)}}
    {\nat{j_1}}
& \Seq{\nat{(\hbox{\sl s}^2 \, j_1)}}
      {j_1 < (\hbox{\sl s}^2 \, j_1)}
\end{array}
\end{displaymath}
\begin{displaymath}
\Seq{\ldots,\seq{j_1}{\nil}{\atom{\natsem{(r \, n)}{v}}}, \ldots}
    {\seq{j_1}{\nil}{\atom{\natsem{(r \, n)}{v}}}}
\end{displaymath}
\begin{displaymath}
\Seq{\ldots, \prv{\atom{\typeof{(r \, n)}{t}}}}
    {\prv{\atom{\typeof{(r \, n)}{t}}}}
\end{displaymath}
\begin{displaymath}
\Seq{\ldots, \prv{\atom{\typeof{v}{t}}}}
    {\prv{\atom{\typeof{v}{t}}}}
\enspace .
\end{displaymath}
The first two sequents follow from Proposition~\ref{prp:nat-other},
and the last three are all immediate.
\end{proof}

\subsection{A language for computable functions}
\label{sec:pcf}

We now extend the encoding of the static and dynamic semantics for
untyped $\lambda$-terms from the previous section to the programming
language PCF \cite{scott69}.
The necessary $\FOLDN$ constants for PCF types are
\begin{displaymath}
\begin{array}{rcl@{\qquad\qquad\qquad}rcl@{\qquad\qquad\qquad}rcl}
\num     & \colon & \tyi
& \bool    & \colon & \tyi
& \hbox{\sl arr}     & \colon & \tyi \rightarrow
			\tyi \rightarrow \tyi
\enspace .
\end{array}
\end{displaymath}
Those for PCF terms are
\begin{displaymath}
\begin{array}{@{}rcl@{\qquad\qquad}rcl@{\qquad\qquad}rcl@{}}
\zero    & \colon & \tmi
& \hbox{\sl succ}    & \colon & \tmi \rightarrow \tmi
& \hbox{\sl if}      & \colon & \tmi \rightarrow
			\tmi \rightarrow \tmi \rightarrow \tmi \\
\true    & \colon & \tmi
& \hbox{\sl pred}    & \colon & \tmi \rightarrow \tmi
& \hbox{\sl abs}     & \colon & \tyi \rightarrow
			 (\tmi \rightarrow \tmi) \rightarrow \tmi \\
\false   & \colon & \tmi
& \hbox{\sl is\_zero} & \colon & \tmi \rightarrow \tmi
& \hbox{\sl app}     & \colon & \tmi \rightarrow \tmi \rightarrow \tmi \\
	&	&
&	&	&
& \hbox{\sl rec}     & \colon & \tyi \rightarrow
			 (\tmi \rightarrow \tmi) \rightarrow \tmi
\enspace .
\end{array}
\end{displaymath}
We have again labeled the type $\itm$ with subscripts to improve
the readability of these declarations.
The first argument to {\sl abs} and {\sl rec} represent the PCF type
tag for the variable bound by the abstraction and recursion constructs.

The object logic predicates representing typability and evaluation are
denoted by the same $\FOLDN$ constants as in
Section~\ref{sec:lambda}, plus the additional constant
$\hbox{\sl value}: \tmi \rightarrow \atm$.
The object-level specification is represented in $\FOLDN$ as
the definition $\Dpcf$ shown in
Tables~\ref{tab:pcf-typeof}, \ref{tab:pcf-natsem}, and
\ref{tab:pcf-transsem}; 
we have again omitted the $\relax\defeq \top$ body of the clauses.
\begin{table}[btp]
\caption{Object logic encoding of typing for PCF} 
\label{tab:pcf-typeof}
\vspace{2pt}
\begin{center}
$\begin{array}{l}
\hline\rule{0pt}{14pt}
\prog{\;\;(\typeof{\zero}{\num})}
     {\qquad\ttrue} \\
\prog{\;\;(\typeof{\true}{\bool})}
     {\qquad\ttrue} \\
\prog{\;\;(\typeof{\false}{\bool})}
     {\qquad\ttrue} \\
\prog{\;\;(\typeof{(\succtm{M})}{\num})}
     {\qquad\atom{\typeof{M}{\num}}} \\
\prog{\;\;(\typeof{(\pred{M})}{\num})}
     {\qquad\atom{\typeof{M}{\num}}} \\
\prog{\;\;(\typeof{(\iszero{M})}{\bool})}
     {\qquad\atom{\typeof{M}{\num}}} \\
\prog{\;\;(\typeof{(\iftm{M}{N_1}{N_2})}{T})}
     {\qquad\atom{\typeof{M}{\bool}}
      \with \atom{\typeof{N_1}{T}}
      \with \atom{\typeof{N_2}{T}}} \\
\prog{\;\;(\typeof{(\tabs{T}{R})}{(\arr{T}{U})})}
     {\qquad\bigwedge n ((\typeof{n}{T})
			 \iimp \atom{\typeof{(R \, n)}{U}})} \\
\prog{\;\;(\typeof{(\app{M}{N})}{T})}
     {\qquad\atom{\typeof{M}{(\arr{U}{T})}} \with
		\atom{\typeof{N}{U}}} \\ 
\prog{\;\;(\typeof{(\rec{T}{R})}{T})}
     {\qquad\bigwedge n ((\typeof{n}{T})
			 \iimp \atom{\typeof{(R \, n)}{T}})}
\\[2pt]
\hline
\end{array}$
\end{center}
\end{table}
\begin{table}[btp]
\caption{Object logic encoding of natural semantics for PCF} 
\label{tab:pcf-natsem}
\vspace{2pt}
\begin{center}
$\begin{array}{l}
\hline\rule{0pt}{14pt}
\prog{\;\;(\natsem{\zero}{\zero})}
     {\qquad\ttrue} \\
\prog{\;\;(\natsem{\true}{\true})}
     {\qquad\ttrue} \\
\prog{\;\;(\natsem{\false}{\false})}
     {\qquad\ttrue} \\
\prog{\;\;(\natsem{(\succtm{M})}{(\succtm{V})})}
     {\qquad\atom{\natsem{M}{V}}} \\
\prog{\;\;(\natsem{(\pred{M})}{\zero})}
     {\qquad\atom{\natsem{M}{\zero}}} \\
\prog{\;\;(\natsem{(\pred{M})}{V})}
     {\qquad\atom{\natsem{M}{(\succtm{V})}}} \\
\prog{\;\;(\natsem{(\iszero{M})}{\true})}
     {\qquad\atom{\natsem{M}{\zero}}} \\
\prog{\;\;(\natsem{(\iszero{M})}{\false})}
     {\qquad\atom{\natsem{M}{(\succtm{V})}}} \\
\prog{\;\;(\natsem{(\iftm{M}{N_1}{N_2})}{V})}
     {\qquad\atom{\natsem{M}{\true}}
      \with \atom{\natsem{N_1}{V}}} \\
\prog{\;\;(\natsem{(\iftm{M}{N_1}{N_2})}{V})}
     {\qquad\atom{\natsem{M}{\false}}
      \with \atom{\natsem{N_2}{V}}} \\
\prog{\;\;(\natsem{(\tabs{T}{R})}{(\tabs{T}{R})})}
     {\qquad\ttrue} \\
\prog{\;\;(\natsem{(\app{M}{N})}{V})}
     {\qquad\atom{\natsem{M}{(\tabs{T}{R})}}
                       \with \atom{\natsem{(R \, N)}{V}}} \\
\prog{\;\;(\natsem{(\rec{T}{R})}{V})}
     {\qquad\atom{\natsem{(R \, (\rec{T}{R}))}{V}}}
\\[2pt]
\hline
\end{array}$
\end{center}
\end{table}
\begin{table}[btp]
\caption{Object logic encoding of transition semantics for PCF} 
\label{tab:pcf-transsem}
\vspace{2pt}
\begin{center}
$\begin{array}{l}
\hline\rule{0pt}{14pt}
\prog{\;\;(\transone{(\succtm{M})}{(\succtm{M'})})}
     {\qquad\atom{\transone{M}{M'}}} \\
\prog{\;\;(\transone{(\pred{\zero})}{\zero})}
     {\qquad\ttrue} \\
\prog{\;\;(\transone{(\pred{(\succtm{V})})}{V})}
     {\qquad\atom{\valuep{V}}} \\
\prog{\;\;(\transone{(\pred{M})}{(\pred{M'})})}
     {\qquad\atom{\transone{M}{M'}}} \\
\prog{\;\;(\transone{(\iszero{\zero})}{\true})}
     {\qquad\ttrue} \\
\prog{\;\;(\transone{(\iszero{(\succtm{V})})}{\false})}
     {\qquad\atom{\valuep{V}}} \\
\prog{\;\;(\transone{(\iszero{M})}{(\iszero{M'})})}
     {\qquad\atom{\transone{M}{M'}}} \\
\prog{\;\;(\transone{(\iftm{\true}{M}{N})}{M})}
     {\qquad\ttrue} \\
\prog{\;\;(\transone{(\iftm{\false}{M}{N})}{N})}
     {\qquad\ttrue} \\
\prog{\;\;(\transone{(\iftm{M}{N_1}{N_2})}{(\iftm{M'}{N_1}{N_2})})}
     {\qquad\atom{\transone{M}{M'}}} \\
\prog{\;\;(\transone{(\app{(\abs{T}{R})}{N})}{(R \, N)})}
     {\qquad\ttrue} \\
\prog{\;\;(\transone{(\app{M}{N})}{(\app{M'}{N})})}
     {\qquad\atom{\transone{M}{M'}}} \\
\prog{\;\;(\transone{(\rec{T}{R})}{(R \, (\rec{T}{R}))})}
     {\qquad\ttrue} \\
\\
\prog{\;\;(\transsem{M}{M})}
     {\qquad\ttrue} \\
\prog{\;\;(\transsem{M}{N})}
     {\qquad(\atom{\transone{M}{M'}} \with \atom{\transsem{M'}{N}})} \\
\\
\prog{\;\;(\valuep{\zero})}
     {\qquad\ttrue} \\
\prog{\;\;(\valuep{\true})}
     {\qquad\ttrue} \\
\prog{\;\;(\valuep{\false})}
     {\qquad\ttrue} \\
\prog{\;\;(\valuep{(\succtm{V})})}
     {\qquad\atom{\valuep{V}}} \\
\prog{\;\;(\valuep{(\abs{T}{R})})}
     {\qquad\ttrue}
\\[2pt]
\hline
\end{array}$
\end{center}
\end{table}
The following theorem lists the properties of PCF that we have derived
in $\FOLDN$.
The type tags in PCF terms allow the unicity of typing
to hold in addition to the determinacy of semantics, equivalence of
semantics and subject reduction.
The $\FOLDN$ derivations again closely follow the informal proofs of
these properties; the only exception is the derivation of the unicity
of typing property, which we discuss below.
\begin{theorem}
\label{thm:pcf-det-sem}
\label{thm:pcf-equiv-sem}
\label{thm:pcf-subj-red}
\label{thm:pcf-unicity}
The following formulas are derivable in $\FOLDN$ from the definition
that accumulates $\Dnat$, $\Dlist{\atm}$, $\Dintuit$, $\Dpcf$ and
the clause $\same{X}{X} \defeq \top$ defining the predicate
$\equiv: \itm \rightarrow \itm \rightarrow \oo$.

{\rm Determinacy of semantics:}
\begin{displaymath}
\begin{array}{c}
\forall m\forall m_1\forall m_2
        (\prv{\atom{\natsem{m}{m_1}}}
         \oimp \prv{\atom{\natsem{m}{m_2}}}
         \oimp \same{m_1}{m_2}) \\
\forall m\forall m_1\forall m_2
        (\prv{\atom{\transone{m}{m_1}}}
         \oimp \prv{\atom{\transone{m}{m_2}}}
         \oimp \same{m_1}{m_2}) \\
\forall m\forall v_1\forall v_2
        (\prv{\atom{\valuep{v_1}}}
         \oimp \prv{\atom{\transsem{m}{v_1}}}
         \oimp \prv{\atom{\valuep{v_2}}}
         \oimp \prv{\atom{\transsem{m}{v_2}}}
         \oimp \same{v_1}{v_2})
\end{array}
\end{displaymath}

{\rm Equivalence of semantics:}
\begin{displaymath}
\begin{array}{c}
\forall m\forall v
        (\prv{\atom{\natsem{m}{v}}}
         \oimp (\prv{\atom{\valuep{v}}}
         \land \prv{\atom{\transsem{m}{v}}})) \\
\forall m\forall v
        ((\prv{\atom{\valuep{v}}}
          \land \prv{\atom{\transsem{m}{v}}})
         \oimp \prv{\atom{\natsem{m}{v}}})
\end{array}
\end{displaymath}

{\rm Subject reduction:}
\begin{displaymath}
\begin{array}{c}
\forall m\forall n
        (\prv{\atom{\natsem{m}{n}}}
         \oimp \forall t(\prv{\atom{\typeof{m}{t}}}
                        \oimp \prv{\atom{\typeof{n}{t}}})) \\
\forall m\forall n
        (\prv{\atom{\transone{m}{n}}}
         \oimp  \forall t(\prv{\atom{\typeof{m}{t}}}
                         \oimp \prv{\atom{\typeof{n}{t}}})) \\
\forall m\forall n
        (\prv{\atom{\transsem{m}{n}}}
         \oimp  \forall t(\prv{\atom{\typeof{m}{t}}}
                         \oimp \prv{\atom{\typeof{n}{t}}}))
\end{array}
\end{displaymath}

{\rm Unicity of typing:}
\begin{displaymath}
\begin{array}{c}
\forall m\forall t_1\forall t_2
        (\prv{\atom{\typeof{m}{t_1}}}
         \oimp \prv{\atom{\typeof{m}{t_2}}}
         \oimp \same{t_1}{t_2})
\end{array}
\end{displaymath}
\end{theorem}

The usual informal proof of the unicity of typing relies on the
requirement that the list of assumptions in the object logic 
sequent contains typing assignments only for variables and no 
more than one assignment for any particular variable.
Since we have encoded the variables of PCF as variables of
our object logic, which in turn are encoded as variables of $\FOLDN$,
we cannot state the first part of this requirement in $\FOLDN$.
Thus our derivation (given in \citeN{mcdowell97phd}) must differ from
the informal proof.
In fact, we make essential use of the PCF recursion construct in the
{\sl abs} case of the derivation; for an arbitrary type $u$, the term
$(\rec{u}{(\lambda y\,y)})$ has the type $u$ and no other type.
As a result, our derivation does not generalize to languages
without this construct.
In the next section we give an encoding of an extension of PCF
in the object logic of Section~\ref{sec:linear}, which
is encoded in $\FOLDN$ using the explicit eigenvariable encoding.
Although this explicit eigenvariable encoding makes the syntax more 
cumbersome, it allows the derivations in $\FOLDN$ to be more natural.
This is illustrated by the fact that we can capture in $\FOLDN$ the
typical proof of the unicity of typing.

\section{Representation and analysis of an imperative programming language}
\label{sec:imperative}

In this section we consider the programming language $\pcfref$,
an extension of PCF with state \cite{gunter92book}.
This language extends PCF with reference types and constructs for
referencing, dereferencing, assignment, and sequential evaluation.
The type $(\refty{\tau})$ is the type of references to values of type
$\tau$.
If $m$ is a term of type $\tau$, then $(\reftm{m})$ has type
$(\refty{\tau})$ and evaluates to a new memory location containing the
value of $m$.
If $m$ is a term of type $(\refty{\tau})$, then the value of $m$ is a
memory location, and $!m$ has type $\tau$ and evaluates to the contents
of that location.
If $m$ has type $(\refty{\tau})$ and $n$ has type $\tau$, then $(m := n)$
has type $\tau$.
The evaluation of $(m := n)$ changes the contents of the value of $m$
to be the value of $n$; its value is the same as the value of $n$.
If $m_1$ and $m_2$ have types $\tau_1$ and $\tau_2$, respectively,
then $(m_1;m_2)$ has type $\tau_2$.
To evaluate $(m_1;m_2)$, we first evaluate $m_1$, then evaluate $m_2$,
and finally return the value of $m_2$.
Clearly the value of a $\pcfref$ term will depend on the state in
which it is evaluated, and the state may be modified in the evaluation
process;
thus evaluation becomes a mapping from a term-state pair to a
value-state pair.

To encode $\pcfref$, we use the linear object logic of
Section~\ref{sec:linear}, since linear logic is well-suited as
a specification logic for programming languages with state
\cite{cervesato96lics,chirimar95phd,miller96tcs}.
For such languages, the order of evaluation becomes important, and so
a continuation-based operational semantics is often used for the
encoding.
In a continuation-based semantics, each rule has at most one premise,
and any additional evaluation steps are encoded in the continuation.
This encoding of the evaluation steps into the continuation makes the
order of evaluation explicit.
A continuation-based semantics for $\pcfref$ is given in
Table~\ref{tab:pcfref-continuation-semantics}; following
\citeN{gunter92book} we specify call-by-value evaluation.
To abbreviate our presentation we omit the rules for the natural number,
boolean, and conditional constructs; a presentation with the full language
is given in \citeN{mcdowell97phd}.
The semantics of Table~\ref{tab:pcfref-continuation-semantics}
and their object logic encoding given below are a
variation of those found in \citeN{cervesato96lics}.
The judgement $\kfocus{\kappa}{M}{\sigma}{\phi}$ represents the idea
that the evaluation of the term $M$ in state $\sigma$ with
continuation $\kappa$ results in the final answer $\phi$.
A continuation is a list whose elements are of the form $\hat{x}.M$,
where $M$ is a term containing the variable $x$.
(We use $\hat{x}$ instead of $\lambda x$ to avoid confusion with
$\lambda$-abstraction in $\pcfref$.)
The answer $\phi$ is a pair including the final value and the final
state.
The judgement $\kreturn{\kappa}{V}{\sigma}{\phi}$ indicates that
passing the value $V$ with state $\sigma$ to the continuation $\kappa$
results in the final answer $\phi$.
In the rules of Table~\ref{tab:pcfref-continuation-semantics}, $c$ is used 
to range over locations (reference cells).
In the rule for the continuation $(\cont{\kappa}{x}{\reftm{x}})$,
$c$ must be a new location, {\it i.e.}, a location that does not occur in the
state $\sigma$.
The expression $\sigma[c \mapsto V]$ represents the state that is the
same as $\sigma$ except that location $c$ contains the value $V$.
\begin{table}[btp]
\caption{Continuation-based natural semantics for $\pcfref$} 
\label{tab:pcfref-continuation-semantics}
\vspace{2pt}
\begin{center}
$\begin{array}{@{}c@{\quad\quad}c@{\quad\quad}c@{}}
\hline
\\
\infer{\kreturn{}{V}{\sigma}{(V,\sigma)}}
      {}
& \infer{\kfocus{\kappa}{\reftm{M}}{\sigma}{\phi}}
      {\kfocus{\cont{\kappa}{x}{\reftm{x}}}{M}{\sigma}{\phi}}
& \infer{\kfocus{\kappa}{!M}{\sigma}{\phi}}
      {\kfocus{\cont{\kappa}{x}{!x}}{M}{\sigma}{\phi}}
\\ \\
\infer{\kfocus{\kappa}{c}{\sigma}{\phi}}
      {\kreturn{\kappa}{c}{\sigma}{\phi}}
& \infer{\kreturn{\cont{\kappa}{x}{\reftm{x}}}{V}{\sigma}{\phi}}
      {\kreturn{\kappa}{c}{\sigma[c\mapsto V]}{\phi}}
& \infer{\kreturn{\cont{\kappa}{x}{!x}}{c}{\sigma}{\phi}}
      {\kreturn{\kappa}{\sigma(c)}{\sigma}{\phi}}
\\ \\
\infer{\kfocus{\kappa}{M:=N}{\sigma}{\phi}}
      {\kfocus{\cont{\kappa}{x}{x:=N}}{M}{\sigma}{\phi}}
& \infer{\kreturn{\cont{\kappa}{x}{x:=N}}{V}{\sigma}{\phi}}
      {\kfocus{\cont{\kappa}{x}{V:=x}}{N}{\sigma}{\phi}}
& \infer{\kreturn{\cont{\kappa}{x}{c:=x}}{V}{\sigma}{\phi}}
      {\kreturn{\kappa}{V}{\sigma[c \mapsto V]}{\phi}}
\\ \\
\multicolumn{3}{c}
{\begin{array}{c@{\quad\quad\quad}c}
\infer{\kfocus{\kappa}{M;N}{\sigma}{\phi}}
      {\kfocus{\cont{\kappa}{x}{x;N}}{M}{\sigma}{\phi}}
& \infer{\kreturn{\cont{\kappa}{x}{x;N}}{V}{\sigma}{\phi}}
      {\kfocus{\kappa}{N}{\sigma}{\phi}}
\end{array}}
\\ \\
\infer{\kfocus{\kappa}{M\,N}{\sigma}{\phi}}
      {\kfocus{\cont{\kappa}{x}{x\,N}}{M}{\sigma}{\phi}}
& \infer{\kreturn{\cont{\kappa}{x}{x\,N}}{V}{\sigma}{\phi}}
      {\kfocus{\cont{\kappa}{x}{V\,x}}{N}{\sigma}{\phi}}
& \infer{\kfocus{\kappa}{\lambda x:\tau.M}{\sigma}{\phi}}
      {\kreturn{\kappa}{\lambda x:\tau.M}{\sigma}{\phi}}
\\ \\
\multicolumn{3}{c}
{\begin{array}{c@{\quad\quad\quad}c}
\infer{\kreturn{\cont{\kappa}{x}{(\lambda y:\tau.M')\,x}}{V}{\sigma}{\phi}}
      {\kfocus{\kappa}{M'[V/y]}{\sigma}{\phi}}
& \infer{\kfocus{\kappa}{\rec{x:\tau.M}}{\sigma}{\phi}}
      {\kfocus{\kappa}{M[\rec{x:\tau.M}/x]}{\sigma}{\phi}}
\end{array}}
\\[2pt]
\hline
\end{array}$
\end{center}
\end{table}

To encode $\pcfref$, we use the constants
\begin{displaymath}
\begin{array}{rcl@{\qquad\qquad}rcl@{\qquad\qquad}rcl}
\hbox{\sl refty} & \colon & \tyi \rightarrow \tyi
& \hbox{\sl ref} & \colon & \tmi \rightarrow \tmi
& \hbox{\sl assign} & \colon & \tmi \rightarrow \tmi \rightarrow \tmi \\
\hbox{\sl cell} & \colon & \lci \rightarrow \tmi
& \hbox{\sl deref} & \colon & \tmi \rightarrow \tmi
& \hbox{\sl sequence} & \colon & \tmi \rightarrow \tmi \rightarrow \tmi
\end{array}
\end{displaymath}
in addition to the constants of Section~\ref{sec:pcf}.
Once again we have labeled the type $\itm$ with subscripts to improve
the readability of these declarations.
The subscript \lc\ indicates that the argument to {\sl cell}
represents a $\pcfref$ location.

The object logic predicate representing typability is denoted by the
same $\FOLDN$ constants as in Section~\ref{sec:functional};
its object-level specification is represented in
$\FOLDN$ as the definition shown in Table~\ref{tab:pcfref-typeof}.
\begin{table}[btp]
\caption{Object logic encoding of typing for $\pcfref$ terms} 
\label{tab:pcfref-typeof}
\vspace{2pt}
\begin{center}
$\begin{array}{l@{\;\;}l}
\hline\rule{0pt}{14pt}
\hbox{\sl prog} & {\typeofl{(\tabsl{T}{R})}{(\arrl{T}{U})})} \\
     	& {\lambda l(\cons{\bigwedge n(\typeof{n}{(T \, l)}
				       \iimp \atom{\typeof{(R \, l \, n)}{(U \, l)}})}
			  {\nil})}
	  {\qquad\nill} \\
\hbox{\sl prog} & {(\typeofl{(\appl{M}{N})}{T})} \\
     	& {(\consl{\atoml{\typeofl{M}{(\arrl{U}{T})}}}
		  {\consl{\atoml{\typeofl{N}{U}}}
			 {\nill}})}
	  {\qquad\nill} \\
\hbox{\sl prog} & {(\typeofl{(\recl{T}{R})}{T})} \\
     	& {\lambda l(\cons{\bigwedge n(\typeof{n}{(T\,l)}
					\iimp \atom{\typeof{(R\,l\,n)}{(T\,l)}})}
			   {\nil})}
	  {\qquad\nill} \\
\hbox{\sl prog} & {(\typeofl{(\reftml{M})}{(\reftyl{T})})} \\
     	& {(\consl{\atoml{\typeofl{M}{T}}}{\nill})}
	  {\qquad\nill} \\
\hbox{\sl prog} & {(\typeofl{(\derefl{M})}{T})} \\
     	& {(\consl{\atoml{\typeofl{M}{(\reftyl{T})}}}{\nill})}
	  {\qquad\nill} \\
\hbox{\sl prog} & {(\typeofl{(\assignl{M}{N})}{T})} \\
     	& {(\consl{\atoml{\typeofl{M}{(\reftyl{T})}}}
		  {\consl{\atoml{\typeofl{N}{T}}}
			 {\nill}})}
	  {\qquad\nill} \\
\hbox{\sl prog} & {(\typeofl{(\sequencel{M}{N})}{T})} \\
     	& {(\consl{\atoml{\typeofl{M}{U}}}
		  {\consl{\atoml{\typeofl{N}{T}}}
			 {\nill}})}
	  {\qquad\nill}
\\[2pt]
\hline
\end{array}$
\end{center}
\end{table}
Recall that
$\lprog{A}{(\cons{C_1}{\ldots\cons{C_n}{\nil}})}
       {(\cons{B_1}{\ldots\cons{B_m}{\nil}})}$
represents the definite clause
\begin{displaymath}
\bigwedge \bar{x}
	(B_1 \iimp \cdots B_m \iimp
	 C_1 \limp \cdots C_n \limp A)
\enspace ,
\end{displaymath}
where the free variables of $A$, $B_1$, \ldots, $B_m$, $C_1$, \ldots,
$C_n$ are included in the list $\bar{x}$.
This means that to derive an instance of $A$, we can instead derive
the corresponding instances of $B_1$, \ldots, $B_m$, $C_1$, \ldots, $C_n$.
To establish $\lprvv{{\it IL}}{{\it LL}}{\atom{A}}$, the rules of linear logic
require that each assumption in ${\it LL}$ be used exactly once in the
derivation of one of the $C_i$'s; it cannot be used in the derivation
of any of the $B_i$'s, or in the derivation of more than one $C_i$.
In the specification of typing, no linear assumptions are introduced,
so ${\it LL}$ will be empty.
In general, we will use linear formulas ($C_1$, \ldots, $C_n$) in the
bodies of specification clauses; we use intuitionistic formulas
($B_1$, \ldots, $B_n$) only where we specifically wish to 
preclude the use of linear assumptions.
This is only done in one clause in the encoding of the operational
semantics, and will be discussed when it is introduced.
We extend the abbreviation convention of
Section~\ref{sec:linear} to the constants of this section.
Thus $(\typeofl{m}{t})$ abbreviates 
$(\lambda l\,\typeof{(m\,l)}{(t\,l)})$,
$(\reftyl{t})$ abbreviates $(\lambda l\,\refty{(t\,l)})$, etc.

The semantics for $\pcfref$ is more complicated than those in the
previous sections.
The constant $\Downarrow$ now has type
$\tmi \rightarrow \sti \rightarrow \ansi \rightarrow \atm$. 
The object logic atom $\natsemst{m}{s}{f}$ represents the evaluation of the term
$m$ in the state $s$ yielding the final answer $f$.
State is encoded using the constants $\nullst\colon\sti$ and
$\hbox{\sl extend\_st}\colon\lci\rightarrow\tmi\rightarrow\sti\rightarrow\sti$;
$\nullst$ represents the state with no locations, and 
$(\extendst{c}{v}{s})$ represents the state obtained by adding the
location $c$ containing value $v$ to the state $s$.
A value and a state are combined into an answer using the constant
$\hbox{\sl answer}\colon\tmi\rightarrow\sti\rightarrow\ansi$;
variables representing new locations are bound using
$\hbox{\sl new}\colon(\lci\rightarrow\ansi)\rightarrow\ansi$.
Our specification of evaluation will also use the predicates
\begin{displaymath}
\begin{array}{rcl}
\hbox{\sl ns\_mach\_1} & \colon & \cntni \rightarrow
			\instri \rightarrow 
			\sti \rightarrow
			\ansi \rightarrow \atm\\
\hbox{\sl ns\_mach\_2} & \colon & \cntni \rightarrow
			\instri \rightarrow
			\ansi \rightarrow \atm \\
\hbox{\sl contains} & \colon & \lci \rightarrow \tmi \rightarrow \atm \\
\hbox{\sl collect\_state} & \colon & \sti \rightarrow \atm
\enspace .
\end{array}
\end{displaymath}
The object logic atom $\nsmachone{\!\!k}{\!\!i}{\!\!s}{\!\!f}$ corresponds to the
two judgements of Table~\ref{tab:pcfref-continuation-semantics}.
Continuations are constructed using $\initk\colon\cntni$ to represent the
initial continuation and
$\succ\colon (\tmi \rightarrow \instri) \rightarrow
		\cntni \rightarrow \cntni$
to extend a continuation.
Instructions, constructed from the constants
\begin{displaymath}
\begin{array}{rcl@{\qquad\qquad}rcl}
\hbox{\sl eval} & \colon & \tmi \rightarrow \instri
& \hbox{\sl new\_ref} & \colon & \tmi \rightarrow \instri \\
\hbox{\sl return} & \colon & \tmi \rightarrow \instri
& \hbox{\sl lookup} & \colon & \tmi \rightarrow \instri \\
\hbox{\sl eval\_arg} & \colon & \tmi \rightarrow \tmi \rightarrow \instri
& \hbox{\sl eval\_rvalue} & \colon & \tmi \rightarrow \tmi \rightarrow \instri \\
\hbox{\sl apply} & \colon & \tmi \rightarrow \tmi \rightarrow \instri
& \hbox{\sl update} & \colon & \tmi \rightarrow \tmi \rightarrow \instri
\enspace ,
\end{array}
\end{displaymath}
are used to indicate the current task in the evaluation of a term.
The object logic atom $\nsmachtwo{\!\!k}{\!\!i}{\!\!f}$ is a variation
of $\nsmachone{\!\!k}{\!\!i}{\!\!s}{\!\!f}$ which does not contain the
state; instead the contents of each location is recorded using the
object logic predicate denoted by the constant {\sl contains}.
The evaluation of terms is specified using this distributed
representation of state; the state portion of the final answer is
constructed again using the predicate {\sl collect\_state}.
The specifications for all of these predicates are represented by the
$\FOLDN$ definition in Tables~\ref{tab:pcfref-natsem1}
and \ref{tab:pcfref-natsem2}.
\begin{table}[btp]
\caption{Object logic encoding of natural semantics for $\pcfref$ (part I)} 
\label{tab:pcfref-natsem1}
\vspace{2pt}
\begin{center}
$\begin{array}{l@{\;\;}l}
\hline\rule{0pt}{14pt}
\hbox{\sl prog} & {(\natsemstl{M}{S}{F})} \\
	& {\nill}
	  {\qquad(\consl{\atoml{\nsmachonel{\initkl}{(\evall{M})}{S}{F}}}
				 {\nill})} \\
\\
\hbox{\sl prog} & {(\nsmachonel{K}{I}{(\extendstl{C}{V}{S})}{F})} \\
	& {(\consl{(\containsl{C}{V} \limpl
		    \atoml{\nsmachonel{K}{I}{S}{F}})}
		  {\nill})}
	  {\qquad\nill} \\
\hbox{\sl prog} & {(\nsmachonel{K}{I}{\nullstl}{F})} \\
	& {(\consl{\atoml{\nsmachtwol{K}{I}{F}}}
		  {\nill})}
	  {\qquad\nill} \\
\\
\hbox{\sl prog} & {(\collectstl{(\extendstl{C}{V}{S})})} \\
	& {(\consl{\atoml{\containsl{C}{V}}}
		  {\consl{\atoml{\collectstl{S}}}
			 {\nill}})}
	  {\qquad\nill} \\
\hbox{\sl prog} & {(\collectstl{\nullstl})} \\
	& {\nill}
	  {\qquad\nill}
\\[2pt]
\hline
\end{array}$
\end{center}
\end{table}
\begin{table}[btp]
\caption{Object logic encoding of natural semantics for $\pcfref$ (part II)} 
\label{tab:pcfref-natsem2}
\vspace{2pt}
\begin{center}
$\begin{array}{l@{\;\;}l}
\hline\rule{0pt}{14pt}
\hbox{\sl prog} & {(\nsmachtwol{\initkl}{(\returnl{V})}{(\answerl{V}{S})})} \\
	& {(\consl{\atoml{\collectstl{S}}}{\nill})}
	  {\qquad\nill} \\
\hbox{\sl prog} & {(\nsmachtwol{(\extendkl{K}{I})}{(\returnl{V})}{F})} \\
	& {(\consl{\atoml{\nsmachtwol{K}{(\lambda l\,I\,l\,(V\,l))}{F}}}{\nill})}
	  {\qquad\nill} \\
\\
\hbox{\sl prog} & {(\nsmachtwol{K}{(\evall{(\locl{C})})}{F})} \\
	& {(\consl{\atoml{\nsmachtwol{K}{(\returnl{(\locl{C})})}{F}}}
		  {\nill})}
	  {\qquad\nill} \\
\\
\hbox{\sl prog} & {(\nsmachtwol{K}{(\evall{(\reftml{M})})}{F})} \\
	& {(\consl{\atoml{\nsmachtwol{(\extendkl{K}{(\lambda l\lambda v\,\newref{v})})}
				     {(\evall{M})}
				     {F}}}
		  {\nill})}
	  {\qquad\nill} \\
\hbox{\sl prog} & {(\nsmachtwol{K}{(\newrefl{V})}{(\newl{F})})} \\
	& {\lambda l(\cons{\bigwedge c(\contains{c}{(V\,l)} \limp
				      \atom{\nsmachtwo{(K\,l)}{(\return{(\loc{c})})}{(F\,l\,c)}})}
			   {\nill})} \\
	& {\nill} \\
\\
\hbox{\sl prog} & {(\nsmachtwol{K}{(\evall{(\derefl{M})})}{F})} \\
	& {(\consl{\atoml{\nsmachtwol{(\extendkl{K}{(\lambda l\lambda v\,\lookup{v})})}
				     {(\evall{M})}
				     {F}}}
		  {\nill})}
	  {\qquad\nill} \\
\hbox{\sl prog} & {(\nsmachtwol{K}{(\lookupl{(\locl{C})})}{F})} \\
	& {(\consl{\atoml{\containsl{C}{V}}}
		  {\consl{(\containsl{C}{V} \limpl
			   \atoml{\nsmachtwol{K}{(\returnl{V})}{F}})}
			 {\nill}})} \\
	& {\nill} \\
\\
\hbox{\sl prog} & {(\nsmachtwol{K}{(\evall{(\assignl{M}{N})})}{F})} \\
	& {(\consl{\atoml{\nsmachtwol{(\extendkl{K}{(\lambda l\lambda v\,\evalrval{v}{(N\,l)})})}
				     {(\evall{M})}
				     {F}}}
		  {\nill})} \\
	& {\nill} \\
\hbox{\sl prog} & {(\nsmachtwol{K}{(\evalrvall{V}{N})}{F})} \\
	& {(\consl{\atoml{\nsmachtwol{(\extendkl{K}{(\lambda l\lambda v\,\update{(V\,l)}{v})})}
				     {(\evall{N})}
				     {F}}}
		  {\nill})}
	  {\qquad\nill} \\
\hbox{\sl prog} & {(\nsmachtwol{K}{(\updatel{(\locl{C})}{V})}{F})} \\
	& {(\consl{\atoml{\containsl{C}{W}}}{}
		  {\consl{(\containsl{C}{V} \limpl
			   \atoml{\nsmachtwol{K}{(\returnl{V})}{F}})}
			 {\nill}})} \\
	& {\nill} \\
\\
\hbox{\sl prog} & {(\nsmachtwol{K}{(\evall{(\sequencel{M}{N})})}{F})} \\
	& {(\consl{\atoml{\nsmachtwol{(\extendkl{K}{(\lambda l\lambda v\,\eval{(N\,l)})})}
				     {(\evall{M})}
				     {F}}}
		  {\nill})}
	  {\qquad\nill}
\\ \\
\hbox{\sl prog} & {(\nsmachtwol{K}{(\evall{(\appl{M}{N})})}{F})} \\
	& {(\consl{\atoml{\nsmachtwol{(\extendkl{K}
					        {(\lambda l\lambda v\,\evalarg{v}{(N\,l)})})}
				     {(\evall{M})}
				     {F}}}
		  {\nill})}
	  {\qquad\nill} \\
\hbox{\sl prog} & {(\nsmachtwol{K}{(\evalargl{V}{N})}{F})} \\
	& {(\consl{\atoml{\nsmachtwol{(\extendkl{K}
					        {(\lambda l\lambda v\,\apply{(V\,l)}{v})})}
				     {(\evall{N})}
				     {F}}}
		  {\nill})}
	  {\qquad\nill} \\
\hbox{\sl prog} & {(\nsmachtwol{K}{(\applyl{(\tabsl{T}{R})}{V})}{F})} \\
	& {(\consl{\atoml{\nsmachtwol{K}{(\evall{(\lambda l\,R\,l\,(V\,l))})}{F}}}
		  {\nill})}
	  {\qquad\nill} \\
\\
\hbox{\sl prog} & {(\nsmachtwol{K}{(\evall{(\tabsl{T}{R})})}{F})} \\
	& {(\consl{\atoml{\nsmachtwol{K}
				     {(\returnl{(\tabsl{T}{R})})}
				     {F}}}
		  {\nill})}
	  {\qquad\nill} \\
\hbox{\sl prog} & {(\nsmachtwol{K}{(\evall{(\recl{T}{R})})}{F})} \\
	& {(\consl{\atoml{\nsmachtwol{K}
				     {(\evall{(\lambda l\,R\,l\,(\rec{(T\,l)}{(R\,l)}))})}
				     {F}}}
		  {\nill})}
	  {\qquad\nill}
\\[2pt]
\hline
\end{array}$
\end{center}
\end{table}

This encoding differs slightly from the continuation semantics in
Table~\ref{tab:pcfref-continuation-semantics}.
The object logic judgement
$\lprvv{\nill}{ll}{\atoml{\nsmachtwo{\!\!k}{\!\!(\return{v})}{\!\!f}}}$
corresponds to the judgement $\kreturn{\kappa}{v'}{\sigma}{\phi}$,
where $\kappa$ is the continuation encoded by $k$, $v'$ is the value
encode by $v$, $\sigma$ is the state encoded by the list $ll$ of {\sl
contains} assumptions, and $\phi$ is the answer encoded by $f$.
However, the specification for $\nsmachtwol{\!\!k}{\!\!(\returnl{v})}{\!\!f}$
takes the first instruction from $k$ and substitutes in the value $v$
to obtain the new instruction.
This new instruction then determines the next step in the evaluation.
On the other hand, the rules of
Table~\ref{tab:pcfref-continuation-semantics} examine the return value
and the first term of the continuation to determine the next
evaluation step.
Other than this small difference, the encoding mirrors the
continuation semantics very closely.

The distributed encoding of state in Tables~\ref{tab:pcfref-natsem1},
and \ref{tab:pcfref-natsem2} makes vital use of linear implication.
Since each assumption of the form $\containsl{c}{v}$ is a linear
assumption, it can only be used once.
This linearity is used, for example, in the clause for
{\sl ns\_mach\_2} with the instruction $(\updatel{(\locl{c})}{v})$;
the desired behavior is that the contents of location $c$ be replaced
by the value $v$.
This clause has two linear formulas in its body,
$\atoml{\containsl{c}{w}}$ and 
$(\containsl{c}{v} \limpl
\atoml{\nsmachtwol{\!\!k}{\!\!(\returnl{v})}{\!\!f}})$. 
Each {\sl contains} assumption must be used exactly once in the
derivation of these two formulas.
Since there is no clause for {\sl contains} in the object logic
theory, the first formula must be derived by the initial rule, and so
will use the one assumption representing the contents of location $c$.
The remainder of the state is then available for the other formula,
which adds a new assumption about the contents of $c$ and then
continues the evaluation encoded in the continuation $k$.
The linearity of the {\sl contains} assumptions is also used in the
clause for {\sl ns\_mach\_2} with the instruction $(\returnl{v})$ and the
continuation $\initkl$.
This clause represents the situation where the evaluation is complete
and we wish to construct the final answer from the value $v$ and the
state encoded in the assumptions.
The clause has the single linear formula $\atoml{\collectstl{s}}$ as its
body.
Thus the derivation of this formula must use all of the {\sl contains}
assumptions;
this ensures that the constructed state includes all of the locations
represented in the assumptions.
Dually, the clause for $\Downarrow$ in Table~\ref{tab:pcfref-natsem1}
has a single intuitionistic formula
$\atoml{\nsmachonel{\!\!\initkl}{\!\!(\evall{m})}{\!\!s}{\!\!f}}$ as
its body.
This clause represents the situation where we wish to evaluate the
term $m$ in the state $s$.
Since the formula in the body is intuitionistic, it must be derived
from an empty set of linear assumptions.
Since there are no linear formulas in the body, this means that
$\natsemstl{m}{s}{f}$ is only derivable from an empty set of linear
assumptions, {\it i.e.}, the state is entirely represented in $s$.

We also introduce typing predicates for continuations, instructions,
and answers:
\begin{displaymath}
\begin{array}{rcl@{\qquad\qquad}rcl}
\hbox{\sl typeof}_{cntn} & \colon & \cntni \rightarrow \tyi \rightarrow \atm
& \hbox{\sl typeof}_{ans} & \colon & \ansi \rightarrow \tyi \rightarrow \atm \\
\hbox{\sl typeof}_{instr} & \colon & \instri \rightarrow \tyi \rightarrow \atm
\enspace .
\end{array}
\end{displaymath}
The object-level specification for these predicates is represented in
$\FOLDN$ by the definition of Table~\ref{tab:pcfref-machtypeof}.
\begin{table}[btp]
\caption{Encoding of typing for $\pcfref$ continuations, instructions, and answers} 
\label{tab:pcfref-machtypeof}
\vspace{2pt}
\begin{center}
$\begin{array}{l@{\;\;}l}
\hline\rule{0pt}{14pt}
\hbox{\sl prog} & {(\typeofcntnl{\initkl}{(\arrl{T}{T})})} \\
     & {\nill}
     {\qquad\nill} \\
\hbox{\sl prog} & {(\typeofcntnl{(\extendkl{K}{I})}{(\arrl{T}{U})})} \\
     & {(\begin{array}[t]{@{}l@{}}
	    \consl{\lambda l\,\bigwedge v(\typeof{v}{(T\,l)}
	                 \iimp \atom{\typeofinstr{(I\,l\,v)}{(T'\,l)}})}{} \\
	    {\consl{\atoml{\typeofcntnl{K}{(\arrl{T'}{U})}}}
		  {\nill}})
	 \end{array}} \\
     & {\nill} \\
\\
\hbox{\sl prog} & {(\typeofinstrl{(\evall{M})}{T})} \\
     & {(\consl{\atoml{\typeofl{M}{T}}}{\nill})}
     {\qquad\nill} \\
\hbox{\sl prog} & {(\typeofinstrl{(\returnl{V})}{T})} \\
     & {(\consl{\atoml{\typeofl{V}{T}}}{\nill})}
     {\qquad\nill} \\
\hbox{\sl prog} & {(\typeofinstrl{(\evalargl{M}{N})}{T})} \\
     & {(\consl{\atoml{\typeofl{M}{(\arrl{U}{T})}}}
	       {\consl{\atoml{\typeofl{N}{U}}}
		      {\nill}})}
     {\qquad\nill} \\
\hbox{\sl prog} & {(\typeofinstrl{(\applyl{M}{N})}{T})} \\
     & {(\consl{\atoml{\typeofl{M}{(\arrl{U}{T})}}}
	       {\consl{\atoml{\typeofl{N}{U}}}
		      {\nill}})}
     {\qquad\nill} \\
\hbox{\sl prog} & {(\typeofinstrl{(\newrefl{M})}{(\reftyl{T})})} \\
     & {(\consl{\atoml{\typeofl{M}{T}}}{\nill})}
     {\qquad\nill} \\
\hbox{\sl prog} & {(\typeofinstrl{(\lookupl{M})}{T})} \\
     & {(\consl{\atoml{\typeofl{M}{(\reftyl{T})}}}{\nill})}
     {\qquad\nill} \\
\hbox{\sl prog} & {(\typeofinstrl{(\evalrvall{M}{N})}{T})} \\
     & {(\consl{\atoml{\typeofl{M}{(\reftyl{T})}}}
	       {\consl{\atoml{\typeofl{N}{T}}}
		      {\nill}})}
     {\qquad\nill} \\
\hbox{\sl prog} & {(\typeofinstrl{(\updatel{M}{N})}{T})} \\
     & {(\consl{\atoml{\typeofl{M}{(\reftyl{T})}}}
	       {\consl{\atoml{\typeofl{N}{T}}}
		      {\nill}})}
     {\qquad\nill} \\
\\
\hbox{\sl prog} & {(\typeofansl{(\answerl{V}{S})}{T})} \\
     & {(\consl{\atoml{\typeofl{V}{T}}}
	       {\consl{\atoml{\welltypedl{S}}}
		      {\nill}})}
     {\qquad\nill} \\
\hbox{\sl prog} & {(\typeofansl{(\newl{F})}{T})} \\
     & {\lambda l\,(\cons{\bigwedge c(\typeof{(\loc{c})}{(\refty{(U\,l)})}
			                 \iimp \atom{\typeofans{(F\,l\,c)}{(T\,l)}})}
			 {\nil})}
     {\qquad\nill} \\
\\
\hbox{\sl prog} & {(\welltypedl{\nullstl})} \\
     & {\nill}
     {\qquad\nill} \\
\hbox{\sl prog} & {(\welltypedl{(\extendstl{C}{V}{S})})} \\
     & {(\consl{\atoml{\typeofl{(\locl{C})}{(\reftyl{T})}}}
	       {\consl{\atoml{\typeofl{V}{T}}}
		      {\consl{\atoml{\welltypedl{S}}}
			     {\nill}}})} \\
     & {\nill}
\\[2pt]
\hline
\end{array}$
\end{center}
\end{table}
A continuation has type $(\arrl{t}{u})$ if it expects a value of type $t$
in order to produce a value of type $u$.
Instructions are typed in the same way as the corresponding terms.
The type of an answer is the same as the type of its value component
under some typing assumptions for any new memory locations.
These assumptions must be consistent with the values stored in those
locations; this consistency is expressed by the predicate
$\hbox{\sl well\_typed} \colon \sti \rightarrow \atm$.

We now present the theorems we have derived in $\FOLDN$ about this
object logic encoding of $\pcfref$.
We will refer to the collected clauses of
Tables~\ref{tab:pcfref-typeof}, \ref{tab:pcfref-natsem1},
\ref{tab:pcfref-natsem2} and \ref{tab:pcfref-machtypeof} as the
definition $\Dpcfref$.
To simplify the presentation of our theorems, we introduce several
$\FOLDN$ predicates:
\begin{displaymath}
\begin{array}[b]{rcl@{\qquad\qquad}rcl}
\hbox{\sl store} & \colon & \atmlstl \rightarrow \oo
& \equiv_{atml} & \colon & \atml \rightarrow \atml \rightarrow \oo \\
\hbox{\sl store\_typing} & \colon & \atmlstl \rightarrow \oo
& \equiv_\itml & \colon & \itml \rightarrow \itml \rightarrow \oo \\
\hbox{\sl store\_typeof} & \colon & \atmlstl	\rightarrow \atmlstl \rightarrow \oo
\enspace .
\end{array}
\end{displaymath}
The {\sl store} predicate indicates that a list of object logic atoms
is a valid distributed encoding of state, that is, its elements are of
the form $\containsl{c}{v}$.
The predicate {\sl store\_typing} holds if its argument is a valid
list of typing assumptions for locations.
The {\sl store\_typeof} predicate holds for a store and store typing if
every location in the store is assigned a type by the store typing
that agrees with a type of the value stored in the location.
Finally, $\equiv_{atml}$ and $\equiv_{itml}$ encode syntactic identity
over the types $\atml$ and $\itml$.
The definition $\Dstore$ for these predicates is presented in
Table~\ref{tab:pcfref-storetypeof}.
\begin{table}[btp]
\caption{Meta-logic predicates for $\pcfref$ stores} 
\label{tab:pcfref-storetypeof}
\vspace{2pt}
\begin{center}
$\begin{array}{rcl}
\hline\rule{0pt}{14pt}
\store{\it LL}
	& \defeq & \tlist{\it LL} \; \land \;
	  \forall a (\begin{array}[t]{@{}l}
		     \element{a}{\it LL} \oimp \\
		     \exists c \exists v (\sameatml{a}{(\containsl{c}{v})}))
		     \end{array} \\
\\
\storetyping{\it IL}
	& \defeq & \tlist{\it IL} \; \land \\
	&	 & \forall a (\begin{array}[t]{@{}l}
			      \element{a}{\it IL} \oimp \\
			      \exists c \exists t
				(\sameatml{a}{(\typeofl{(\locl{c})}
							    {(\reftyl{t})})})) \; \land
			      \end{array} \\
	&	 & \forall c \forall t_1 \forall t_2
			(\begin{array}[t]{@{}l}
			 \element{(\typeofl{(\locl{c})}{(\reftyl{t_1})})}
				 {\it IL} \oimp \\
			 \element{(\typeofl{(\locl{c})}{(\reftyl{t_2})})}
				 {\it IL} \oimp \\
			 \sameitml{t_1}{t_2})
			 \end{array} \\
\\
\storetypeof{\it LL}{\it IL}
	& \defeq & \forall c \forall v
			(\begin{array}[t]{@{}l}
			 \element{(\containsl{c}{v})}{\it LL} \oimp \\
			 \exists t(\begin{array}[t]{@{}l}
				   \element{(\typeofl{(\locl{c})}
						     {(\reftyl{t})})}
					   {\it IL}	\; \land \\
				   \lprvv{\it IL}{\nill}
					     {\atoml{\typeofl{v}{t}}}))
				   \end{array}
			 \end{array} \\
\\
\sameatml{A}{A}
	& \defeq & \top \\
\\
\sameitml{X}{X}
	& \defeq & \top
\\[2pt]
\hline
\end{array}$
\end{center}
\end{table}
The following theorem states that we have derived the subject
reduction and unicity of typing properties for $\pcfref$ in $\FOLDN$.
The $\FOLDN$ derivations again closely follow the informal proofs of
these properties.
We expect that the determinacy of semantics is also derivable, but
have not yet shown this.
We use the following abbreviations from
Section~\ref{sec:linear}: 
$\Dalllists$ for
\begin{displaymath}
\Dlistl{\atm} \cup \Dlistl{\prp}
 \cup \Dlistll{\atm} \cup \Dlistll{\prp}
\enspace ,
\end{displaymath}
and $\Dallevars$ for
\begin{displaymath}
\Devars{\atm} \cup \Devars{\prp} \cup
 \Devars{\atmlst} \cup \Devars{\prplst}
\enspace .
\end{displaymath}
\begin{theorem}
\label{thm:pcfref-subj-red}
\label{thm:pcfref-unicity}
The following formulas are derivable in $\FOLDN$ from the definition
that accumulates $\Dnat$, $\Dalllists$, $\Dallevars$, $\Dlinear$,
$\Dpcfref$, and $\Dstore$:

{\rm Subject reduction:}
\begin{displaymath}
\begin{array}{c}
\forall m\forall s\forall f
        (\begin{array}[t]{@{}l}
         \prv{(\atoml{\natsemstl{m}{s}{f}})} \; \oimp \; \\
	 \forall ts\forall t
	        (\begin{array}[t]{@{}l}
		 \storetyping{il} \; \oimp \;
		 \lprvv{il}{\nill}{\atoml{\welltypedl{s}}} \; \oimp \; \\
		 \lprvv{il}{\nill}
			   {\atoml{\typeofl{m}{t}}} \; \oimp \; \\
		 \lprvv{il}{\nill}
			   {\atoml{\typeofansl{\!f}{\!t}}}))
		 \end{array}
	 \end{array}
\\ \\
\forall ll\forall k\forall i\forall f
        (\begin{array}[t]{@{}l}
	 \store{ll} \; \oimp \;
	 \lprvv{\nill}{ll}
	       {\atoml{\nsmachtwol{\!\!k}{\!\!i}{\!\!f}}} \; \oimp \; \\
	 \forall il\forall t\forall u
	        (\begin{array}[t]{@{}l}
		 \storetyping{il} \; \oimp \;
		 \storetypeof{ll}{il} \; \oimp \; \\
		 \lprvv{il}{\nill}
		       {\atoml{\typeofcntnl{\!k}{\!(\arrl{t}{u})}}} \; \oimp \; \\
		 \lprvv{il}{\nill}
		       {\atoml{\typeofinstrl{\!i}{\!t}}} \; \oimp \; \\
		 \lprvv{il}{\nill}
		       {\atoml{\typeofansl{\!f}{\!u}}}))
		 \end{array}
	 \end{array}
\end{array}
\end{displaymath}

{\rm Unicity of typing:}
\begin{displaymath}
\begin{array}{c}
\forall m\forall t_1\forall t_2
        (\prv{\atoml{\typeofl{m}{t_1}}}
         \; \oimp \; \prv{\atoml{\typeofl{m}{t_2}}}
         \; \oimp \; \sameitml{t_1}{t_2})
\end{array}
\end{displaymath}
\end{theorem}
\begin{proof}
The derivation of the unicity of typing is by complete induction on the
height of the first typing derivation $\prv{\atoml{\typeofl{m}{t_1}}}$.
Let $\hbox{\sl P}_1$ be the predicate
\begin{displaymath}
\lambda il \forall a
	(\element{a}{il} \; \oimp \;
	 \exists x \exists t
		(\sameatml{a}{(\typeofl{(\tfstl{\itm}{x})}{t})}))
\end{displaymath}
and $\hbox{\sl P}_2$ the predicate
\begin{displaymath}
\lambda il \forall x \forall t_1 \forall t_2
	(\element{(\typeofl{x}{t_1})}{il} \; \oimp \;
	 \element{(\typeofl{x}{t_2})}{il} \; \oimp \;
	 \sameitml{t_1}{t_2})
\enspace .
\end{displaymath}
These predicates encode the requirements that the list of assumptions
contains only typing assignments for variables and assigns only one
type to any one variable.
Our induction predicate {\sl IP} is then
\begin{displaymath}
\lambda j\forall il
	(\begin{array}[t]{@{}l@{}}
	 \tlist{il} \oimp \hbox{\sl P}_1 \, il \oimp \hbox{\sl P}_2 \, il \; \oimp \\
	 \forall m\forall t_1\forall t_2
		(\begin{array}[t]{@{}l@{}}
	         \lseq{j}{il}{\nill}
			 {\atoml{\typeofl{m}{t_1}}} \; \oimp \\
		 \lprvv{il}{\nill}
			     {\atoml{\typeofl{m}{t_2}}} \; \oimp \;
	         \sameitml{t_1}{t_2}))
\enspace .
		 \end{array}
	 \end{array}
\end{displaymath}
The details of the proof are presented in \citeN{mcdowell97phd}.
\end{proof}

\section{Related work}
\label{sec:related2}

There are several approaches others have taken to reason about
higher-order abstract syntax encodings directly in a formalized
meta-language.
Despeyroux, Felty, and Hirschowitz in
\citeyear{despeyroux94lpar,despeyroux95tlca} show that induction
principles for a restricted form of second-order abstract syntax can
be derived in the Coq proof development system.
To keep the definitions monotone, they introduce a separate type for 
variables and explicit coercions from variables to other types.
For example, their constructors for $\lambda$-terms would be
\begin{displaymath}
\begin{array}{rcl@{\qquad\qquad}rcl@{\qquad\qquad}rcl}
\hbox{\sl var}  &\colon   & \hbox{\sl vr} \rightarrow \tm
& \hbox{\sl abs}  &\colon   & (\hbox{\sl vr} \rightarrow \tm) \rightarrow \tm
& \hbox{\sl app}  &\colon   & \tm \rightarrow \tm \rightarrow \tm
\enspace ,
\end{array}
\end{displaymath}
and the corresponding definition of {\sl typeof} would be
\begin{displaymath}
\begin{array}{rcl@{\qquad\qquad}rcl}
\hbox{\sl typeof}_{vr} &\colon & \hbox{\sl vr} \rightarrow \ty \rightarrow \oo
& \hbox{\sl typeof} &\colon & \tm \rightarrow \ty \rightarrow \oo \\
\end{array}
\end{displaymath}
\begin{displaymath}
\begin{array}{rcl}
\typeof{(\hbox{\sl var} \; X)}{T}
  & \defeq & {\hbox{\sl typeof}_{vr} \; X \; T} \\
\typeof{(\abs{M})}{(\arr{T}{U})}
  & \defeq & {\forall x(\hbox{\sl typeof}_{vr} \; x \; T
                        \oimp \typeof{(M \, x)}{U})} \\
\typeof{(\app{M}{N})}{T}
  & \defeq & {\exists u(\typeof{M}{(\arr{u}{T})} \land
                                \typeof{N}{u})}
\enspace .
\end{array}
\end{displaymath}
This is similar to our use of the two predicates {\sl hyp} and
{\sl conc} in our encoding of intuitionistic logic in  
Section~\ref{sec:impl-seq}.
Notice that the type $\tm$ does not occur negatively in the type of
any of its constructors, nor does the predicate {\sl typeof} occur
negatively in its definition.
This allows Coq to automatically construct induction principles for
$\tm$ and {\sl typeof}.
Since object-level variable binding is still represented by meta-level
$\lambda$-abstraction, the object language still inherits 
$\alpha$-equivalence from the meta-language.
Because the abstraction is over the type {\sl vr}, however, meta-level
$\beta$-reduction cannot be used for substitution.\footnote{Here we are
comparing the object system encodings.
It is true that our explicit eigenvariable encoding style requires an 
explicit definition of substitution {\em for the specification logic}.
So at the specification logic level of our framework, we too lose some of 
the benefits of higher-order abstract syntax.
However, at the level of the object system, we use a true higher-order 
abstract syntax encoding with all of its benefits.
Since we expect there to be only a few specification logics, but many
object systems, it seems worth putting the extra effort into the
specification logic to reap the benefit for the object systems.}
These approaches also lessen the power of the meta-level cut rule
as a reasoning tool.
Suppose that
$\forall x(\hbox{\sl typeof}_{vr} \; x \; T \oimp \typeof{(M \, x)}{U})$ and
$\typeof{N}{T}$ are derivable.
In contrast to our encoding, it is not immediate that substituting $N$
for $(\hbox{\sl var} \, x)$ in $(M \, x)$ yields a term $M'$ such that
$\typeof{M'}{U}$ is derivable.
Thus of the three key benefits to higher-order abstract syntax, they
only retain $\alpha$-conversion.
In addition, the Coq type $(\hbox{\sl vr} \rightarrow \tm)$ includes
functions besides those expressible as $\lambda$-terms, so the type
$\tm$ includes expressions that do not encode terms of the
object language.
They avoid these {\em exotic} terms through the definition and use of
a validation predicate.
The term language of $\FOLDN$, unlike that of Coq, does not include
primitive recursion, so these exotic terms do not arise in our framework.

Despeyroux, Pfenning, and Sch\"urmann \citeyear{despeyroux97tlca} address
the problem of exotic terms by using a modal operator to distinguish
the types of parametric functions (expressible as $\lambda$-terms)
from the types of arbitrary functions.
As a result, their calculus allows primitive recursive functionals
while preserving the adequacy of higher-order abstract syntax
encodings.
This represents a start toward a logical framework supporting
meta-theoretic reasoning, higher-order abstract syntax, and the
judgments-as-types principle.
In such a framework a derivation would be represented as a function
whose type is the derived property.
Thus the $\rightarrow$ type constructor must be rich enough to include
the mappings from derivations to derivations such as the realizations
of case analysis and induction.
Their work is orthogonal to our work presented in this paper.
We are not attempting to support the judgments-as-types principle, so
the types of our meta-logic are only used to encode syntactic
structure.
Thus we can restrict these types to include only $\lambda$-terms,
ensuring the adequacy of encodings in higher-order abstract syntax.
They, on the other hand, do not address the issue of induction
principles for higher-order abstract syntax, or more generally, the
issue of formal reasoning about higher-order abstract syntax
encodings.

Sch\"urmann and Pfenning \citeyear{schurmann98cade} construct a meta-logic
${\cal M}_2$ to reason about deductive systems represented in LF.
Their approach is similar in spirit to ours in that there are three
levels: the deductive system(s) under consideration, the logic in
which the deductive systems are encoded, and the logic in which
meta-theoretic analysis takes place.
The meta-logic ${\cal M}_2$ includes a case-analysis rule comparable to
our $\defL$ rule and a recursion rule that generalizes our $\natL$ rule.
Their intermediate logic, LF, includes dependent types, and so is 
richer than the intermediate logics we consider.
On the other hand, our meta-logic is a general framework capable of
supporting a variety of intermediate logics (such as
intuitutionistic and linear logics), whereas
${\cal M}_2$ is designed for the specific, fixed intermediate logic LF.

Still another strategy for meta-theoretic reasoning about higher-order
abstract syntax encodings is to perform each case of a proof in the
meta-logic, but verify the completeness of the proof outside the
logical framework.
Rohwedder and Pfenning \citeyear{pfenning92cade,rohwedder96esop}
investigate the design and implementation of such external validity
conditions.

Matthews seeks to reconcile the advantages of LF-style encodings with
the facilities for meta-theoretic analysis found in theories of
inductive definitions \cite{matthews97cade}.
His approach has some similarity to our own, in that he creates a
three-level hierarchy, with each level being encoded in the previous.
As in our approach, his top level contains a definition facility and
induction principles for reasoning about encodings at the next level.
However, his logic at the intermediate level contains only an
implication connective and no quantifiers.
Thus he does not address the treatment of object-level bound
variables, a major feature of higher-order abstract syntax and,
consequently, of our work.

\section{Conclusion}\label{sec:conclusion}

In this paper we have presented a single and simply motivated
meta-logic $\FOLDN$.
We used this meta-logic as the basis of a framework for formal reasoning
about systems expressed in higher-order abstract syntax, avoiding the apparent
tradeoff between the benefits of this representation technique and the
ability to perform meta-theoretic analyses of encodings.
We demonstrated this framework on encodings of three programming
languages encompassing both functional and imperative paradigms.
A number of significant theorems about these languages were derived in
this framework, including unicity of typing and subject reduction.
The flexibility of the framework was also shown through the use of
intuitionistic and linear specification logics.

The meta-logic $\FOLDN$ has also been used to reason about simulation
and bisimulation in abstract transition systems and CCS
\cite{mcdowell00tcsb}.
These transition systems did not contain binding operators, and 
so both the specification and reasoning was done in the meta-logic.
We have already begun using the techniques presented in the current
paper to extend that work to the setting of applicative bisimulation
\cite{abramsky90rtfp}.
It would also be interesting to use Howe's technique \cite{howe96ic}
to prove the congruence of bisimulation in our framework.

Additional work in analysis of programming languages along the lines
of \pcfpart\ could also be done.
Time precluded us from proving the determinacy of evaluation for
$\pcfref$, for example, and a transition semantics for the language
could be constructed and shown to be equivalent to the natural
semantics we constructed.
It would also be interesting to formalize other analyses;
\citeN{hannan92mscs}, for example, construct abstract machines from
operational semantics by applying a series of transformations and argue
informally that the transformations preserve correctness.
Richer languages could also be considered, including features such as
concurrency, exceptions, and polymorphism.
Linear logic has been used to specify such features in a manner that is 
suitable for use in our setting \cite{chirimar95phd,miller96tcs}.

The formal derivations described in this paper have been checked using
the Pi derivation editor of Lars-Henrik Eriksson
\cite{eriksson94cade}; see \citeN{mcdowell97phd} for a discussion of
the effectiveness of this editor for constructing $\FOLDN$ proofs.
An important next step in this line of work is to 
implement a theorem prover that provides
semi-automated assistance in proving $\FOLDN$ theorems.
Miller and Wajs are building
a prototype theorem prover named Iris \cite{wajs00mse} within
$\lambda$Prolog.

Finally, alternatives to the explicit eigenvariable encoding of
Section~\ref{sec:expl-eigen} could be explored.
Although this encoding supports the higher-order abstract syntax
representation of bound variables and allows substantial
meta-theoretic analysis, it does have some drawbacks.
The pervasive presence of the $\evars$ parameter representing the free
variable list is somewhat cumbersome, and numerous lemmas must be
proved to show that various properties are preserved by extensions of
this list or substitution for free variables.
The obvious alternative, a de Bruin-style encoding of free variables,
would require a similar amount of work and would not support the
higher-order abstract syntax representation for bound variables.
It is important to point out that this issue relates to the encoding
of the specification logic, not the object systems, of our framework.
Thus these lemmas need to be proved only once for any specification
logic, {\em not} for every object system, and so the representational
advantage of higher-order abstract syntax for the object systems is
preserved.

\begin{acks}

We would like to thank Frank Pfenning for helpful feedback on early
drafts of this work and Lars-Henrik Eriksson for making his Pi
derivation editor \cite{eriksson94cade} available to help check the
formal derivations described here.  Two anonymous referees provided
extensive comments that helped improve the presentation of this
paper.

\end{acks}

\begin{received}
Received Month Year;
revised Month Year;
accepted Month Year.
\end{received}

\end{document}